\documentclass[12pt]{article}
\pdfoutput=1
\usepackage{jheppub}

\usepackage{amsmath,bbm,array,amsfonts,graphicx,wrapfig,lscape,float,mathtools,multirow,longtable,tikz}

\newcommand{\be}{\begin{equation}}
\newcommand{\ee}{\end{equation}}
\newcommand{\beq}{\begin{equation}}
\newcommand{\beql}[1]{\begin{equation}\label{#1}}
\newcommand{\eeq}{\end{equation}}
\newcommand{\ba}{\begin{array}}
\newcommand{\ea}{\end{array}}
\newcommand{\bea}{\begin{eqnarray}}
\newcommand{\beal}[1]{\begin{eqnarray}\label{#1}}
\newcommand{\eea}{\end{eqnarray}}
\newcommand{\ben}{\begin{enumerate}}
\newcommand{\een}{\end{enumerate}}
\newcommand{\bean}{\begin{eqnarray*}}
\newcommand{\eean}{\end{eqnarray*}}
\newcommand{\eref}[1]{(\ref{#1})}
\newcommand{\sref}[1]{\S\ref{#1}}
\newcommand{\tref}[1]{Table~\ref{#1}}
\newcommand{\nn}{\nonumber}

\newcommand{\fref}[1]{Figure \ref{#1}}
\newcommand{\btab}[1]{\begin{tabular}{#1}}
\newcommand{\etab}{\end{tabular}}

\newcommand{\master}{\mathcal{F}^\flat}

\newcommand{\comment}[1]{}

\newcommand{\qed}{\nobreak \ifvmode \relax \else
      \ifdim\lastskip<1.5em \hskip-\lastskip
      \hskip1.5em plus0em minus0.5em \fi \nobreak
      \vrule height0.75em width0.5em depth0.25em\fi}

\newcommand{\Tr}{\text{Tr}}

\newcommand{\drawsquare}[2]{\hbox{%
\rule{#2pt}{#1pt}\hskip-#2pt
\rule{#1pt}{#2pt}\hskip-#1pt
\rule[#1pt]{#1pt}{#2pt}}\rule[#1pt]{#2pt}{#2pt}\hskip-#2pt
\rule{#2pt}{#1pt}}

\newcommand\encircle[1]{%
  \tikz[baseline=(X.base)] 
    \node (X) [draw, shape=circle, inner sep=0] {\strut #1};}

\newcommand{\fund}{~\raisebox{-.5pt}{\drawsquare{6.5}{0.4}}~}
\newcommand{\antifund}{~\overline{\raisebox{-.5pt}{\drawsquare{6.5}{0.4}}}~}

\title{2d (0,2) Quiver Gauge Theories and D-Branes}

\author[a,b]{Sebasti\'an Franco,}
\author[c]{Dongwook Ghim,}
\author[c,d,e,f]{Sangmin Lee,}
\author[f]{Rak-Kyeong Seong,}
\author[g]{Daisuke Yokoyama}

\affiliation[a]{
Physics Department, The City College of the CUNY \\
160 Convent Avenue, New York, NY 10031, USA}

\affiliation[b]{The Graduate School and University Center, The City University of New York  \\
365 Fifth Avenue, New York NY 10016, USA }

\affiliation[c]{
Department of Physics and Astronomy, Seoul National University, Seoul 151-747, Korea
}

\affiliation[d]{
Center for Theoretical Physics, Seoul National University, Seoul 151-747, Korea
}

\affiliation[e]{
College of Liberal Studies, Seoul National University, Seoul 151-742, Korea
}

\affiliation[f]{
School of Physics, Korea Institute for Advanced Study, Seoul 130-722, Korea
}

\affiliation[g]{
Department of Mathematics, King's College London, The Strand, London WC2R 2LS, United Kingdom.
}

\emailAdd{sfranco@ccny.cuny.edu}
\emailAdd{sg1841@snu.ac.kr}
\emailAdd{sangmin@snu.ac.kr}
\emailAdd{rkseong@kias.re.kr}
\emailAdd{daisuke.yokoyama@kcl.ac.uk}

\preprint{
{\footnotesize
\begin{flushright}
CCNY-HEP-15-03 \\
SNUTP15-003\\
KIAS-P15026 \\
KCL-MTH-15-04
\end{flushright}
}
}

\abstract{We initiate a systematic study of $2d$ $(0,2)$ quiver gauge theories on the worldvolume of D1-branes probing singular toric Calabi-Yau 4-folds. We present an algorithm for efficiently calculating the classical mesonic moduli spaces of these theories, which correspond to the probed geometries. We also introduce a systematic procedure for constructing the gauge theories for arbitrary toric singularities by means of partial resolution, which translates to higgsing in the field theory. Finally, we introduce Brane Brick Models, a novel class of brane configurations that consist of D4-branes suspended from an NS5-brane wrapping a holomorphic surface, tessellating a 3-torus. Brane Brick Models are the $2d$ analogues of Brane Tilings and allow a direct connection between geometry and gauge theory.
}

\begin{document}

\maketitle

\section{Introduction}

Engineering gauge theories in various dimensions and with different amounts of supersymmetry in terms of branes in string and M-theory is a fruitful approach for studying their dynamics.

$2d$ $\mathcal{N}=(0,2)$ theories are interesting for various reasons. It is reasonable to expect that although they have only two supercharges, it is possible to make considerable progress in understanding their dynamics, thanks to the control provided by chirality, holomorphy and anomalies. In addition, they are central in the worldsheet description of heterotic models. 

This work concentrates on the realization of $2d$ $\mathcal{N}=(0,2)$ theories in terms of branes. Our main goals are to understand in detail the gauge theories on D1-branes probing arbitrary toric singular Calabi-Yau 4-folds and to develop T-dual brane setups for them analogous to brane tilings. The study of $4d$  $\mathcal{N}=1$ gauge theories on D3-branes probing toric singular Calabi-Yau 3-folds in terms of brane tilings \cite{Hanany:2005ve,Franco:2005rj} has been an extreme success until now, laying out a path to follow for $2d$ $(0,2)$ theories. 

In order to guide our quest for $2d$ $(0,2)$ theories, let us recount the main developments that culminated in the discovery of brane tilings. At first, understanding gauge theories living on D3-branes probing singularities of abelian orbifolds of $\mathbb{C}^3$ \cite{Douglas:1996sw,Douglas:1997zj,Douglas:1997de} led to the identification of gauge theories for non-orbifold singularities via partial resolution \cite{Morrison:1998cs,Beasley:1999uz,Feng:2000mi}. The resulting large catalogue of explicit examples paved the way towards understanding basic structures of quiver gauge theories corresponding to various toric Calabi-Yau 3-folds \cite{Klebanov:1998hh,Beasley:2001zp,Feng:2000mi,Feng:2002fv,Feng:2004uq}. In parallel, it was argued in \cite{Hanany:1997tb,Hanany:1998it} that brane boxes, which are periodic arrays of orthogonal NS5-branes on $T^2$ from which stacks of D5-branes are suspended, are related by T-duality to D3-branes on $\mathbb{C}^3/\mathbb{Z}_n \times \mathbb{Z}_m$ orbifolds. Brane boxes provided valuable insights towards better understanding the brane constructions under T-duality, but can be seen now as little more than efficient bookkeeping devices for the restricted set of orbifold theories. 

The true breakthrough came with the discovery of brane tilings \cite{Hanany:2005ve,Franco:2005rj}. Brane tilings are the actual configurations of NS5- and D5-branes that are T-dual to D3-branes on {\it arbitrary} toric CY$_3$ singularities.\footnote{Brane boxes can be regarded as certain degenerate limits of the brane tilings associated with orbifolds.} Moreover, they have shed light on the connection between the geometry of the toric singularities and the corresponding gauge theories, in both directions \cite{Hanany:2005ve,Franco:2005rj,Hanany:2005ss,Feng:2005gw,Franco:2006gc,Gulotta:2008ef}.

In contrast, the understanding of $2d$ $(0,2)$ theories in terms of branes remains considerably underdeveloped. Our present knowledge is limited to D1-branes over abelian orbifolds of $\mathbb{C}^4$ \cite{Mohri:1997ef} and the T-dual configurations of brane boxes on $T^3$ \cite{GarciaCompean:1998kh}.\footnote{While inspiring, $2d$ brane boxes suffer from limitations that are similar to the ones of their $4d$ counterparts.} In recent years there has been substantial progress in our understanding of the field theory side, making the quest for a brane realization of these theories even timelier. An incomplete list of recent developments includes $c$-extremization \cite{Benini:2012cz,Benini:2013cda}, triality \cite{Gadde:2013lxa} and new ideas on dimensional reduction from $4d$ $\mathcal{N}=1$ theories \cite{Kutasov:2013ffl,Kutasov:2014hha}.

This paper presents the first step in our program devoted to filling this gap, introducing the tools for constructing the $2d$ theories on D1-branes over arbitrary toric Calabi-Yau 4-folds. Additional progress in this program will be presented in future publications \cite{topub1,topub2}.

This work is organized as follows.  Section \sref{stheory} contains a brief review of $2d$ $(0,2)$ theories and section \sref{section_D1_cones} outlines the basic features of the setups of D1-branes over toric Calabi-Yau 4-folds that we study. Section \sref{sec:orbifold} is devoted to gauge theories for abelian orbifolds of $\mathbb{C}^4$ and discusses periodic quivers in their context. Section \sref{section_geometry_from_quivers} introduces the forward algorithm which is a systematic method for computing the classical mesonic moduli spaces of the $2d$ $(0,2)$ theories under consideration. In fact, there are no moduli spaces of vacua in $2d$. The mesonic moduli spaces we compute should be regarded, in the spirit of the Born-Oppenheimer approximation, as target spaces of non-linear sigma models \cite{Diaconescu:1997gu}. They correspond to the Calabi-Yau 4-folds probed by the D1-branes. Section \sref{section_CY3xC} investigates $2d$ $(2,2)$ theories for D1-branes on singularities of the form $\mathrm{CY}_3 \times \mathbb{C}$. We obtain them via dimensional reduction from the $4d$ $\mathcal{N}=1$ theories for the CY$_3$, verify that the CY$_3\times \mathbb{C}$ arises as the mesonic moduli space using the forward algorithm and introduce a lifting algorithm for constructing the corresponding periodic quivers from those of the $4d$ parent theories. Section \sref{section_higgsing} explains how partial resolution of singularities translates into higgsing of the corresponding gauge theories. We also explain how to systematically use higgsing in order to obtain gauge theories for arbitrary toric CY 4-folds. Explicit examples of theories obtained via partial resolutions are presented in section \sref{section_beyond_orbifolds}, including theories for singularities that are neither orbifolds nor of the form CY$_3\times \mathbb{C}$. Section \sref{section_brane_bricks} previews the results of \cite{topub1}, where the brane configurations that are T-dual to the D1-branes at toric singularities are going to be presented in full detail. These configurations, named brane brick models, establish a direct connection between CY$_4$ geometry and $2d$ $(0,2)$ quiver gauge theories. Conclusions and future directions are presented in section \sref{section_conclusions}.

\bigskip

\section{$2d$ (0,2) Field Theories \label{stheory}}

This section briefly reviews the general structure of $2d$ $(0,2)$ theories, mainly to establish notations for later sections. We will not discuss all terms in the Lagrangian but only some of its most salient features. We refer the reader to \cite{Witten:1993yc,GarciaCompean:1998kh,Gadde:2013lxa,Kutasov:2013ffl} for details. 

\subsection{Constructing $2d$ $(0,2)$ Theories}

\label{section_constructing_(0,2)_theories}

We describe these theories in terms of $2d$ $(0,2)$ superspace $(x^\alpha,\theta^+,\bar{\theta}^+)$, $\alpha=0,1$. Three types of multiplets are needed to construct a $2d$ $(0,2)$ 
gauge theory.\footnote{
We follow the conventions of \cite{GarciaCompean:1998kh}, 
except  
$\sqrt{2} \psi_+^\mathrm{there} = \psi_+^\mathrm{here}$, 
$\lambda_-^\mathrm{there}/\sqrt{2} = \lambda_-^\mathrm{here}$, 
$\Lambda^\mathrm{there}/\sqrt{2} = \Lambda^\mathrm{here}$.
}
The first one is the gauge multiplet, which contains the gauge boson $v_\alpha$ $(\alpha =0,1)$, the adjoint chiral fermions $\chi_-$, $\bar{\chi}_-$ 
and an auxiliary field $D$. We are not going to need the detailed 
structure of the multiplet. The second type is the chiral multiplet, 
\beal{chiralm}
\Phi = \phi + \theta^+ \psi_+ -i \theta^+ \bar{\theta}^+ D_+ \phi \,, 
\quad \overline{\mathcal{D}}_+ \Phi = 0 \,, 
\eea
where the on-shell degrees of freedom are a complex scalar $\phi$ and a chiral fermion $\psi_+$, and $\overline{\mathcal{D}}_+$ is a super-covariant derivative. 
The third type is the Fermi multiplet whose chirality condition 
may be deformed by a holomorphic function of chiral fields $E(\Phi_i)$, which introduces interactions among matter fields. We have 
\beal{fermim}
\Lambda = \lambda_- - \theta^+ G -i \theta^+ \bar{\theta}^+ D_+ \lambda_- 
- \bar{\theta}^+ E \,, 
\quad \overline{\mathcal{D}}_+ \Lambda = E(\Phi_i)\,.
\eea
Here, $G$ is an auxiliary field and the chiral fermion $\psi_-$ is the only on-shell degree of freedom. 

The kinetic terms for the Fermi multiplets and some interactions are included in 
\beal{LF}
L_F= \int d^2y \, d^2\theta \sum_a \left( \bar{\Lambda}_a \Lambda_a\right) \, ,
\eea
where $a$ runs over all Fermi fields in the theory.

Another way to add interactions is to use a $(0,2)$ analog of the superpotential, 
\beal{LJ}
L_J = -\int d^2y \, d\theta^+ \sum_a \left( \Lambda_a J_a(\Phi_i)|_{\bar{\theta}^+=0}\right)-h.c. \,,
\eea
where the $J_a(\Phi_i)$ are holomorphic functions of chiral fields. In a gauge theory, $E_a$ has the same gauge quantum numbers  
as $\Lambda_a$ while $J_a$ has the conjugate gauge quantum numbers. 
The deformed chirality condition \eqref{fermim} and 
the chirality of $L_J$ \eqref{LJ} requires $J$ and $E$-terms 
to satisfy an overall constraint, 
\beq
\sum_a \mathrm{tr} \left[ E_a(\Phi_i) J_a(\Phi_i) \right]= 0 \,.
\eeq
For the $2d$ $(0,2)$ theories which we are considering in the following sections, the above constraint is explicitly checked and confirmed. There is a full symmetry under the individual exchanges $J_a \leftrightarrow E_a$, which corresponds to exchanging $\Lambda_a$ for $\bar{\Lambda}_a$.

Upon integrating out the auxiliary fields $G_a$, $L_F$ and $L_J$ produce the scalar potential
\beq
V= \sum_a \left( \mathrm{tr}|E_a(\phi)|^2 +  \mathrm{tr}|J_a(\phi)|^2 \right)\,,
\label{V_JE}
\eeq
as well as interactions between scalars and pairs of fermions
\beq
V_Y = - \sum_{a,i} \mathrm{tr} \left( \bar{\lambda}_{-a}\frac{\partial E_a}{\partial \phi_j}\psi_{+j} + \lambda_{-a}\frac{\partial J_a}{\partial \phi_j}\psi_{+j} +   \mathrm{h.c.} \right) \,,
\label{VY}
\eeq 
which include the usual Yukawa couplings.

\bigskip

\subsection{Anomalies \label{section_anomalies}}

\begin{figure}[h]
	\centering
	\includegraphics[width=5.5cm]{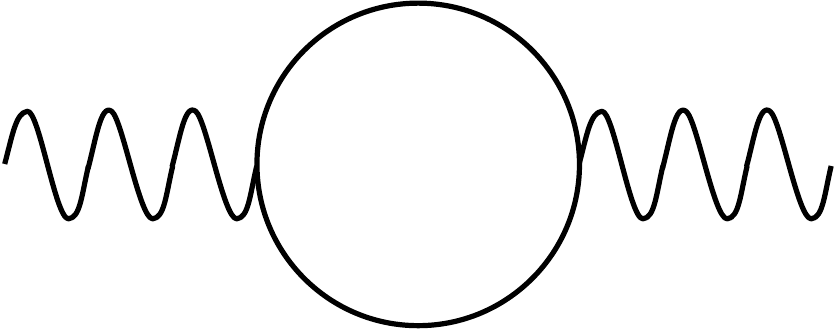}
\caption{A generic 1-loop diagram to compute anomalies in $2d$.}
	\label{2d-anomaly}
\end{figure}

As usual, anomalies play a central role in the analysis of quantum field theories. This section briefly reviews the subject of anomalies in $2d$ $(0,2)$ theories, both in gauge and global symmetries. In $2d$, anomalies follow from 1-loop diagrams of the general form shown in \fref{2d-anomaly}. 

Consistency of the theories at the quantum level requires cancellation of gauge anomalies. The gauge groups of the worldvolume theories on the probe D1-branes are $U(N_i)=SU(N_i)\times U(1)_i$. Below we consider the anomalies that do not automatically vanish, focusing on those groups and representations that appear in the $2d$ quiver theories of this work. 

\paragraph{Non-Abelian Anomalies.}

Let us first consider $SU(N_i)^2$ anomalies, where $SU(N_i)$ might be global or gauged. The corresponding anomaly is given by
\beq
\Tr[\gamma^3 J_{SU(N_i)} J_{SU(N_i)}],
\label{SU(N)^2_anomaly}
\eeq
where $\gamma^3$ is the chirality matrix in $2d$ and $J_{SU(N_i)}$ indicates the generator of the symmetry group in the representations in which each of the fields transforms.

The contributions of different fields and representations to \eref{SU(N)^2_anomaly} are given by:

\bigskip
\beq
\begin{array}{|c|c|c|}
\hline
\ \ \ \mbox{{\bf Multiplet}} \ \ \ & \ \ \ \mbox{{\bf Rep.}} \ \ \ & \ \ \ \mbox{{\bf Contribution}} \ \ \ \\ \hline
\mbox{{\bf Chiral}} & \fund \mbox{ or } \antifund & 1/2  \ \ \\
& \mbox{adj} & N_i  \\ \hline
\mbox{{\bf Fermi}} & \fund \mbox{ or } \antifund & -1/2  \\
& \mbox{adj} & -N_i  \\ \hline
\mbox{{\bf Vector}} &  \mbox{adj} & -N_i  \\ \hline
\end{array} 
\label{2d_anomaly_contributions}
\eeq

\medskip

\noindent Not surprisingly, given that anomalies are quadratic in 2d, things work quite differently than in $4d$. Two of the most notable differences we observe in \eref{2d_anomaly_contributions} are that the sign of the anomalies does not depend on the orientation of arrows and that the contributions from chiral adjoints and vector multiplets are non-vanishing.

\paragraph{Abelian Anomalies.}

In addition, it might be possible to have $U(1)_i^2$ anomalies, given by
\beq
\Tr[\gamma^3 Q_i^2],
\eeq
and mixed $U(1)_i U(1)_j$ anomalies, given by
\beq
\Tr[\gamma^3 Q_i Q_j] .
\eeq
As before, the $U(1)$ groups under consideration might be either global or gauged. 

Indeed, the theories we study generically have non-vanishing abelian gauge anomalies. In theories on D1-branes at singularities, we expect them to be cancelled by a generalized Green-Schwarz mechanism via interactions with bulk RR fields, as shown in \cite{Mohri:1997ef} for orbifolds of $\mathbb{C}^4$.\footnote{An alternative, field theoretic, approach to the cancellation of abelian gauge anomalies is the addition of appropriate matter, as explained in \cite{Gadde:2013lxa}. We are not going to pursue this direction.}

\paragraph{Anomalies for Quivers.}

The specific theories we are interested in are quiver theories. Nodes in these quivers correspond to $U(N_i)$ gauge groups. 
Let us now study in further detail the cancellation of non-abelian anomalies in the case of quivers. In order to allow for the possibility of multiple fields between a given pair of nodes, it is convenient to define: $n_{ij}^\chi :=$ number of chiral arrows from node $i$ to node $j$, $n_{ij}^F:=$ number of Fermi lines stretching between node $i$ and node $j$,\footnote{Due to the symmetry between $\Lambda_{a}$ and $\bar{\Lambda}_a$, Fermi fields are represented in the quiver diagram by lines without orientation.} $a_i^{\chi/F} :=$ number of adjoint chiral/Fermi lines attached to node $i$. Cancellation of $SU(N)_i^2$ anomalies then requires

\beq
\sum_{j\neq i} \left( n_{ji}^\chi N_j  +  n_{ij}^\chi N_j -  n_{ij}^F N_j \right) + 2 (a^{\chi}_i -  a^F_i)N_i = 2N_i ,
\label{anomaly_cancellation}
\eeq
for every node $i$.

In this article, we focus on the case in which all ranks are equal, i.e. $N_i=N$, which corresponds to a stack of $N$ regular D1-branes.\footnote{It would be very interesting to investigate more general anomaly-free rank assignments and their correspondence to fractional D1-branes. We leave this question for future work.} In this case, \eref{anomaly_cancellation} simplifies to
\beq
n^\chi_i-n^F_i=2 .
\label{simple_anomaly_cancellation}
\eeq
In this expression, $n^\chi_i$ and $n^F_i$ indicate the total number of incoming plus outgoing chiral and Fermi fields at node $i$, respectively. Since adjoint fields are represented in the quiver by lines that start and end on the same node, each of them contributes 2 to these numbers. 

It is interesting to notice that constructing quiver theories for which \eref{simple_anomaly_cancellation} is satisfied for all gauge groups {\it without} the addition of flavors (i.e. of matter in e.g. purely fundamental or anti-fundamental representations of gauge groups) and the associated flavor symmetries they would introduce is a challenging combinatorial problem. Remarkably, \eref{simple_anomaly_cancellation} is automatically satisfied by all the theories we construct.

\bigskip

\subsection{Triality and Dynamical SUSY Breaking}

As studied in \cite{Gadde:2013lxa}, $2d$ $(0,2)$ theories can dynamically break SUSY at low energies. A powerful tool for elucidating the IR dynamics of these theories is the {\it triality} introduced in \cite{Gadde:2013lxa}, which for this purpose plays a role similar to that of Seiberg duality in $4d$. We are not going to discuss the details of triality at all in this article. In fact its implementation in string theory constructions will be the subject of one of our future publications \cite{topub2}.

A basic property of triality is that the original theory is recovered only after acting with it three times on the same gauge group. Of course the space of dual theories grows considerably when there are multiple gauge groups.

A possible diagnostic for SUSY preserving theories advocated in \cite{Gadde:2013lxa} is that no negative ranks are generated in any duality frame. In the notation of \eref{anomaly_cancellation}, this occurs for nodes without adjoints for which $\sum_{j\neq i} n_{ji}^\chi N_j < N_i$ or $\sum_{j\neq i} n_{ij}^\chi N_j < N_i$.\footnote{Due to the form of the anomaly cancellation conditions in $2d$, it is possible for the numbers of incoming and outgoing chiral fields to be different.} This is analogous to what happens in $4d$. There, reaching a negative rank by Seiberg duality really means that the duality should not have been performed because it involved an $N_f<N_c$ gauge group which, in turn, generates an ADS superpotential that spontaneously breaks SUSY \cite{Affleck:1983mk}. As a cross-check, whenever this criterion is not met in $2d$, the equivariant index of the theory vanishes, also indicating the absence of a SUSY vacuum. The detailed dynamical process that triggers SUSY breaking is not yet known. 

Arbitrary $2d$ $(0,2)$ quivers generically break SUSY spontaneously for the reasons discussed above. Given this challenge, as discussed in \cite{Gadde:2013lxa}, it is desirable to come up with a (combinatorial) prescription for generating SUSY preserving theories. The approach discussed in this article, of realizing theories with regular D1-branes at singular toric CY 4-folds, gives rise to an infinite class of theories for which its natural to expect that SUSY is preserved.\footnote{This might no longer be true when considering fractional D1-branes} 

\bigskip

\section{$2d$ (0,2) Theories from D1-Branes over Toric CY$_4$ Cones}

\label{section_D1_cones}

The main goal of this article is to understand the gauge theories arising in the low energy limit of a stack of D1-branes probing a singular toric CY 4-fold. Among other things, we want to know how the gauge theory is determined by the CY 4-fold and, conversely, how the CY 4-fold is captured by the gauge theory.

Type IIB string theory on $\mathbb{R}^{1,1}\times\mathrm{CY}_4$ with parallel D1-branes spanning $\mathbb{R}^{1,1}$ preserves $(0,2)$ SUSY on the worldvolume of the D1-branes.\footnote{An alternative interesting approach for constructing $2d$ $(0,2)$ theories involves realizing them as compactifications of $6d$ theories on 4-manifolds \cite{Gadde:2013sca}.} 
Non-chiral SUSY enhancement occurs when the putative CY$_4$ 
contains $\mathbb{C}$ factors; $\mathbb{C}^4$, CY$_2\times\mathbb{C}^2$, 
CY$_3\times \mathbb{C}$ preserve $(8,8)$, $(4,4)$, $(2,2)$ SUSY, respectively. 
The D1-brane theory in these cases can be obtained from the
dimensional reduction of a higher dimensional D-brane theory. 
Chiral enhancement to $(0,4)$ SUSY arises from $\mathrm{CY}_2\times \mathrm{CY}_2$. Further chiral enhancement to $(0,6)$ or $(0,8)$ 
is possible for particular orbifold geometries. 

Among all CY$_4$ geometries, we restrict our attention to toric CY$_4$ cones. The tools from toric geometry make it easier to find the map between gauge theories and CY$_4$ geometries. 
The methods we use are similar to those from previous work on D3-branes probing toric CY$_3$ cones or M2-branes probing toric CY$_4$ cones. A notable difference is that the D3-brane or M2-brane theories 
at large $N$ offer examples of AdS/CFT in the weakly coupled gravity 
description, whereas the D1-brane theories 
under consideration here do not. 

The main diagnostic we use to check the map between gauge theories and geometries is the classical mesonic moduli space of the gauge theory. 
The classical mesonic moduli space is the geometry underlying 
the chiral ring of gauge-invariant single-trace operators 
modulo the relations coming from vanishing $J$- and $E$-terms. The moduli space is expected to reproduce the probed Calabi-Yau geometry. 

Our analysis is going to be entirely classical. The underlying Calabi-Yau cone is expected to capture robust properties of the quantum field theory. At present, we do not have a full understanding of the quantum effects in these theories in terms of branes. Early attempts in this direction can be found in \cite{Karch:1998sj}. This is certainly a direction worth studying in the future.

By definition, the isometry group of a toric CY$_4$ contains $U(1)^4$. This isometry translates into a global symmetry of the gauge theory. A linear combination of the four $U(1)$'s accounts for the R-symmetry of $(0,2)$ SUSY. The non-R $U(1)^3$ symmetry, which we call the {\it mesonic flavor symmetry}, is going to be exploited in later sections and in a forthcoming paper \cite{topub1} in order to introduce the notion of a periodic quiver. 

For some geometries, the isometry group is enhanced to a non-abelian group. For instance, the isometry of the $Q^{1,1,1}$ geometry, 
to be discussed in section \sref{sq111}, is $U(1)_R\times SU(2)^3$. In this example, the non-abelian global symmetry is not manifest at the level of the Lagrangian, but the chiral ring elements 
can be organized in multiplets of the global symmetry. It is convenient to enumerate the elements of the chiral ring 
by the Hilbert series \cite{Benvenuti:2006qr}. 
The Hilbert series is a function in terms of fugacities for the 
$U(1)^4$ global symmetry. When the global symmetry is non-abelian, 
the Hilbert series can be expanded in characters of 
the non-abelian group. Hilbert series will be presented in a forthcoming paper \cite{topub1}.

In order to fully understand the theories on D1-branes over general toric singularities, we first work our way through geometries for which the corresponding gauge theories are already well understood.  
First, we consider orbifold theories in section \sref{sec:orbifold}. Next, in section \sref{section_CY3xC} we study the $2d$ $(2,2)$ theories obtained by dimensional reduction of $4d$ $\mathcal{N}=1$ theories on D3-branes over toric CY 3-folds. Understanding both classes of theories is going to help us to develop the main ideas that apply to the generic case. In addition, these theories are connected to the ones for general toric singularities through RG flows.

\bigskip

\subsection{Basic Structure}

\label{section_basic_structure}

Below we mention some additional general properties of the theories under consideration.

\paragraph{Gauge Group.}

One piece of information that can directly be derived from the toric diagram of the probed geometry is the total number of nodes in the quiver, which corresponds to the number of independent ways of wrapping D-branes at the singularity. This is given by the number of minimal tetrahedra that fit into the toric diagram, i.e. $6 \times V$, with $V$ being the volume of the toric diagram.\footnote{The toric diagram of a toric CY$_4$ cone lives in a hyperplane and hence can be projected to $3d$. Here and throughout the paper we consider such $3d$ toric diagrams.}

All the examples we consider in this article satisfy this property. It is true by construction for the orbifolds and dimensionally reduced theories. It is also true for the new models that are obtained by partial resolution, i.e. higgsing from the gauge theory point of view. However, we note that it is possible to find theories that seem to violate this rule by higgsing. Our expectation is that such theories suffer from inconsistencies similar to the ones observed in some otherwise healthy looking $4d$ $\mathcal{N}=1$ theories for D3-branes over toric CY 3-folds. In $4d$, such inconsistencies have been understood from a variety of perspectives, including: global symmetries \cite{Hanany:2005ss}, zig-zag paths in the associated brane tilings \cite{Hanany:2005ss} and algebraic conditions \cite{2011arXiv1104.1592B,2010arXiv1012.5449I}. We are not going to consider such theories any further in this work. In $4d$, inconsistencies can be eliminated by further higgsing. A more thorough investigation of consistency conditions is certainly worthwhile but exceeds the scope of this work.

\paragraph{$J$- and $E$-Terms.}

Another special property of the gauge theories associated with toric CY 4-folds is the structure of the $J$- and $E$-terms. All of them take the form of differences between two monomials in chiral fields. We refer to this property as the {\it toric form} of these functions. Once again, this property is present in orbifolds and dimensionally reduced theories and preserved by the higgsing associated with partial resolutions that connects them to other theories. This property is the $2d$ analogue of the {\it toric superpotentials} of $4d$ $\mathcal{N}=1$ gauge theories for D3-branes on toric CY 3-folds \cite{Feng:2002zw}. It plays an important role in the emergence of the probed geometry as the classical moduli space of the gauge theory.

\section{Orbifold Theories \label{sec:orbifold}} 

This section studies the gauge theories on D1-branes over orbifolds of $\mathbb{C}^4$, which have been originally considered in \cite{GarciaCompean:1998kh,Mohri:1997ef}. This infinite family of theories gives a nice class of examples to start developing our ideas in the following sections. In sections \sref{section_orbifold_exmaples0} and \sref{section_orbifold_exmaples}, we investigate orbifolds in further detail, explaining how the geometry arises as the mesonic moduli space of the corresponding gauge theories.

\bigskip

\subsection{D1-Branes over $\mathbb{C}^4$ \label{section_orbifold_exmaples0}}

\begin{figure}[h]
	\centering
	\includegraphics[height=4cm]{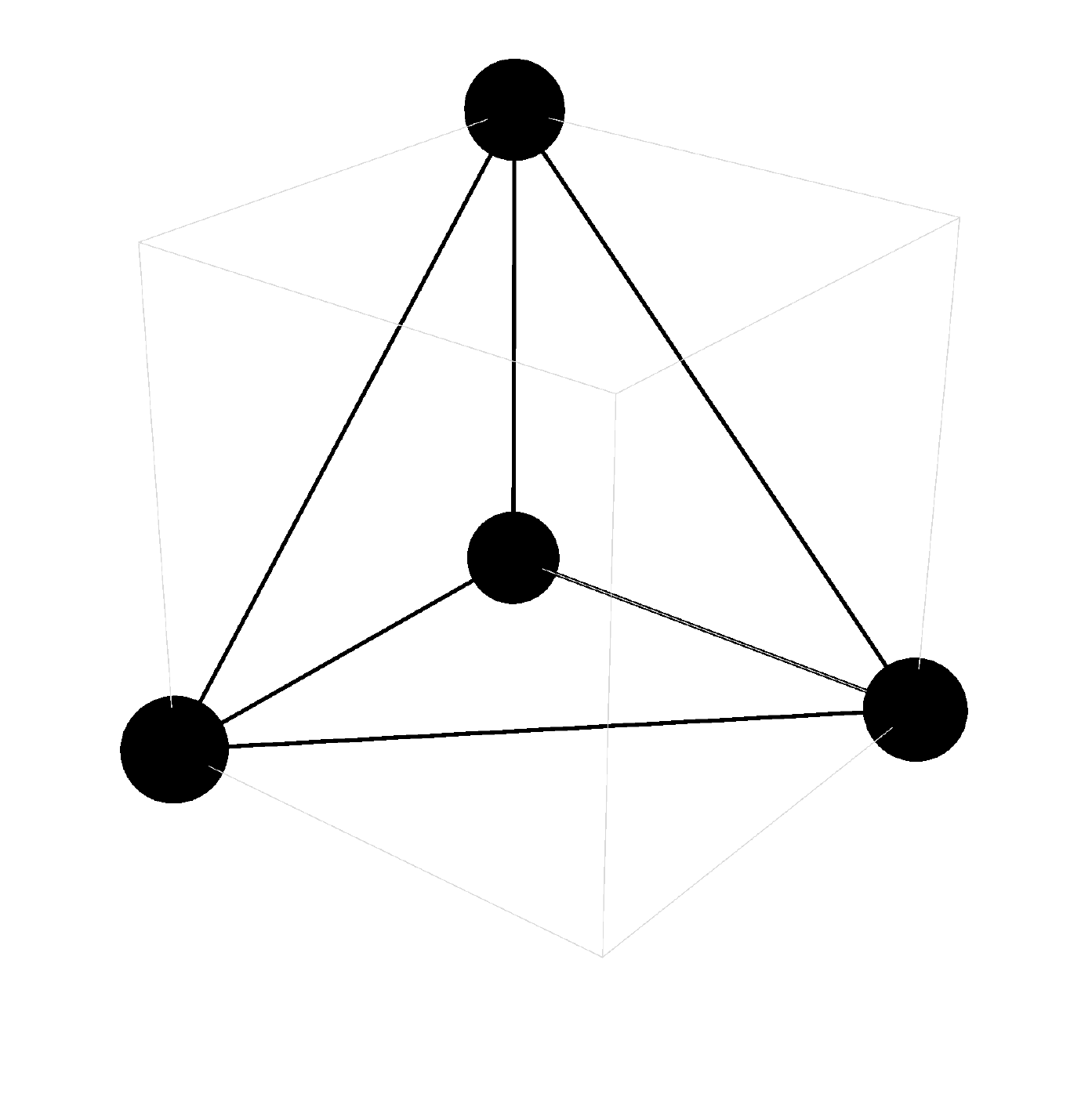}
\vspace{-1cm}\caption{Toric diagram of $\mathbb{C}^4$.}
	\label{fig:C4-toric}
\end{figure}

Let us first consider the simplest gauge theory engineered with D1-branes, that of a stack of D1-branes on $\mathbb{C}^4$, whose toric diagram is shown in \fref{fig:C4-toric}. 

\begin{figure}[h]
	\centering
	\includegraphics[width=5.5cm]{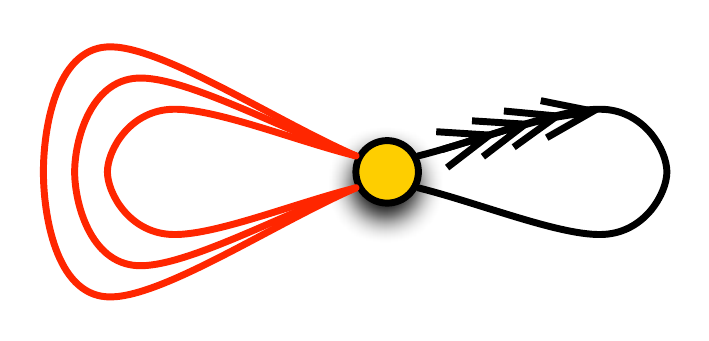}
\caption{The quiver diagram for $N$ D1-branes over $\mathbb{C}^4$. It consists of a single $U(N)$ gauge node, four adjoint chiral fields (shown in black) and three Fermi fields (shown in red).}
	\label{c4quiver}
\end{figure}

This theory can be obtained by dimensional reduction of $4d$ $\mathcal{N}$=4 SYM.\footnote{Dimensional reduction is discussed at length in \sref{section_dimensional_reduction}, here we just present the result.} The $2d$ theory has an enhanced $(8,8)$ non-chiral SUSY. In $(0,2)$ language, the theory contains the vector multiplet associated with a single $U(N)$ gauge group, four chiral fields ($X$, $Y$, $Z$ and $D$) and three Fermi fields ($\Lambda^{(i)}$, $i=1,2,3$), all transforming in the adjoint representation of the gauge group. All this information can be summarized in the quiver shown in \fref{c4quiver}.

The corresponding $J$- and $E$-terms are as follows
\beq
\begin{array}{rclccclcc}
& & \ \ \ \ \ \ \ \ \ \ \ \ J & & & & \ \ \ \ \ \ \ \ \ \ \ \ E & & \\
\Lambda^{(1)} : & \ \ \ & Y \cdot Z - Z \cdot Y & = & 0 & \ \ \ \ & D \cdot X - X \cdot D & = & 0 \\ 
\Lambda^{(2)} : & \ \ \ & Z \cdot X - X \cdot Z & = & 0 & \ \ \ \ & D \cdot Y - Y \cdot D & = & 0 \\ 
\Lambda^{(3)} : & \ \ \ & X \cdot Y - Y \cdot X & = & 0 & \ \ \ \ & D \cdot Z - Z \cdot D & = & 0 
\end{array}
\label{EJ_C^4}
\eeq
We observe that this theory nicely agrees with our general discussion in section \sref{section_basic_structure}. The single gauge group is in agreement with the fact that the toric diagram in \fref{fig:C4-toric} consists of a single unit tetrahedron. In addition, the relations in \eref{EJ_C^4} have the general structure anticipated for theories corresponding to toric Calabi-Yau 4-folds.

\bigskip

\subsection{Orbifolds \label{section_orbifold_exmaples}}

Abelian orbifolds of $\mathbb{C}^D$ have been extensively studied from a geometric viewpoint for various dimensions $D$ in \cite{Muto:1997pq,Kachru:1998ys,Lawrence:1998ja,Davey:2010px,Hanany:2010ne}. Here we are interested in determining the gauge theories on D1-branes over orbifolds of $\mathbb{C}^4$. These take the most general form $\mathbb{C}^4/\mathbb{Z}_{n_1}\times \mathbb{Z}_{n_2}\times \mathbb{Z}_{n_3}$ where the order of the orbifold group is $n=n_1 n_2 n_3$. $\mathbb{C}^4$ is parameterized by complex coordinates $x_k$ with $k=1,\dots,4$ and the orbifold action is determined by three integer 4-vectors
\beal{es200a1}
(a_1,a_2,a_3,a_4)~,~(b_1,b_2,b_3,b_4)~,~ (c_1,c_2,c_3,c_4)~,~
\eea
where $a_k$, $b_k$ and $c_k$ relate to the coordinate action on $\mathbb{C}^4$ as follows
\beal{es200a2}
x_k \mapsto e^{2\pi i \left(\frac{a_k}{n_1}+\frac{b_k}{n_2}+\frac{c_k}{n_3}  \right)} x_k ~.~
\eea
The action satisfies
\beal{mod-n}
\sum_{k=1}^4 a_k = 0 \bmod n_1 \,, 
\quad 
\sum_{k=1}^4 b_k = 0 \bmod n_2 \,,
\quad 
\sum_{k=1}^4 c_k = 0 \bmod n_3 \,.
\eea
A simple sub-class of orbifolds of $\mathbb{C}^4$ is $\mathbb{C}^4/\mathbb{Z}_n$, with orbifold action $(a_1,a_2,a_3,a_4)$. The action on the $\mathbb{C}^4$ coordinates $x_k$ is given by 
\beal{es200a3}
x_k \mapsto e^{2\pi i \frac{a_k}{n}} x_k~.~
\eea

Let us now determine how the orbifold action on $\mathbb{C}^4$ translates into the corresponding gauge theory.
First, there are four types of bifundamental chiral fields in the quiver of the $2d$ $(0,2)$ theory which we are considering. These are related to the four complex coordinates of $\mathbb{C}^4$,
\beal{es200a4}
x_1 \leftrightarrow X_{i,j} ~,~
x_2 \leftrightarrow Y_{i,j} ~,~
x_3 \leftrightarrow Z_{i,j} ~,~
x_4 \leftrightarrow D_{i,j} ~,~
\eea
where the subindices $i$ and $j$ indicate that the corresponding field transforms in the fundamental representation of $U(N)_i$ and in the anti-fundamental representation of $U(N)_j$. For the most general orbifolds with $n_1,n_2,n_3 >1$, we take 
the gauge group indices $i$ and $j$ to carry three components, $i=(i_1,i_2,i_3)$, $j=(j_1,j_2,j_3)$ where $i_k,j_k = 0,\dots,n_k-1$. The four types of chiral fields connect gauge groups according to 
the following rule:
\beal{es200a5}
X_{i,j} ~:~ && j = i + (a_1,b_1,c_1) \bmod (n_1,n_2,n_3)  ~,~
\nn\\
Y_{i,j} ~:~  && j = i + (a_2,b_2,c_2) \bmod (n_1,n_2,n_3)  ~,~
\nn\\
Z_{i,j} ~:~ &&  j = i + (a_3,b_3,c_3) \bmod (n_1,n_2,n_3)  ~,~
\nn\\
D_{i,j} ~:~ && j = i + (a_4,b_4,c_4) \bmod (n_1,n_2,n_3)  ~.~
\eea
One can always map the 3-vector labels $i,j$ to single integer labels such that $i,j=1,\dots,n$. Note that this map is trivial for $\mathbb{C}^4/\mathbb{Z}_{n}$ orbifolds. In addition, there are three types of Fermi fields in bifundamental representations, which can be labeled
\beal{es200a5b}
\Lambda_{i,j}^{(1)} ~:~ && j=  i - (a_2+ a_3,b_2+b_3 ,c_2 + c_3) \bmod (n_1,n_2,n_3) ~,~
\nn\\
\Lambda_{i,j}^{(2)} ~:~ && j= i - (a_3+ a_1,b_3+b_1 , c_3 + c_1) 
\bmod (n_1,n_2,n_3) ~,~
\nn\\
\Lambda_{i,j}^{(3)} ~:~ && j= i - (a_1+ a_2,b_1+b_2 , c_1 + c_2) 
\bmod (n_1,n_2,n_3) ~.
\eea
It is also possible to write the $J$- and $E$-terms associated with Fermi fields for an arbitrary orbifold in a closed form. For simplicity, let us take $\mathbb{C}^4/\mathbb{Z}_n$ with orbifold action $(a_1,a_2,a_3,a_4)$ with $\sum_{k} a_k = 0 \bmod n$, and a single component index $i$ defined modulo $n$. The $J$- and $E$-terms are
\beal{es200a6}
\Lambda_{i,i-a_2-a_3}^{(1)} :
&~J_{i-a_2-a_3,i}^{(1)} &=
Y_{i-a_2-a_3, i-a_3} \cdot Z_{i-a_3, i} - Z_{i-a_2-a_3, i-a_2} \cdot Y_{i-a_2, i} ~,~
\nn\\
&~E_{i,i-a_2-a_3}^{(1)} ~&= 
D_{i,i+a_4} \cdot X_{i+a_4,i-a_2-a_3} - X_{i,i+a_1} \cdot D_{i+a_1,i-a_2-a_3}
~,~
\nn\\
\Lambda_{i,i-a_3-a_1}^{(2)} :
&~J_{i-a_3-a_1,i}^{(2)} &=
Z_{i-a_1- a_3, i-a_1} \cdot X_{i-a_1, i} - X_{i-a_1-a_3, i-a_3} \cdot Z_{i-a_3,i}  ~,~
\nn\\
&~E_{i,i-a_3-a_1}^{(2)} &=
D_{i,i+a_4} \cdot Y_{i+a_4,i-a_3-a_1} - Y_{i,i+a_2} \cdot D_{i+a_2, i-a_3-a_1}
~,~
\nn\\
\Lambda_{i,i-a_1-a_2}^{(3)} :
&~J_{i-a_1-a_2,i}^{(3)} &=
X_{i-a_1-a_2,i-a_2} \cdot Y_{i-a_2,i} - Y_{i-a_1-a_2,i-a_1} \cdot X_{i-a_1,i}  ~,~
\nn\\
&~E_{i,i-a_1-a_2}^{(3)} &=
D_{i,i+a_4} \cdot Z_{i+a_4, i -a_1-a_2} - Z_{i,i+a_3} \cdot D_{i+a_3, i-a_1-a_2}
~.~
\eea
Extending these expressions to general orbifolds of $\mathbb{C}^4$ is straightforward. Clearly, all abelian orbifolds of $\mathbb{C}^4$ satisfy the condition \eref{simple_anomaly_cancellation} for the cancellation of non-abelian anomalies.

\bigskip

\subsection{Periodic Quivers: a First Encounter}

Periodic quivers for $2d$ theories were originally introduced in \cite{GarciaCompean:1998kh} in the context of orbifolds of $\mathbb{C}^4$. They are standard quivers living on a 3-torus. For orbifolds, they are motivated by the Type IIA brane box constructions consisting of D4-branes and three kinds of NS5-branes \cite{GarciaCompean:1998kh} that are, roughly speaking, dual to the quiver. 

In this section we review the periodic quivers for orbifolds of $\mathbb{C}^4$ following \cite{GarciaCompean:1998kh} and add some new insights to their construction.  We will later see that all gauge theories on D1-branes over toric CY 4-folds are associated with a periodic quiver on $T^3$. In fact, the three periodic directions of the torus correspond to the $U(1)^3$ mesonic flavor symmetry of the gauge theory that follows from three of the $U(1)$ isometries of the toric CY$_4$, with the fourth one being related to the R-symmetry of the gauge theory. In a forthcoming paper  \cite{topub1}, we will fully explore how periodic quivers are dual to brane configurations that generalize, and improve, brane box models. These brane setups are previewed in section \sref{section_brane_bricks}. T-dualizing such brane configuration along the three periodic directions of the 3-torus, one obtains the stack of D1-branes at the toric CY$_4$ singularity. 

\bigskip

\subsubsection{$\mathbb{C}^4$ and General Orbifolds \label{sorbifolds}}

\begin{figure}[h]
	\centering
	\includegraphics[height=5.5cm]{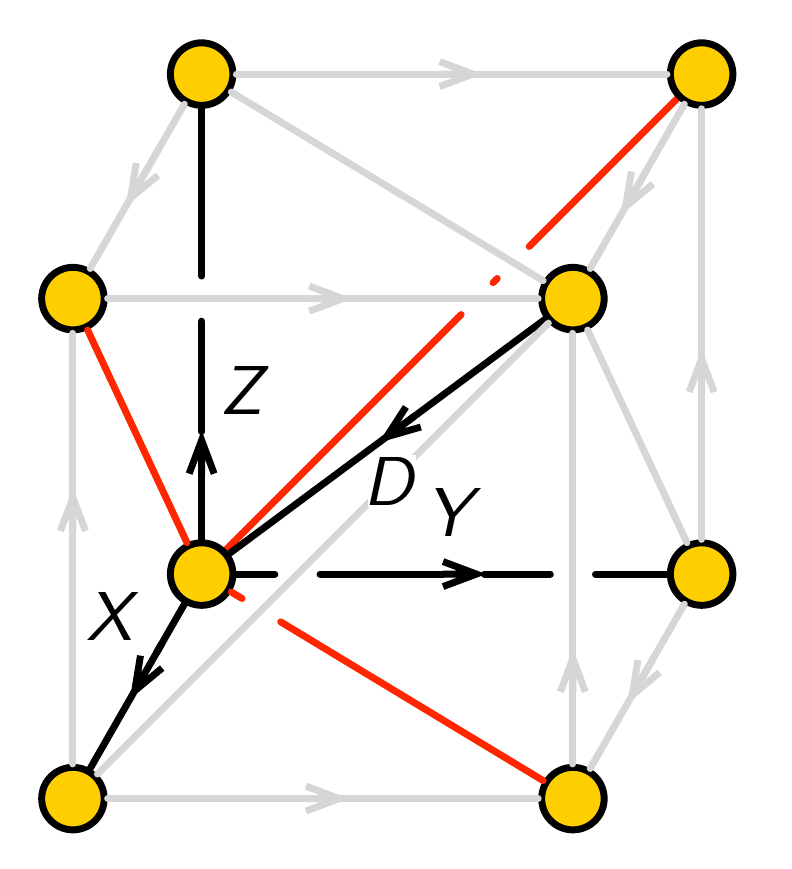}
\caption{A unit cell of the periodic quiver of $\mathbb{C}^4$, which is periodically identified along the three axes.}
	\label{fig:bcc}
\end{figure}

Let us first consider the periodic quiver for D1-branes over $\mathbb{C}^4$. The corresponding $2d$ theory has been reviewed in section \sref{section_orbifold_exmaples0} with the corresponding $J$- and $E$-terms given in \eref{EJ_C^4}. The standard quiver in \fref{c4quiver} can be turned into the periodic one shown in \fref{fig:bcc}.

A remarkable illustration of the power of periodic quivers is that the gauge theories for arbitrary abelian orbifolds of $\mathbb{C}^4$ can be constructed by appropriately stacking $n$ copies of the quiver in \fref{fig:bcc}, with $n$ being the order of the orbifold group. The precise way in which these cubes are stacked is determined by the action of the orbifold group given in \eref{es200a5} and \eref{es200a5b}. 

\begin{figure}[ht!!]
\begin{center}
\resizebox{0.5\hsize}{!}{
\includegraphics[trim=0cm 0cm 0cm 0cm,totalheight=10 cm]{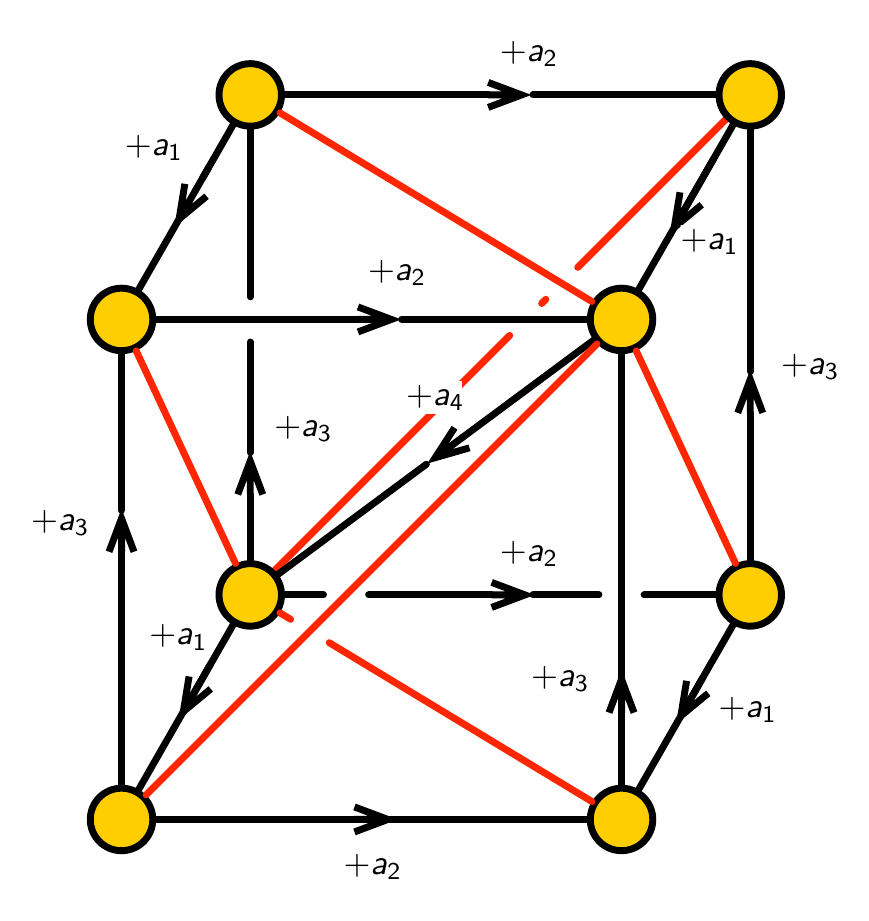}
}  
\caption{The local structure of the periodic quiver for an arbitrary $\mathbb{C}^4/\mathbb{Z}_{n}$ orbifold with action $(a_1,a_2,a_3,a_4)$, where $a_4 = -a_1-a_2-a_3$.  All integer labels on the gauge nodes of the periodic quiver are considered modulo $n$. On the chiral field arrows, we indicate the shift in the node label between the two endpoints.
\label{forbifold}}
 \end{center}
 \end{figure}

For illustration, \fref{forbifold} shows the local structure of the periodic quiver for an arbitrary $\mathbb{C}^4/\mathbb{Z}_{n}$ orbifold with action $(a_1,a_2,a_3,a_4)$, where $a_4 = -a_1-a_2-a_3$. All integer labels on the gauge nodes of the periodic quiver are considered modulo $n$. A unit cell in this quiver contains $n$ nodes. Its precise shape depends on the orbifold action, and is determined by the periodicity of the resulting node labels.

\bigskip 

\begin{figure}[ht!!]
\begin{center}
\resizebox{0.4\hsize}{!}{
\includegraphics[trim=0cm 0cm 0cm 0cm,totalheight=10 cm]{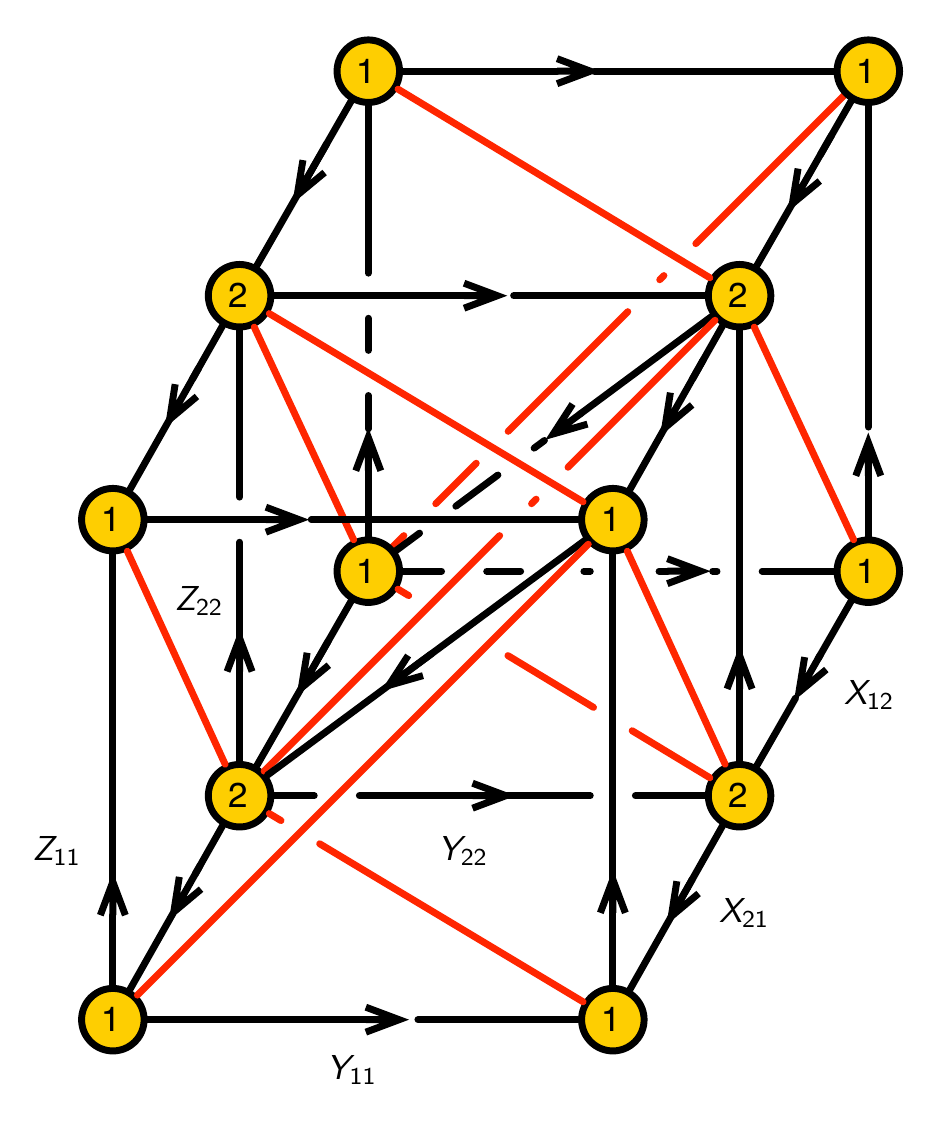}
}  
\caption{
A unit cell in the periodic quiver for $\mathbb{C}^4/\mathbb{Z}_{2}$ with action $(1,0,0,1)$.
\label{forbifoldZ2}}
 \end{center}
 \end{figure}

\paragraph{Example.} In order to show how the general local structure of the orbifold periodic quiver shown in \fref{forbifold} determines a unit cell, let us consider the example of $\mathbb{C}^4/\mathbb{Z}_{2}$ with action $(1,0,0,1)$. It is equally straightforward to apply our results to more general orbifold actions. This example is particularly simple, since the orbifold group acts only on two of the complex planes. In other words, the orbifold takes the form $\mathbb{C}^2/\mathbb{Z}_{2} \times \mathbb{C}^2$, which has $SU(2)$ holonomy, leading to enhanced $(4,4)$ SUSY. In fact, the theory can be obtained by dimensional reduction from a $4d$ $\mathcal{N}=2$ orbifold theory. The corresponding periodic quiver is shown in \fref{forbifoldZ2}.

The corresponding $J$- and $E$-terms are given by 
\beq
\begin{array}{rclccclcc}
& & \ \ \ \ \ \ \ \ \ \ \ \ \ J & & & & \ \ \ \ \ \ \ \ \ \ \ \ \ E & & \\
\Lambda_{11 } : & \ \ \ &  Y_{11} \cdot Z_{11} - Z_{11} \cdot Y_{11} & = & 0 & \ \ \ \ & D_{12} \cdot X_{21} - X_{12} \cdot D_{21} & = & 0 \\
\Lambda_{22 } : & \ \ \ & Y_{22} \cdot Z_{22} - Z_{22} \cdot Y_{22}& = & 0 & \ \ \ \ & D_{21} \cdot X_{12} - X_{21} \cdot D_{12} & = & 0 \\
\Lambda_{12}^{1} : & \ \ \ & Z_{22} \cdot X_{21} - X_{21} \cdot Z_{11}  & = & 0 & \ \ \ \ & D_{12} \cdot Y_{22} - Y_{11} \cdot D_{12} & = & 0 \\
\Lambda_{21}^{1} : & \ \ \ & Z_{11} \cdot X_{12} - X_{12} \cdot Z_{22} & = & 0 & \ \ \ \ & D_{21} \cdot Y_{11} - Y_{22} \cdot D_{21} & = & 0 \\  
\Lambda_{12}^{2} : & \ \ \ &  X_{21} \cdot Y_{11} - Y_{22} \cdot X_{21} & = & 0 & \ \ \ \ & D_{12} \cdot Z_{22} - Z_{11} \cdot D_{12} & = & 0 \\
\Lambda_{21}^{2}  : & \ \ \ & X_{12} \cdot Y_{22} - Y_{11} \cdot X_{12} & = & 0 & \ \ \ \ & D_{21} \cdot Z_{11} - Z_{22} \cdot D_{21} & = & 0 
\end{array}
\label{es200a7}
\eeq

\bigskip

\subsubsection{$J$- and $E$-Terms from Plaquettes}

$J$- and $E$-terms are conveniently encoded in terms of {\it plaquettes}. We define a plaquette as a closed loop in the quiver consisting of an arbitrary number of chiral fields and a single Fermi field. The chiral fields in a plaquette form an oriented path with two endpoints, which are connected by the Fermi field, closing the loop. 

The relative orientation between the path of chiral fields in a plaquette and the gauge quantum numbers of the corresponding Fermi field depend on whether it corresponds to a $J$- or an $E$-term. Since $E$-terms are auxiliary components of Fermi fields, they share the same gauge quantum numbers. Hence, the path of chiral fields in a plaquette representing a contribution to an $E_{ij}$ term has the orientation of the corresponding Fermi field $\Lambda_{ij}$. Conversely, $J$-terms transform in the conjugate representation of the corresponding Fermi field. As a result, a contribution to a $J_{ji}$ term associated with a Fermi field $\Lambda_{ij}$ corresponds to a plaquette in which the path of chiral fields goes from node $j$ to $i$.

\begin{figure}[h]
	\centering
	\includegraphics[width=6.5cm]{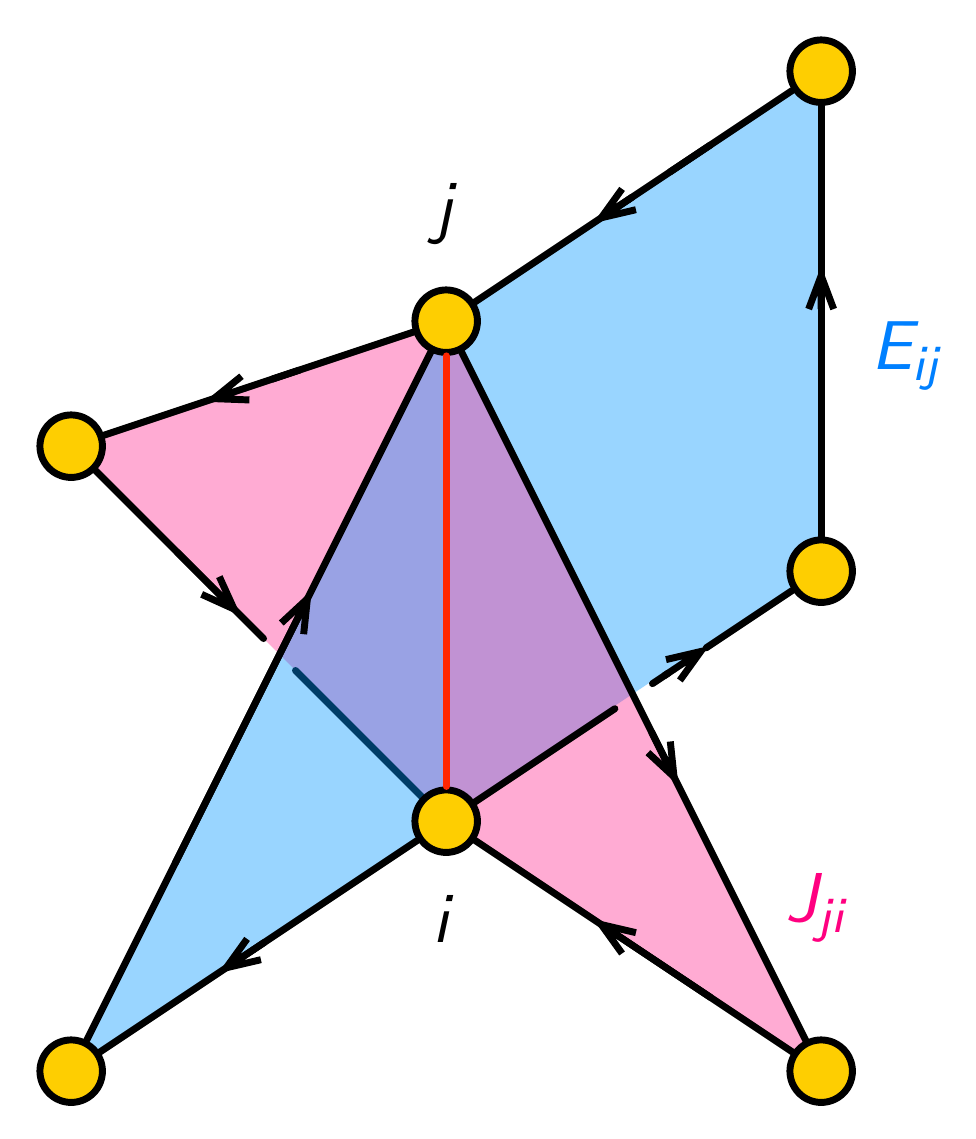}
\caption{Schematic representation of the four plaquettes giving rise to the $J$- and $E$-terms associated with a Fermi field $\Lambda_{ij}$.}
	\label{general_EJ_plaquettes}
\end{figure}

Keeping our discussion general, we allow plaquettes to have arbitrary numbers of chiral fields. For the special case of orbifolds, every plaquette has two chiral fields.
As we discussed in section \sref{section_basic_structure}, the $J$- and $E$-terms in toric theories have a special form: they are differences of two monomials in chiral fields. This implies that in these theories every Fermi field participates in four plaquettes, which separate into two pairs with opposite orientations. \fref{general_EJ_plaquettes} schematically shows the basic structure of plaquettes for a given Fermi field. 

A remarkable property of periodic quivers, which we study in the following sections, is that periodic quivers beautifully encode the $J$- and $E$-terms of the theory in terms of certain ``minimal" plaquettes. This is the $2d$ analogue of the fact that plaquettes in periodic quivers on a 2-torus precisely encode the superpotential of $4d$ toric theories \cite{Franco:2005rj}.\footnote{For $4d$ theories, plaquettes are defined as closed oriented loops of chiral fields.}

\bigskip

\section{Geometry from $2d$ (0,2) Quivers}

\label{section_geometry_from_quivers}

A central goal of this work is to identify $2d$ gauge theories which arise on the worldvolume of D1-branes probing toric Calabi-Yau 4-folds. The primary tool we use in order to establish this correspondence is the classical mesonic moduli space of the gauge theory. For a stack of $N$ D1-branes, we take the expectation values of the scalars in chiral fields to be proportional to the identity. The resulting mesonic moduli space takes the form $\mathcal{M}_N=\mbox{Sym}^N \mathcal{M}_1$, namely the symmetric product of $N$ copies of the moduli space associated with a single D1-brane $\mathcal{M}_1$. $\mathcal{M}_1$ corresponds to the probed CY$_4$. For this reason, in what follows, we focus on the moduli space of abelian theories. 

This section presents an algorithm for extracting the information about the geometry of the toric CY$_4$. As emphasized above, the toric CY$_4$ is the vacuum moduli space of the $2d$ gauge theories which have been discussed in section \sref{stheory}. This algorithm is a direct generalization of the so called {\it forward algorithm} for $4d$ $\mathcal{N}=1$ quiver gauge theories on the worldvolume of D3-branes probing toric Calabi-Yau 3-folds \cite{Feng:2000mi,Franco:2005rj}.

\bigskip

\subsection{The Forward Algorithm for $2d$ (0,2) Theories \label{sfa} \label{sec:geometry}}

\begin{figure}[h]
\begin{center}
\resizebox{0.4\hsize}{!}{
\includegraphics[trim=0cm 0cm 0cm 0cm,totalheight=10 cm]{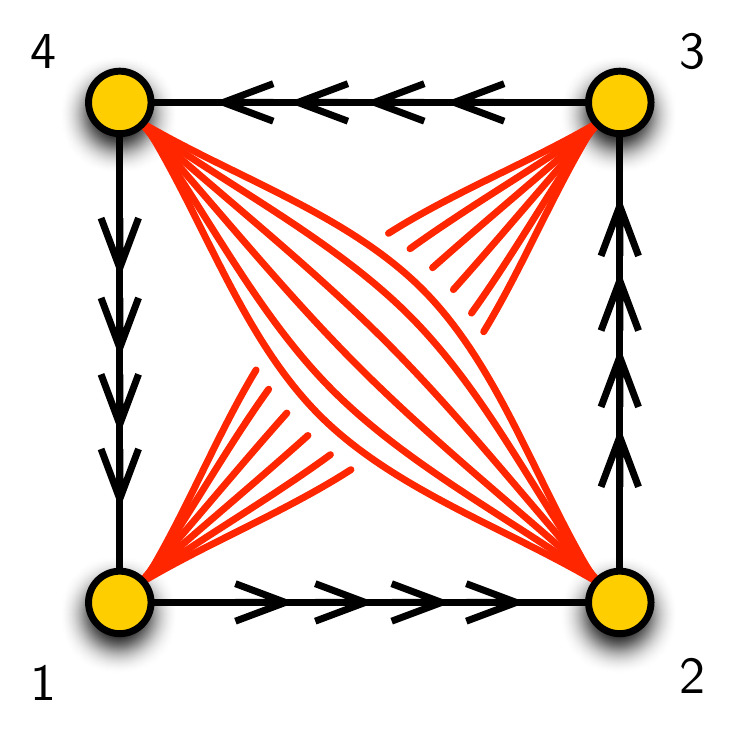}
}  
\caption{Quiver diagram for the $\mathbb{C}^4/\mathbb{Z}_{4}$ orbifold with action $(1,1,1,1)$.
\label{fz20011b}}
 \end{center}
 \end{figure}

This section presents the forward algorithm for $2d$ $(0,2)$ theories.\footnote{Our discussion also applies to theories with enhanced SUSY.} It is convenient to illustrate its implementation in terms of an explicit example. To do so, we consider $\mathbb{C}^4/\mathbb{Z}_{4}$ $(1,1,1,1)$ whose quiver diagram is shown in \fref{fz20011b}. The corresponding $J$- and $E$-terms are presented in \eref{es80e1}.\footnote{For brevity, we provide standard quivers instead of periodic ones for this model and the orbifold examples of \sref{section_orbifold_examples}. We will return to periodic quivers in \sref{section_CY3xC}.}

The forward algorithm involves the following steps:

\medskip

\begin{itemize}
\item {\bf \underline{$J$- and $E$-terms and the $K$-matrix}:} The first step in the algorithm is to take the $J$- and $E$-terms associated with all Fermi fields. A special feature of the theories under consideration, as discussed in section \sref{section_basic_structure}, is that these relations equate a monomial to a monomial. 

Let us denote by $X_{m}$, with $m=1,\dots,n^\chi$, all the chiral fields in the quiver.\footnote{For simplicity, in what follows, we often use chiral fields as synonyms of their scalar components.}  
Remarkably, the space of solutions to the vanishing $J$- and $E$-terms can be expressed in terms of $G+3$ independent chiral fields $v_k$, with $G$ being the number of gauge groups in the quiver. $X_m$ can thus be expressed as 
\beal{es80e0c}
X_{m} = \prod_{k} v_{k}^{K_{mk}} ~,~
\eea
where $m=1,\dots,n^\chi$ and $k=1,\dots,G+3$. $K$ is an $n^\chi \times(G+3)$-dimensional matrix which precisely encodes the relations from the vanishing $J$- and $E$-terms. In toric geometry, the $K$-matrix defines a cone $\textbf{M}^{+}$ generated by non-negative linear combinations of $n^\chi$ $\vec{K}_m$ vectors in $\mathbb{Z}^{G+3}$. 

For the example $\mathbb{C}^4/\mathbb{Z}_{4}$ $(1,1,1,1)$, the $J$- and $E$-terms are given by
\beq
\begin{array}{rclccclcc}
& & \ \ \ \ \ \ \ \ \ \ \ \ \ J & & & & \ \ \ \ \ \ \ \ \ \ \ \ \ E & & \\
\Lambda_{13}^{1}  : & \ \ \ &  X_{34}\cdot Y_{41} - Y_{34}\cdot X_{41} & = & 0 & \ \ \ \ & D_{12}\cdot Z_{23} - Z_{12}\cdot D_{23}& = & 0 \\
\Lambda_{13}^{2}  : & \ \ \ &  Y_{34}\cdot Z_{41} - Z_{34}\cdot Y_{41} & = & 0 & \ \ \ \ & D_{12}\cdot X_{23} - X_{12}\cdot D_{23}& = & 0 \\
\Lambda_{13}^{3}  : & \ \ \ &  Z_{34}\cdot X_{41} -X_{34}\cdot Z_{41} & = & 0 & \ \ \ \ &  D_{12}\cdot Y_{23} - Y_{12}\cdot D_{23}& = & 0 \\
\Lambda_{31}^{1}  : & \ \ \ &  X_{12}\cdot Y_{23} - Y_{12}\cdot X_{23} & = & 0 & \ \ \ \ & D_{34}\cdot Z_{41} - Z_{34}\cdot D_{41}& = & 0 \\
\Lambda_{31}^{2}  : & \ \ \ &  Y_{12}\cdot Z_{23} - Z_{12}\cdot Y_{23} & = & 0 & \ \ \ \ & D_{34}\cdot X_{41} - X_{34}\cdot D_{41}& = & 0 \\
\Lambda_{31}^{3}  : & \ \ \ &  Z_{12}\cdot X_{23} -X_{12}\cdot Z_{23} & = & 0 & \ \ \ \ &  D_{34}\cdot Y_{41} - Y_{34}\cdot D_{41}& = & 0 \\
\Lambda_{24}^{1}  : & \ \ \ &  Z_{41}\cdot X_{12} -X_{41}\cdot Z_{12} & = & 0 & \ \ \ \ &  D_{23}\cdot Y_{34} - Y_{23}\cdot D_{34}& = & 0 \\
\Lambda_{24}^{2}  : & \ \ \ &  X_{41}\cdot Y_{12} - Y_{41}\cdot X_{12} & = & 0 & \ \ \ \ &  D_{23}\cdot Z_{34} - Z_{23}\cdot D_{34}& = & 0 \\ 
\Lambda_{24}^{3}  : & \ \ \ &  Y_{41}\cdot Z_{12} - Z_{41}\cdot Y_{12} & = & 0 & \ \ \ \ &  D_{23}\cdot X_{34} - X_{23}\cdot D_{34}& = & 0 \\
\Lambda_{42}^{1}  : & \ \ \ & X_{23}\cdot Y_{34} - Y_{23}\cdot X_{34} & = & 0 & \ \ \ \ &  D_{41}\cdot Z_{12} - Z_{41}\cdot D_{12}& = & 0 \\
\Lambda_{42}^{2}  : & \ \ \ &  Y_{23}\cdot Z_{34} - Z_{23}\cdot Y_{34} & = & 0 & \ \ \ \ &  D_{41}\cdot X_{12} - X_{41}\cdot D_{12}& = & 0 \\
\Lambda_{42}^{3}  : & \ \ \ &  Z_{23}\cdot X_{34} -X_{23}\cdot Z_{34} & = & 0 & \ \ \ \ &  D_{41}\cdot Y_{12} - Y_{41}\cdot D_{12}& = & 0 \\
\end{array}
\label{es80e1}
\eeq

\smallskip

\noindent where on the left there are the corresponding Fermi fields.

All chiral fields can be expressed in terms of $4+3=7$ variables $v_k$ as follows
\beq
\begin{array}{cclccclccclcccl}
D_{12} & = & v_1 & \ \ \ \ \ & D_{23} & = & v_2 & \ \ \ \ \ & D_{34} & = & v_3 & \ \ \ \ \ & D_{41} & = & v_4 \\ [.2cm]
X_{12} & = & v_5 & & X_{23} & = & \dfrac{v_2 v_5}{v_1} & & X_{34} & = & \dfrac{v_3 v_5}{v_1} & & X_{41} & = & \dfrac{v_4 v_5}{v_1} \\ [.4cm]
Y_{12} & = & v_6 & & Y_{23} & = & \dfrac{v_2 v_6}{v_1} & & Y_{34} & = & \dfrac{v_3 v_6}{v_1} & & Y_{41} & = & \dfrac{v_4 v_6}{v_1} \\ [.4cm]
Z_{12} & = & v_7 & & Z_{23} & = & \dfrac{v_2 v_7}{v_1} & & Z_{34} & = & \dfrac{v_3 v_7}{v_1} & & Z_{41} & = & \dfrac{v_4 v_7}{v_1}
\end{array}
\label{es80e3}
\eeq
We then define $K$ following \eref{es80e0c} and obtain
\beal{es80e2}
\resizebox{0.9\columnwidth}{!}{$
(K^t)_{(G+3)\times E} = 
\left(
\begin{array}{c|cccccccccccccccc}
\;& D_{12}& D_{23}& D_{34}& D_{41}& X_{12}& X_{23}& X_{34}& X_{41}& Y_{12}& Y_{23}& Y_{34}& Y_{41}& Z_{12}& Z_{23}& Z_{34}& Z_{41}
\\
\hline
v_1 = D_{12} & 1 & 0 & 0 & 0 & 0 & -1 & -1 & -1 & 0 & -1 & -1 & -1 & 0 & -1 & -1 & -1 \\
v_2 = D_{23} & 0 & 1 & 0 & 0 & 0 & 1 & 0 & 0 & 0 & 1 & 0 & 0 & 0 & 1 & 0 & 0 \\
v_3 = D_{34} & 0 & 0 & 1 & 0 & 0 & 0 & 1 & 0 & 0 & 0 & 1 & 0 & 0 & 0 & 1 & 0 \\
v_4 = D_{41} & 0 & 0 & 0 & 1 & 0 & 0 & 0 & 1 & 0 & 0 & 0 & 1 & 0 & 0 & 0 & 1 \\
v_5 = X_{12} & 0 & 0 & 0 & 0 & 1 & 1 & 1 & 1 & 0 & 0 & 0 & 0 & 0 & 0 & 0 & 0 \\
v_6 = Y_{12} & 0 & 0 & 0 & 0 & 0 & 0 & 0 & 0 & 1 & 1 & 1 & 1 & 0 & 0 & 0 & 0 \\
v_7 = Z_{12} & 0 & 0 & 0 & 0 & 0 & 0 & 0 & 0 & 0 & 0 & 0 & 0 & 1 & 1 & 1 & 1 \\
\end{array}
\right)
$}
~.~
\nn\\
\eea

\bigskip

\item {\bf \underline{The dual cone and the $P$-matrix}:} Entries in the $K$-matrix are integers but can be negative. This means that the chiral fields can involve negative powers of the independent fields $v_k$. Such negative powers can be avoided by constructing a new set of variables as follows. We define the dual cone $\mathbf{N}^{+}$ in terms of vectors $\vec{T}_m$, $m=1,\ldots,c$, which can be combined into a $(G+3)\times c $-dimensional matrix that we call $T$.\footnote{Notice that the number of independent vectors $c$ spanning the dual cone is a priori unknown and results from the actual computation.} This matrix has positive integer entries and is defined through the condition
\beal{es81e1}
\vec{K} \cdot \vec{T} \geq 0 ~.~
\eea
We can now use $T$ to trade the independent chiral fields $v_k$ for a new set of fields $p_\alpha$, $\alpha = 1,\dots, c$, such that only positive powers are involved, 
\beal{es81e2}
v_k = \prod_{\alpha} p_{\alpha}^{T_{k\alpha}} ~.~
\eea
The $p_\alpha$ are interpreted as GLSM fields \cite{Witten:1993yc} in the toric description of the moduli space.\footnote{We will reserve the term GLSM for the theories constructed via the forward algorithm which, unlike the original $2d$ quivers, involve unconstrained fields.}

Combining \eref{es80e0c} and \eref{es81e2}, all chiral fields can be expressed in terms of the GLSM fields according to
\beq
X_{m} = \prod_{\alpha} p_{\alpha}^{P_{m\alpha}} ~,~
\label{map_X_p}
\eeq
where we have defined the $n^\chi \times c$-dimensional $P$-matrix as
\beal{es81e3}
P_{n^\chi \times c} = K_{n^\chi \times (G+3)} \cdot  T_{(G+3) \times c} ~.~
\eea
Above, the labels on the matrices indicate their dimensions. The map in \eref{map_X_p} only involves positive powers and is fully controlled by the $P$-matrix, which is going to play a central role in connecting gauge theory and geometry in the following sections.

For the $\mathbb{C}^4/\mathbb{Z}_{4}$ example, $T$ is given by
\beal{es81e5}
T= 
\left(
\ba{l|cccccccc}
\; & p_1 & p_2 & p_3 & p_4 & q_1 & q_2 & q_3 & q_4 \\
\hline
v_1 = D_{12} & 1 & 0 & 0 & 0 & 1 & 0 & 0 & 0 \\
v_2 = D_{23} & 1 & 0 & 0 & 0 & 0 & 1 & 0 & 0 \\
v_3 = D_{34} & 1 & 0 & 0 & 0 & 0 & 0 & 1 & 0 \\
v_4 = D_{41} & 1 & 0 & 0 & 0 & 0 & 0 & 0 & 1 \\
v_5 = X_{12} & 0 & 1 & 0 & 0 & 1 & 0 & 0 & 0 \\
v_6 = Y_{12} & 0 & 0 & 1 & 0 & 1 & 0 & 0 & 0 \\
v_7 = Z_{12} & 0 & 0 & 0 & 1 & 1 & 0 & 0 & 0 \\
\ea
\right)~.~
\eea
This matrix encodes the following map
\beq
\begin{array}{cclccclcccl}
v_1 & = & p_1 q_1 & \ \ \ \ \ \ & v_2 & = & p_1 q_2 & \ \ \ \ \ \ & v_3 & = & p_1 q_3 \\ [.2cm]
v_4 & = & p_1 q_4 & \ \ \ \ \ & v_5 & = & p_2 q_1 & \ \ \ \ \ & v_6 & = & p_3 q_1 \\ [.2cm]
 & & & & v_7 & = & p_4 q_1 & \ \ \ \ \ & & & 
 \end{array}
\label{es81e6}
\eeq

The resulting $P$-matrix is given by
\beal{es81e7}
P=
\left(
\begin{array}{c|cccccccc}
\; & p_1 & p_2 & p_3 & p_4 & q_1 & q_2 & q_3 & q_4 \\
\hline
D_{12} & 1 & 0 & 0 & 0 & 1 & 0 & 0 & 0 \\
D_{23} & 1 & 0 & 0 & 0 & 0 & 1 & 0 & 0 \\
D_{34} & 1 & 0 & 0 & 0 & 0 & 0 & 1 & 0 \\
D_{41} & 1 & 0 & 0 & 0 & 0 & 0 & 0 & 1 \\
X_{12} & 0 & 1 & 0 & 0 & 1 & 0 & 0 & 0 \\
X_{23} & 0 & 1 & 0 & 0 & 0 & 1 & 0 & 0 \\
X_{34} & 0 & 1 & 0 & 0 & 0 & 0 & 1 & 0 \\
X_{41} & 0 & 1 & 0 & 0 & 0 & 0 & 0 & 1 \\
Y_{12} & 0 & 0 & 1 & 0 & 1 & 0 & 0 & 0 \\
Y_{23} & 0 & 0 & 1 & 0 & 0 & 1 & 0 & 0 \\
Y_{34} & 0 & 0 & 1 & 0 & 0 & 0 & 1 & 0 \\
Y_{41} & 0 & 0 & 1 & 0 & 0 & 0 & 0 & 1 \\
Z_{12} & 0 & 0 & 0 & 1 & 1 & 0 & 0 & 0 \\
Z_{23} & 0 & 0 & 0 & 1 & 0 & 1 & 0 & 0 \\
Z_{34} & 0 & 0 & 0 & 1 & 0 & 0 & 1 & 0 \\
Z_{41} & 0 & 0 & 0 & 1 & 0 & 0 & 0 & 1 \\
\end{array}
\right) ~.~
\eea
This implies the following map between GLSM and chiral fields
\beq
\begin{array}{cclccclccclcccl}
D_{12} & = & p_1 q_1 & \ \ \ \ \ &
D_{23} & = & p_1 q_2 & \ \ \ \ \ &
D_{34} & = & p_1 q_3 & \ \ \ \ \ &
D_{41} & = & p_1 q_4 \\ [.2cm]
X_{12} & = & p_2 q_1 & \ \ \ \ \ &
X_{23} & = & p_2 q_2 & \ \ \ \ \ &
X_{34} & = & p_2 q_3 & \ \ \ \ \ &
X_{41} & = & p_2 q_4 \\ [.2cm]
Y_{12} & = & p_3 q_1 & \ \ \ \ \ &
Y_{23} & = & p_3 q_2 & \ \ \ \ \ &
Y_{34} & = & p_3 q_3 & \ \ \ \ \ &
Y_{41} & = & p_3 q_4 \\ [.2cm]
Z_{12} & = & p_4 q_1 & \ \ \ \ \ &
Z_{23} & = & p_4 q_2 & \ \ \ \ \ &
Z_{34} & = & p_4 q_3 & \ \ \ \ \ &
Z_{41} & = & p_4 q_4
\end{array}
\label{es81e8}
\eeq

\bigskip

\item {\bf\underline{GLSM charges from $J$- and $E$-terms}:} As already mentioned, the GLSM fields $p_\alpha$ encoded in the $P$-matrix can be considered as a new basis of fields parameterizing the space of solutions to the vanishing $J$- and $E$-terms. The resulting relations between chiral fields can be neatly implemented by introducing $U(1)$ gauge symmetries to the GLSM and assigning appropriate charges to its fields. These charges are encoded in a $(c-(G+3))\times c$ -dimensional charge matrix $Q_{JE}$ which is the kernel of the $P$-matrix,
\beal{es82e1}
(Q_{JE})_{(c-(G+3))\times c} = \ker(P) ~.~
\eea

For our $\mathbb{C}^4/\mathbb{Z}_4$ example, we get
\beal{es82e2}
Q_{JE} = \left(
\ba{cccccccc}
p_1 & p_2 & p_3 & p_4 & q_1 & q_2 & q_3 & q_4 \\
\hline
 1 & 1 & 1 & 1 & -1 & -1 & -1 & -1 \\
\ea
\right)~.~
\eea
Above, a single $U(1)$ symmetry is necessary in order to capture the $J$- and $E$-terms.

\bigskip

\item {\bf \underline{GLSM charges from D-terms}:}  The final step in the computation of the moduli space is to impose the vanishing D-terms. Recall, we are focusing on the abelian theory. The $U(1)$ gauge charges of the chiral fields are encoded by the $G\times n^\chi$-dimensional incidence matrix $d$ of the quiver, where $G$ is the number of nodes. The incidence matrix is defined as follows
\beal{es83e1}
d_{ai} = \left\{
\ba{cl}
+1 & \text{if $X_i$ is a fundamental of node } a
\\
-1 & \text{if $X_i$ is an anti-fundamental of node } a
\\
0 & \text{if $X_i$ is an adjoint of node } a
\ea
\right.
\eea

Since in the theories under consideration all chiral fields always transform in bifundamental or adjoint representations, the incidence matrix satisfies
\beal{es83e2}
\sum_a d_{ai} = 0 ~.~
\eea
As a result, only $G-1$ rows of the incidence matrix are independent. It is thus sufficient to focus on a $(G-1)\times n^\chi$-dimensional matrix $\Delta$, obtained from $d$ by deleting any of its rows.

Next, it is possible to establish $U(1)$ charges for the GLSM fields $p_\alpha$ that would result in the desired charges for the chiral fields via the map in \eref{map_X_p}.
The $U(1)$ charges of the $p_\alpha$'s are summarized in a $(G-1)\times c$-dimensional matrix $Q_{D}$ that satisfies
\beal{es83e3}
\Delta_{(G-1)\times n^\chi} = (Q_{D})_{(G-1)\times c}\cdot P^{t}_{c\times n^\chi} ~.~
\eea
Notice that, in general, this equation does not fix $Q_D$ uniquely. The final moduli space is however independent of which solution is used.

From \fref{fz20011b}, we determine the following quiver incidence matrix 
\beal{es83e4}
\resizebox{0.85\columnwidth}{!}{$
d= \left(
\ba{l|cccccccccccccccc}
\;&D_{12} & D_{23} & D_{34} & D_{41} & X_{12} & X_{23} & X_{34} & X_{41} & Y_{12} & Y_{23} & Y_{34} & Y_{41} & Z_{12} & Z_{23} & Z_{34} & Z_{41}
\\
\hline
\encircle{1} & 1 & 0 & 0 & -1 & 1 & 0 & 0 & -1 & 1 & 0 & 0 & -1 & 1 & 0 & 0 & -1 \\
\encircle{2} & -1 & 1 & 0 & 0 & -1 & 1 & 0 & 0 & -1 & 1 & 0 & 0 & -1 & 1 & 0 & 0 \\
\encircle{3} & 0 & -1 & 1 & 0 & 0 & -1 & 1 & 0 & 0 & -1 & 1 & 0 & 0 & -1 & 1 & 0 \\
\encircle{4} & 0 & 0 & -1 & 1 & 0 & 0 & -1 & 1 & 0 & 0 & -1 & 1 & 0 & 0 & -1 & 1 \\
\ea
\right) ~.~
$}
\nn\\
\eea
Combining \eref{es81e7} and \eref{es83e4}, we obtain the following possible $Q_D$ matrix for the GLSM fields
\beal{es83e5}
Q_D
= \left(
\ba{cccccccc}
p_1 & p_2 & p_3 & p_4 & q_1 & q_2 & q_3 & q_4 \\
\hline
 0 & 0 & 0 & 0 & -1 & 1 & 0 & 0 \\
 0 & 0 & 0 & 0 & 0 & -1 & 1 & 0 \\
 0 & 0 & 0 & 0 & 0 & 0 & -1 & 1 \\
\ea
\right) ~.~
\eea

\bigskip

\item {\bf \underline{CY$_4$ moduli space and its toric diagram}:}  We have translated the relations associated with vanishing $J$- and $E$-terms into $U(1)$ charges given by the matrix $Q_{JE}$. Similarly, $D$-terms are captured by the charges in $Q_D$. We define the total charge matrix 
\beal{es84e1}
(Q_{t})_{(c-4)\times c} = (
(Q_{JE})_{(c-(G+3))\times c} ~,~
(Q_{D})_{(G-1)\times c}
)
\eea
as the concatenation of $Q_{JE}$ and $Q_D$. The total charge matrix is $(c-4)\times c$-dimensional, with $c$ being the number of GLSM fields. 

We can now determine the moduli space of the gauge theory. The kernel of $Q_D$ is a $4\times c$-dimensional matrix,
\beal{es84e2}
G = \ker(Q_t) ~.~
\eea
This matrix should be interpreted as determining the toric diagram of the moduli space, which is a toric Calabi-Yau 4-fold. Every column in $G$ corresponds to a GLSM field and determines the position in $\mathbb{Z}^4$ of the corresponding point in the toric diagram. 

\begin{figure}[h]
\begin{center}
\resizebox{0.5\hsize}{!}{
\includegraphics[trim=0cm 0cm 0cm 0cm,totalheight=10 cm]{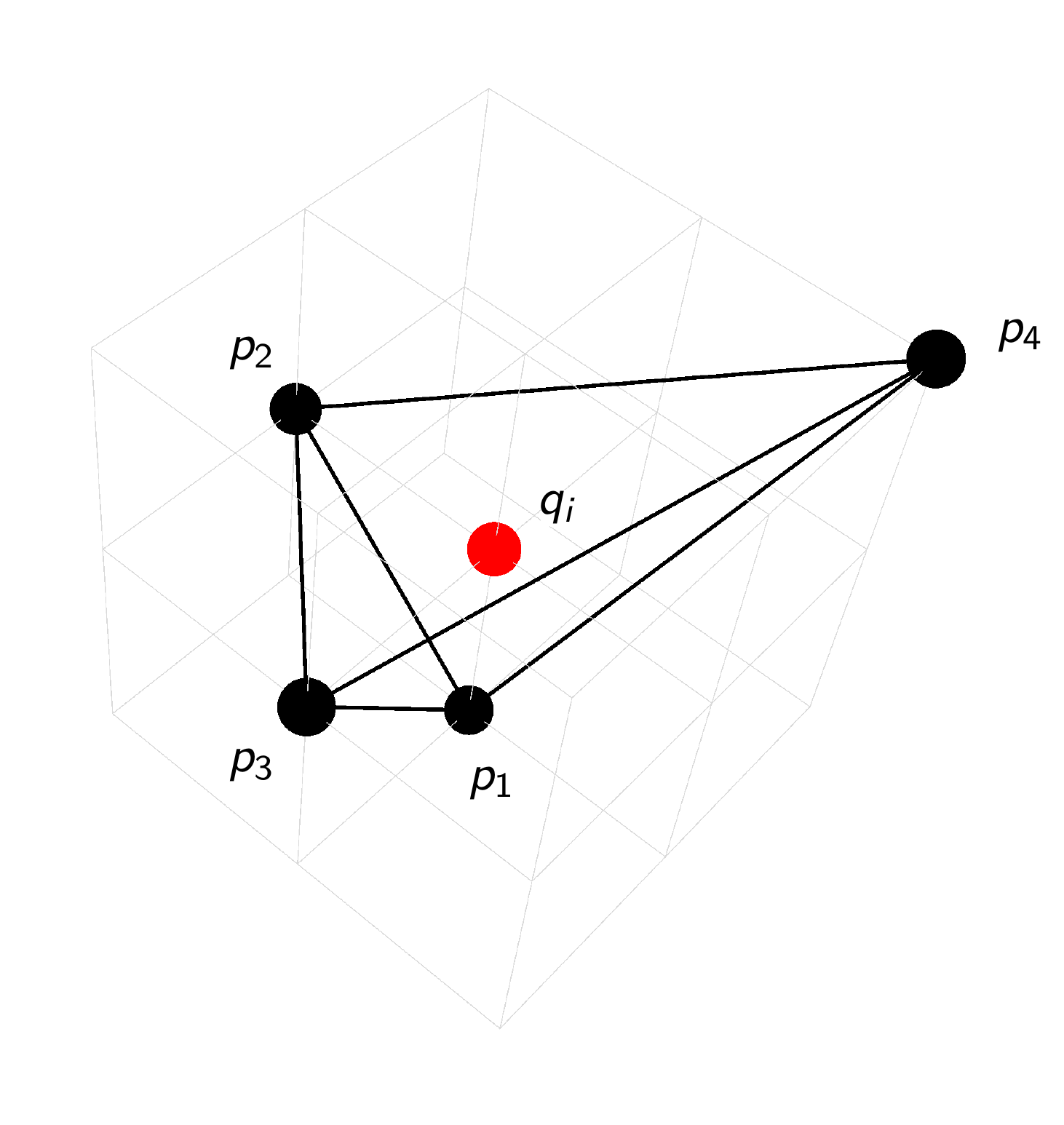}
}  
\caption{Toric diagram for the $\mathbb{C}^4/\mathbb{Z}_{4}$ orbifold with action $(1,1,1,1)$. This geometry was obtained as the mesonic moduli space of the corresponding gauge theory.
GLSM fields $q_1$, $q_2$, $q_3$ and $q_4$ correspond to a single internal point, shown in red.
\label{ftoricc4z4a1111}}
 \end{center}
 \end{figure}

For the $\mathbb{C}^4/\mathbb{Z}_4$ $(1,1,1,1)$ gauge theory, we obtain
\beal{es84e3}
G= \left(
\ba{cccccccc}
p_1 & p_2 & p_3 & p_4 & q_1 & q_2 & q_3 & q_4
\\
\hline
 1 & 1 & 1 & 1 & 1 & 1 & 1 & 1 \\
 -1 & 0 & 0 & 1 & 0 & 0 & 0 & 0 \\
 0 & 1 & 0 & -1 & 0 & 0 & 0 & 0 \\
 0 & 0 & 1 & -1 & 0 & 0 & 0 & 0 \\
\ea
\right)~.~
\eea
We note that the GLSM fields $q_1$, $q_2$, $q_3$ and $q_4$ correspond to the same point in the toric diagram. We say that this point has multiplicity 4. The toric diagram associated with \eref{es84e3} is shown in \fref{ftoricc4z4a1111}. Since all 4-vectors corresponding to toric points live on a 3-dimensional hyperplane, it can be drawn in $\mathbb{Z}^3$. The toric diagram in \fref{ftoricc4z4a1111} corresponds to the $\mathbb{C}^4/\mathbb{Z}_4$ $(1,1,1,1)$ orbifold. Remarkably, we obtained it as the moduli space of the corresponding gauge theory. Notice that four unit tetrahedra fit into this toric diagram, in agreement with the fact that the corresponding quiver has four nodes.

\end{itemize}

\bigskip

\subsection{Algebraic Varieties, Extra GLSM Fields and Hilbert Series \label{sextrahs}}

The previous section introduced the forward algorithm, which gives a toric description of the mesonic moduli space of the $2d$ abelian theory. This section gives an overview of the algebraic structure of the Calabi-Yau moduli spaces. Furthermore, it also sheds light on a new phenomenon exhibited by some $2d$ theories: the appearance of extra GLSM fields.

\bigskip

\paragraph{The Master Space and the Mesonic Moduli Space.}

The chiral fields $X_{ij}$, which are subject to the vanishing $J$- and $E$-term relations, form an affine algebraic variety 
\beal{es150e1}
\master = \mathbb{C}^{n^\chi}[X_1,\dots X_{n^\chi}] / \langle J_{a},E_{a}\rangle ~.~
\eea
The quotienting ideal $ \langle J_{a},E_{a}\rangle $ is over the relations $J_{a}=0$ and $E_{a}=0$ corresponding to all Fermi fields $\Lambda_{a}$. We refer to this variety as the {\it master space}, in analogy to the master space of $4d$ $\mathcal{N}=1$ theories, which is the space of vanishing F-terms \cite{Forcella:2008bb}. For the $2d$ theories under consideration, the master space is a $G+3$-dimensional toric Calabi-Yau, where $G$ is the number of gauge groups.

The mesonic moduli space $\mathcal{M}$ is obtained by further quotienting the master space by the $U(1)^G$ gauge charges carried by the chiral fields
\beal{es150e2}
\mathcal{M}= \master // U(1)^{G} ~.~
\eea
As explained in section \sref{sfa}, note that one of the $U(1)$ gauge symmetries is redundant in the $2d$ quiver gauge theory. $\mathcal{M}$ is a toric Calabi-Yau 4-fold and it is precisely the space probed by the D1-brane whose worldvolume theory is the $2d$ quiver gauge theory.

\bigskip

\paragraph{The Mesonic Moduli Space as a K\"ahler Quotient.} 

The forward algorithm, reformulates the vanishing $J$- and $E$-terms as a set of complexified $U(1)^{c-G-3}$ charges on GLSM fields $p_a$. These charges are summarized in the matrix $Q_{EJ~(c-G-3)\times c}$ as defined in \eref{es82e1}. Combined with the $U(1)^G$ gauge charges on the GLSM fields, which are summarized in the matrix $Q_{D~(G-1)\times c}$ as shown in \eref{es83e3}, the Calabi-Yau 4-fold mesonic moduli space can be expressed as the following K\"ahler quotient
\beal{es150e3}
\mathcal{M} =( \mathbb{C}[p_1,\dots,p_c] // Q_{JE})//Q_{D} ~,~
\eea
where $\mathbb{C}[p_1,\dots,p_c]$ is the freely generated $c$-dimensional space of GLSM fields $p_a$ defined in \eref{es81e3}. In other words, the mesonic moduli space is the space of $U(1)^{c-4}$ invariant operators in terms of GLSM fields $p_a$.

From \eref{es150e2} and \eref{es150e3}, we conclude that $\mathcal{M}$ can be expressed in terms of chiral fields subject to $J$- and $E$-term relations and $U(1)^G$ invariance, or in terms of GLSM fields invariant under $Q_{JE}$ and $Q_D$.

\bigskip

\paragraph{Extra GLSM Fields.} 

The two constructions \eref{es150e2} and \eref{es150e3} of the mesonic moduli space give rise to the same geometry. However, it is interesting to point out a new phenomenon exhibited by $2d$ theories: in some cases, the forward algorithm makes use of additional GLSM fields that are {\it redundant} for describing $\mathcal{M}$. 

Such extra GLSM fields manifest as additional points in the toric diagram of $\mathcal{M}$ that, however, can be neglected when identifying the corresponding geometry. In physical terms, they can be understood as follows. $\mathcal{M}$ is parameterized by mesonic gauge invariant operators, which form the spectrum of the quotients in \eref{es150e2} and \eref{es150e3}. These operators can be expressed in terms of either chiral fields or GLSM fields. When extra GLSM fields are present, they do not affect the spectrum of operators. Instead, they can be regarded as an over-parameterization of the mesonic moduli space, where generators and relations amongst generators remain unaffected by their presence. Hilbert series provide an efficient tool for verifying that this is actually the case. Hilbert series are partition functions that count gauge invariant operators and have been extensively used in $4d$ theories corresponding to brane tilings to study the algebraic structure of their vacuum moduli spaces \cite{Benvenuti:2006qr,Hanany:2012hi,Hanany:2012vc}.

In all the examples presented in this paper, we have used Hilbert series to verify that the presence of extra GLSM fields does not affect the algebraic properties of the mesonic moduli space. We postpone a more detailed study and presentation of this approach for the future \cite{topub1}. The following sections concentrate on toric diagrams, after appropriately identifying extra GLSM fields when they are present, as a tool for characterizing the mesonic moduli space.

We have empirically found various characteristic properties that can be exploited for identifying points in the toric diagram corresponding to possible extra GLSM fields. They are:

\smallskip

\begin{itemize}
\item {\bf Property 1:} The points do not lie on the same 3-dimensional hyperplane as the rest of the toric diagram of the CY$_4$.
\item {\bf Property 2:} The points have multiplicity greater than 1. While this seems to be the case for extra GLSM fields in all examples we considered, it is also a rather common feature of normal GLSM fields, so it is not a particularly restrictive condition.
\item {\bf Property 3:} According to the examples we studied, it appears that extra GLSM fields contain more chiral fields than other GLSM fields. This seems to be a necessary but not sufficient condition.
\end{itemize}

\smallskip
\noindent Let us reiterate that the above are observations. In the following sections, once candidate extra GLSM fields are found, we verify the actual geometry of the moduli space with its generators and relations using Hilbert series. For explicit computations of the Hilbert series, the reader is referred to our future work \cite{topub1}.

The following sections present several explicit examples of the forward algorithm applied to $2d$ theories both with and without extra GLSM fields. 
\bigskip

\subsection{Orbifold Examples}

\label{section_orbifold_examples}

We now illustrate the forward algorithm with additional orbifold examples. As expected, the mesonic moduli space of the gauge theory nicely reproduces the probed geometry. We have picked examples that show the phenomenon of extra GLSM fields. Additional orbifold theories are presented in appendix \ref{sappc4}. The forward algorithm is applied to non-orbifold theories in sections \sref{section_CY3xC} and \sref{section_beyond_orbifolds}.

\bigskip

\subsubsection{$\mathbb{C}^4/\mathbb{Z}_{2}$ $(1,1,1,1)$ with extra GLSM fields}

\begin{figure}[ht!!]
\begin{center}
\resizebox{0.6\hsize}{!}{
\includegraphics[trim=0cm 0cm 0cm 0cm,totalheight=10 cm]{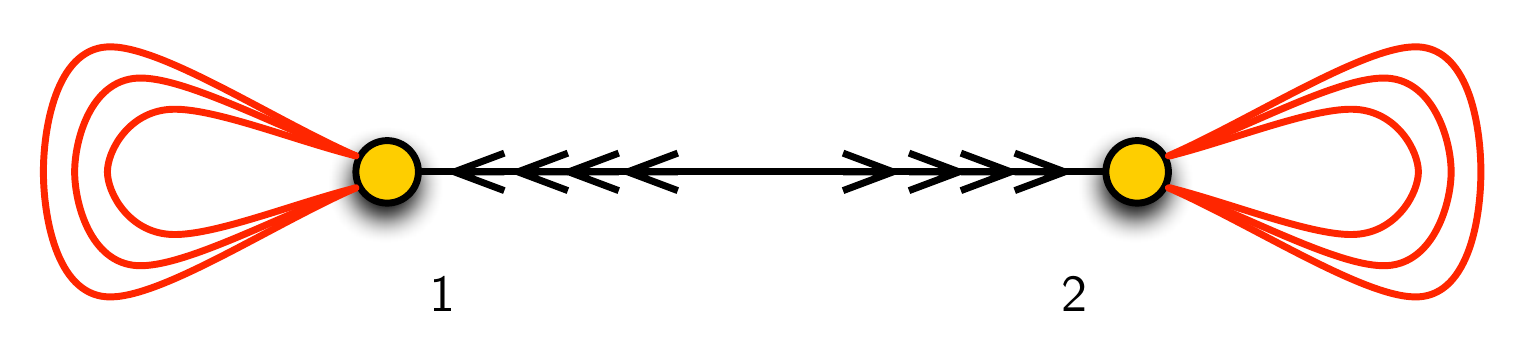}
}  
\caption{
Quiver diagram for $\mathbb{C}^4/\mathbb{Z}_2$ $(1,1,1,1)$.
\label{fz31111}}
 \end{center}
 \end{figure}

The quiver diagram for $\mathbb{C}^4/\mathbb{Z}_2$ $(1,1,1,1)$ is shown in \fref{fz31111}.

\noindent The $J$- and $E$-terms with their corresponding Fermi fields are
\beq
\begin{array}{rclccclcc}
& & \ \ \ \ \ \ \ \ \ \ \ \ \ J & & & & \ \ \ \ \ \ \ \ \ \ \ \ \ E & & \\
\Lambda_{11}^{1} : & \ \ \ & Y_{12}\cdot Z_{21} - Z_{12}\cdot Y_{21} & = & 0 & \ \ \ \ & D_{12}\cdot X_{21} - X_{12}\cdot D_{21} & = & 0 \\
\Lambda_{22}^{1} : & \ \ \ & Y_{21}\cdot Z_{12} - Z_{21}\cdot Y_{12} & = & 0 & \ \ \ \ & D_{21}\cdot X_{12} - X_{21}\cdot D_{12} & = & 0 \\
\Lambda_{11}^{2} : & \ \ \ &  Z_{12}\cdot X_{21} -X_{12}\cdot Z_{21} & = & 0 & \ \ \ \ & D_{12}\cdot Y_{21} - Y_{12}\cdot D_{21} & = & 0 \\
\Lambda_{22}^{2} : & \ \ \ & Z_{21}\cdot X_{12} -X_{21}\cdot Z_{12} & = & 0 & \ \ \ \ & D_{21}\cdot Y_{12} - Y_{21}\cdot D_{12}  & = & 0\\
\Lambda_{11}^{3} : & \ \ \ & X_{12}\cdot Y_{21} - Y_{12}\cdot X_{21} & = & 0 & \ \ \ \ & D_{12}\cdot Z_{21} - Z_{12}\cdot D_{21} & = & 0 \\
\Lambda_{22}^{3} : & \ \ \ & X_{21}\cdot Y_{12} - Y_{21}\cdot X_{12} & = & 0 & \ \ \ \ & D_{21}\cdot Z_{12} - Z_{21}\cdot D_{12} & = & 0 \\
\label{es101e1}
\end{array}
\eeq
The quiver incidence matrix for the chiral fields is 
 \beal{es101e2}
 d=
 \left(
\begin{array}{c|cccccccc}
\; & D_{12} & D_{21} & X_{12} & X_{21} & Y_{12} & Y_{21} & Z_{12} & Z_{21}\\
\hline
\encircle{1} & 1 & -1 & 1 & -1 & 1 & -1 & 1 & -1 \\
\encircle{2} & -1 & 1 & -1 & 1 & -1 & 1 & -1 & 1 \\
\end{array}
\right)~.~
 \eea
The relations in \eref{es101e1} can be reduced to independent relations that give rise to the following $K$-matrix
 \beal{es101e3}
 K=
 \left(
\begin{array}{c|cccccccc}
\; & D_{12} & D_{21} & X_{12} & X_{21} & Y_{12} & Y_{21} & Z_{12} & Z_{21}\\
\hline
D_{12} & 1 & 0 & 0 & -1 & 0 & -1 & 0 & -1 \\
D_{21} & 0 & 1 & 0 & 1 & 0 & 1 & 0 & 1 \\
X_{12} & 0 & 0 & 1 & 1 & 0 & 0 & 0 & 0 \\
Y_{12} & 0 & 0 & 0 & 0 & 1 & 1 & 0 & 0 \\
Z_{12} & 0 & 0 & 0 & 0 & 0 & 0 & 1 & 1 \\
\end{array}
\right)
~.~
 \eea
The $P$-matrix connecting chiral fields to GLSM fields is given by
\beal{es101e4}
P=
\left(
\begin{array}{c|cccccc}
\; & p_1 & p_2 & p_3 & p_4 & q_1 & q_2
\\
\hline
D_{12} & 1 & 0 & 0 & 0 & 1 & 0 \\
D_{21} & 1 & 0 & 0 & 0 & 0 & 1 \\
X_{12} & 0 & 1 & 0 & 0 & 1 & 0 \\
X_{21} & 0 & 1 & 0 & 0 & 0 & 1 \\
Y_{12} & 0 & 0 & 1 & 0 & 1 & 0 \\
Y_{21} & 0 & 0 & 1 & 0 & 0 & 1 \\
Z_{12} & 0 & 0 & 0 & 1 & 1 & 0 \\
Z_{21} & 0 & 0 & 0 & 1 & 0 & 1 \\
\end{array}
\right)~.~
\eea
The charge matrices encoding vanishing $J$- and $E$-terms and the vanishing $D$-terms are respectively
\beal{es101e5}
Q_{JE} = 
\left(
\begin{array}{cccccc}
 p_1 & p_2 & p_3 & p_4 & q_1 & q_2
\\
\hline
 1 & 1 & 1 & 1 & -1 & -1 \\
\end{array}
\right)
~~,~~
Q_D =
\left(
\begin{array}{cccccc}
 p_1 & p_2 & p_3 & p_4 & q_1 & q_2\\
 \hline
 0 & 0 & 0 & 0 & -1 & 1 \\
\end{array}
\right)
~.~
\eea
The resulting toric data is given by the following matrix
\beal{es101e6}
G =
\left(
\begin{array}{cccccc}
 p_1 & p_2 & p_3 & p_4 & q_1 & q_2\\
 \hline
 2 & 0 & 0 & 0 & 1 & 1 \\
 -1 & 0 & 0 & 1 & 0 & 0 \\
 -1 & 0 & 1 & 0 & 0 & 0 \\
 -1 & 1 & 0 & 0 & 0 & 0 \\
\end{array}
\right)
~,~
\eea
where column vectors should be regarded as the positions of the corresponding points in the toric diagram of the CY$_4$. GLSM fields $p_1$, $p_2$, $p_3$ and $p_4$ correspond to toric points with multiplicity 1 whereas GLSM fields $q_1$ and $q_2$ correspond to the same toric point which has multiplicity 2. 

Let us explore whether some of these points correspond to extra GLSM fields. By performing an $SL(4,\mathbb{Z})$ transformation with
\beal{es101e8}
M = 
\left(
\begin{array}{cccc}
 \, 2 \, & \, 1 \, & \, 1 \, & \, 1 \, \\
 1 & 1 & 0 & 0 \\
 1 & 1 & 1 & 0 \\
 1 & 1 & 0 & 1 \\
\end{array}
\right)
~,~
\eea
on the coordinates of the toric points.\footnote{In the examples that follow, we will often perform similar $SL(4,\mathbb{Z})$ transformations on the $G$-matrices in order to obtain better looking toric diagrams.} The new toric diagram matrix becomes
\beal{es101e9}
\widetilde{G} = M.G = 
\left(
\begin{array}{cccccc}
 p_1 & p_2 & p_3 & p_4 & q_1 & q_2\\
 \hline
 1 & 1 & 1 & 1 & 2 & 2 \\
 1 & 0 & 0 & 1 & 1 & 1 \\
 0 & 0 & 1 & 1 & 1 & 1 \\
 0 & 1 & 0 & 1 & 1 & 1 \\
\end{array}
\right)
~.~
\eea
It can be seen that the $x$-plane corresponding to the first row of the toric diagram matrix above is a 3-dimensional hyperplane on which GLSM fields $p_1$, $p_2$, $p_3$ and $p_4$ lie, whereas GLSM fields $q_1$ and $q_2$ are not on the hyperplane. As a result, $q_1$ and $q_2$ satisfy property 1, which is a strong indication that they are extra GLSM fields. In this case, $q_1$ and $q_2$ also exhibit the other two properties which are outlined in section \sref{sextrahs}. First, they correspond to a point with multiplicity 2, satisfying property 2. Finally, the $P$-matrix in \eref{es101e4} shows that $q_1$ and $q_2$ contain more chiral fields than the remaining GLSM fields, thus also satisfying property 3.

\begin{figure}[ht!!]
\begin{center}
\resizebox{0.4\hsize}{!}{
\includegraphics[trim=0cm 0cm 0cm 0cm,totalheight=10 cm]{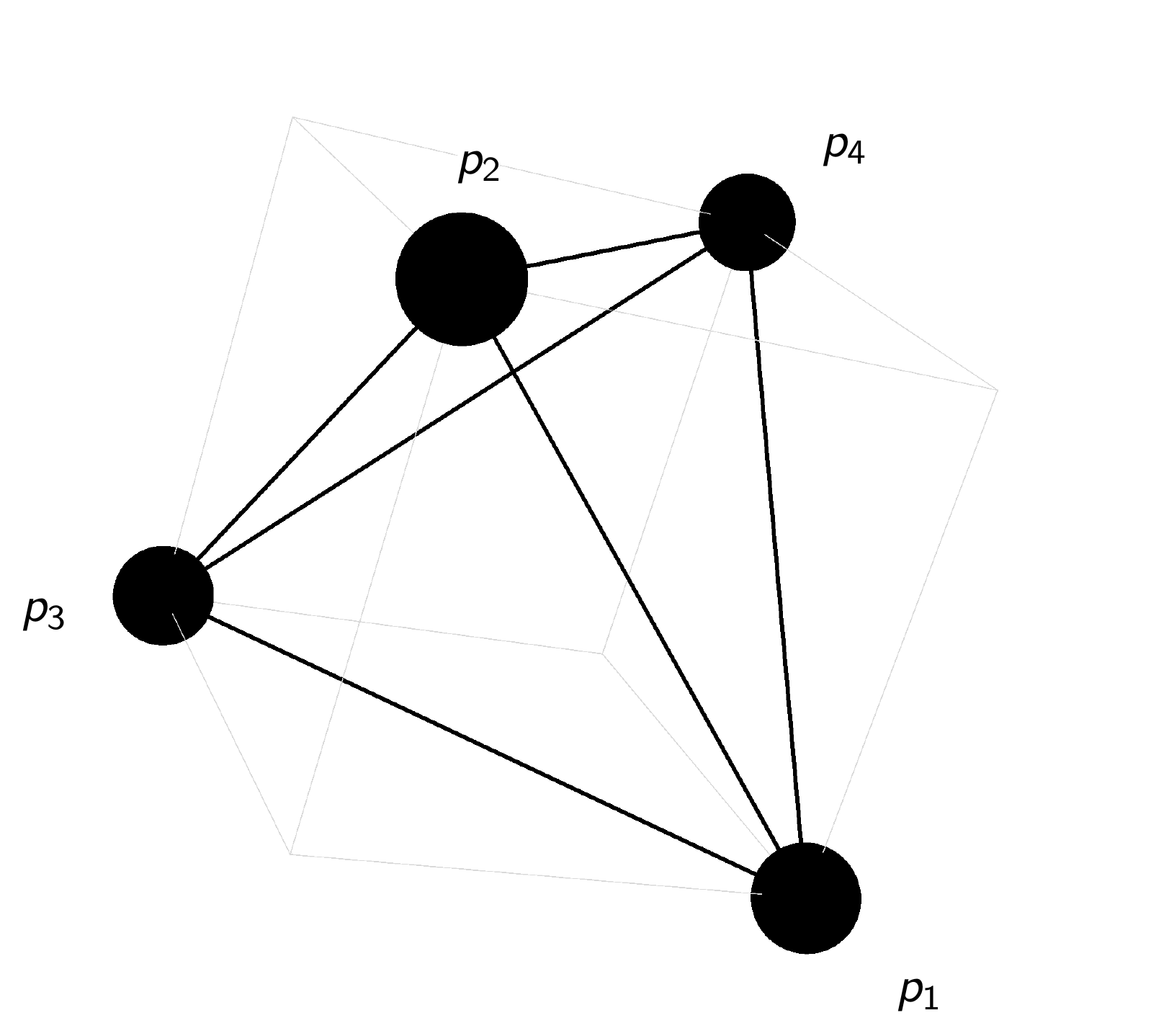}
}  
\caption{
After removing the extra GLSM fields, we obtain the toric diagram for $\mathbb{C}^4/\mathbb{Z}_2$ $(1,1,1,1)$.
\label{ftoricc4z2a11011}}
 \end{center}
 \end{figure}

After identifying $q_1$ and $q_2$ as potential extra GLSM fields, it is straightforward to confirm that this is indeed the case by verifying that the generators of the mesonic moduli space and their relations are independent of them. We can thus remove $q_1$ and $q_2$ from the toric diagram matrix \eref{es101e9}. \fref{ftoricc4z2a11011} shows the resulting toric diagram, which is precisely the one for $\mathbb{C}^4/\mathbb{Z}_{2}$ $(1,1,1,1)$ \cite{Davey:2010px,Hanany:2010ne}.

\bigskip

\subsubsection{$\mathbb{C}^4/\mathbb{Z}_{3}$ $(1,1,2,2)$ with extra GLSM fields}

\begin{figure}[h]
\begin{center}
\resizebox{0.5\hsize}{!}{
\includegraphics[trim=0cm 0cm 0cm 0cm,totalheight=10 cm]{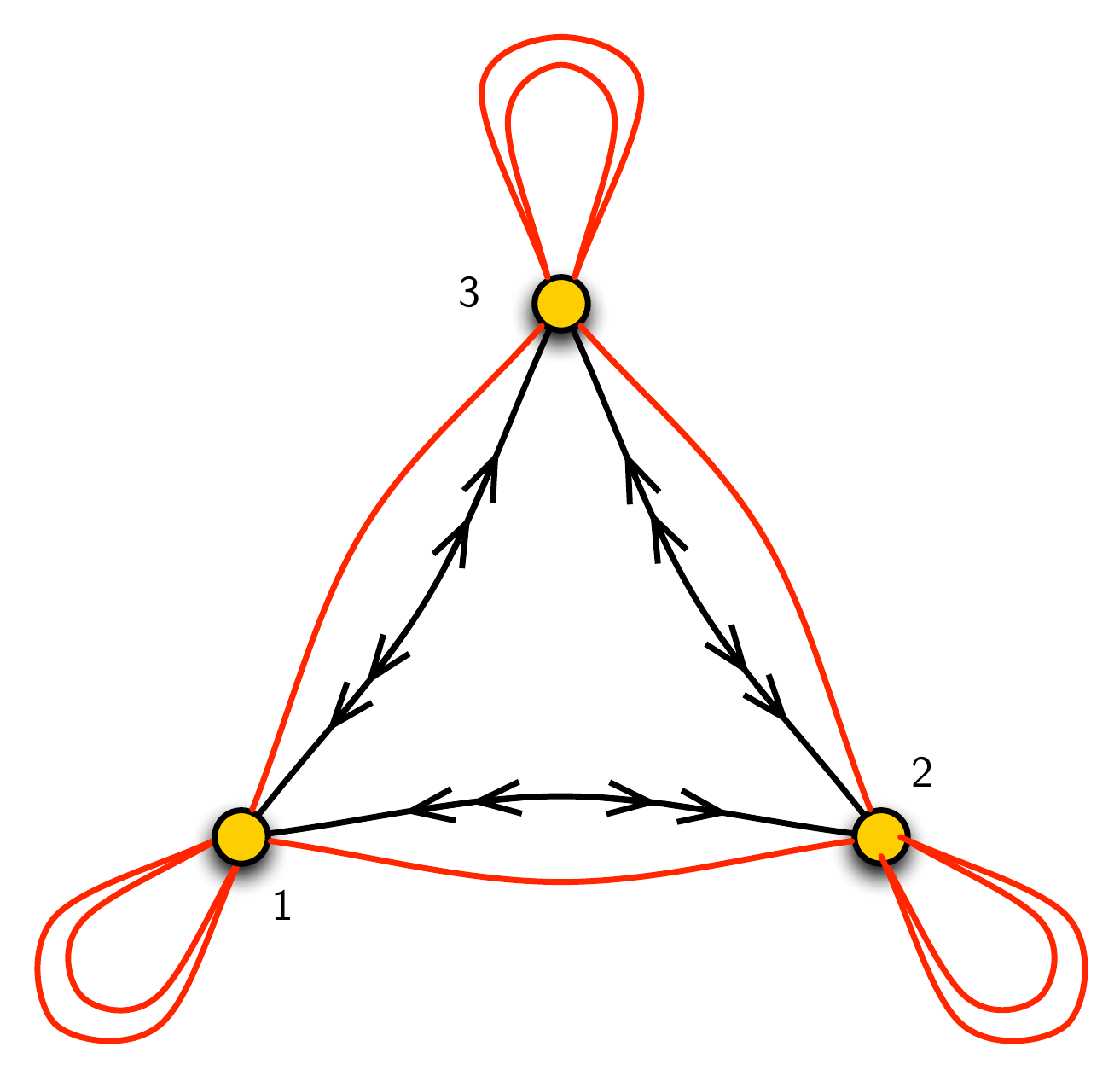}
}  
\caption{
Quiver diagram for $\mathbb{C}^4/\mathbb{Z}_3$ $(1,1,2,2)$.
\label{fz31122}}
 \end{center}
 \end{figure}

Let us now consider the $\mathbb{C}^4/\mathbb{Z}_{3}$ orbifold with action $(1,1,2,2)$, whose corresponding quiver is shown in \fref{fz31122}.

\noindent The $J$- and $E$-terms read
{\footnotesize
\beq
\begin{array}{rclccclcc}
& & \ \ \ \ \ \ \ \ \ \ \ \ \ J & & & & \ \ \ \ \ \ \ \ \ \ \ \ \ E & & \\
\Lambda_{11}^{1} : & \ \ \ & Y_{13}\cdot Z_{31} - Z_{12}\cdot Y_{21}& = & 0 & \ \ \ \ & D_{13}\cdot X_{31} - X_{12}\cdot D_{21}& = & 0 \\
\Lambda_{22}^{1} : & \ \ \ & Y_{21}\cdot Z_{12} - Z_{23}\cdot Y_{32}& = & 0 & \ \ \ \ & D_{21}\cdot X_{12} - X_{23}\cdot D_{32}& = & 0 \\
\Lambda_{33}^{1} : & \ \ \ & Y_{32}\cdot Z_{23} - Z_{31}\cdot Y_{13} & = & 0 & \ \ \ \ & D_{32}\cdot X_{23} - X_{31}\cdot D_{13}& = & 0 \\
\Lambda_{11}^{2} : & \ \ \ & X_{12}\cdot Y_{21} - Y_{13}\cdot X_{31} & = & 0 & \ \ \ \ & D_{13}\cdot Z_{31} - Z_{12}\cdot D_{21}& = & 0 \\ 
\Lambda_{22}^{2} : & \ \ \ & X_{23}\cdot Y_{32} - Y_{21}\cdot X_{12}& = & 0 & \ \ \ \ & D_{21}\cdot Z_{12} - Z_{23}\cdot D_{32}& = & 0 \\ 
\Lambda_{33}^{2} : & \ \ \ & X_{31}\cdot Y_{13} - Y_{32}\cdot X_{23}& = & 0 & \ \ \ \ & D_{32}\cdot Z_{23} - Z_{31}\cdot D_{13}& = & 0 \\
\Lambda_{12} : & \ \ \ & Z_{23}\cdot X_{31} -X_{23}\cdot Z_{31}& = & 0 & \ \ \ \ & D_{13}\cdot Y_{32} - Y_{13}\cdot D_{32}& = & 0 \\ 
\Lambda_{23} : & \ \ \ & Z_{31}\cdot X_{12} -X_{31}\cdot Z_{12}& = & 0 & \ \ \ \ & D_{21}\cdot Y_{13} - Y_{21}\cdot D_{13}& = & 0 \\
\Lambda_{31} : & \ \ \ &  Z_{12}\cdot X_{23} -X_{12}\cdot Z_{23}& = & 0 & \ \ \ \ & D_{32}\cdot Y_{21} - Y_{32}\cdot D_{21}& = & 0 \\
\end{array}
\label{es104e1}
\eeq}
The quiver incidence matrix is 
\beal{es104e2}
d=\left(
\begin{array}{c|cccccccccccc}
\; &D_{13}& D_{21}& D_{32}& X_{12}& X_{23}& X_{31}& Y_{13}& Y_{21}& Y_{32}& Z_{12}& Z_{23}& Z_{31} \\
\hline
\encircle{1} & 1 & -1 & 0 & 1 & 0 & -1 & 1 & -1 & 0 & 1 & 0 & -1 \\
\encircle{2} & 0 & 1 & -1 & -1 & 1 & 0 & 0 & 1 & -1 & -1 & 1 & 0 \\
\encircle{3} & -1 & 0 & 1 & 0 & -1 & 1 & -1 & 0 & 1 & 0 & -1 & 1 \\
\end{array}
\right)
~.~
\eea
From \eref{es104e1}, we obtain
\beal{es104e3}
K=\left(
\begin{array}{c|cccccccccccc}
\; &D_{13}& D_{21}& D_{32}& X_{12}& X_{23}& X_{31}& Y_{13}& Y_{21}& Y_{32}& Z_{12}& Z_{23}& Z_{31} \\
\hline
D_{13} & 1 & 0 & 0 & 0 & 0 & -1 & 0 & -1 & -1 & 0 & 0 & -1 \\
D_{32} & 0 & 1 & 1 & 0 & 0 & 1 & 0 & 1 & 1 & 0 & 0 & 1 \\
X_{12} & 0 & -1 & 0 & 1 & 0 & 0 & 0 & -1 & 0 & 0 & -1 & -1 \\
X_{23} & 0 & 1 & 0 & 0 & 1 & 1 & 0 & 1 & 0 & 0 & 1 & 1 \\
Y_{13} & 0 & 0 & 0 & 0 & 0 & 0 & 1 & 1 & 1 & 0 & 0 & 0 \\
Z_{12} & 0 & 0 & 0 & 0 & 0 & 0 & 0 & 0 & 0 & 1 & 1 & 1 \\
\end{array}
\right)
~.~
\eea
The $P$-matrix becomes
\beal{es104e4}
P=\left(
\begin{array}{c|cccccccccc}
\; & p_1 & p_2 & p_3 & p_4 & q_1 & q_2 & q_3 & q_4 & r_1 & r_2 \\
\hline
D_{13}& 1 & 0 & 0 & 0 & 1 & 0 & 1 & 1 & 0 & 0 \\
D_{21}& 1 & 0 & 0 & 0 & 1 & 1 & 0 & 0 & 1 & 0 \\
D_{32}& 1 & 0 & 0 & 0 & 0 & 1 & 1 & 0 & 0 & 1 \\
X_{12}&  0 & 1 & 0 & 0 & 0 & 0 & 1 & 1 & 0 & 1 \\
X_{23}& 0 & 1 & 0 & 0 & 1 & 0 & 0 & 1 & 1 & 0 \\
X_{31}& 0 & 1 & 0 & 0 & 0 & 1 & 0 & 0 & 1 & 1 \\
Y_{13}& 0 & 0 & 1 & 0 & 1 & 0 & 1 & 1 & 0 & 0 \\
Y_{21}& 0 & 0 & 1 & 0 & 1 & 1 & 0 & 0 & 1 & 0 \\
Y_{32}& 0 & 0 & 1 & 0 & 0 & 1 & 1 & 0 & 0 & 1 \\
Z_{12}& 0 & 0 & 0 & 1 & 0 & 0 & 1 & 1 & 0 & 1 \\
Z_{23}& 0 & 0 & 0 & 1 & 1 & 0 & 0 & 1 & 1 & 0 \\
Z_{31}& 0 & 0 & 0 & 1 & 0 & 1 & 0 & 0 & 1 & 1 \\
\end{array}
\right)
~.~
\eea

\begin{figure}[ht!]
\begin{center}
\resizebox{0.3\hsize}{!}{
\includegraphics[trim=0cm 0cm 0cm 0cm,totalheight=10 cm]{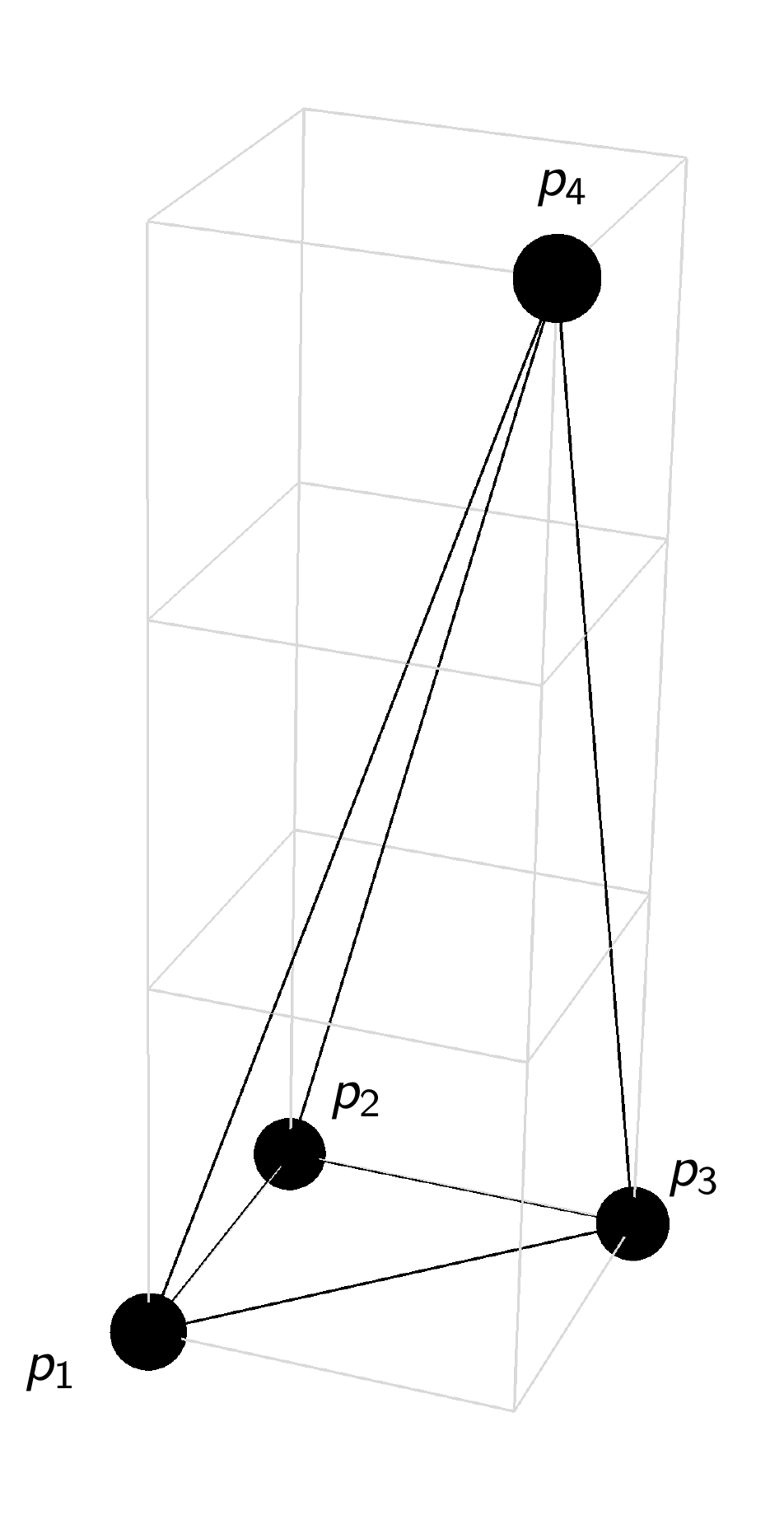}
}  
\caption{
Toric diagram for $\mathbb{C}^4/\mathbb{Z}_3$ $(1,1,2,2)$.
\label{ftoricz3a1122}}
 \end{center}
 \end{figure}

\noindent Finally, the charge matrices of the forward algorithm are
\beal{es104e5}
Q_{JE}=
\left(
\begin{array}{cccccccccc}
p_1 & p_2 & p_3 & p_4 & q_1 & q_2 & q_3 & q_4 & r_1 & r_2 \\
\hline
 1 & 2 & 1 & 2 & 0 & 0 & 0 & -1 & -1 & -1 \\
 0 & 1 & 0 & 1 & 1 & 0 & 0 & -1 & -1 & 0 \\
 0 & 1 & 0 & 1 & 0 & 0 & 1 & -1 & 0 & -1 \\
 0 & 1 & 0 & 1 & 0 & 1 & 0 & 0 & -1 & -1 \\
\end{array}
\right)~~,~~
Q_D=
\left(
\begin{array}{cccccccccc}
p_1 & p_2 & p_3 & p_4 & q_1 & q_2 & q_3 & q_4 & r_1 & r_2 \\
\hline
 0 & 0 & 0 & 0 & 0 & 0 & 0 & 0 & 1 & -1 \\
 0 & 0 & 0 & 0 & 0 & 0 & 0 & -1 & 0 & 1 \\
\end{array}
\right)
~~,~~
\nn\\
\eea
from which we obtain the toric diagram matrix
\beal{es104e6}
G=
\left(
\begin{array}{cccccccccc}
p_1 & p_2 & p_3 & p_4 & q_1 & q_2 & q_3 & q_4 & r_1 & r_2 \\
\hline
 1 & 1 & 1 & 1 & 2 & 2 & 2 & 2 & 2 & 2 \\
 0 & 1 & 1 & 0 & 1 & 1 & 1 & 1 & 1 & 1 \\
 0 & 0 & 1 & 1 & 1 & 1 & 1 & 1 & 1 & 1 \\
 0 & 0 & 0 & 3 & 1 & 1 & 1 & 2 & 2 & 2 \\
\end{array}
\right)
~.~
\eea
    Proceeding as in the previous example, we can determine that $q_i$ and $r_i$ are extra GLSM fields. After removing them, we obtain the toric diagram in \fref{ftoricz3a1122}, which is precisely the one expected for $\mathbb{C}^4/\mathbb{Z}_{3}$ $(1,1,2,2)$ \cite{Davey:2010px,Hanany:2010ne}.

\bigskip

\section{CY$_3\times \mathbb{C}$ Theories}

\label{section_CY3xC}

In this section, we continue our journey beyond orbifolds and study the class of $2d$ theories on the worldvolume of D1-branes over toric CY 4-folds of the form 
$\mathrm{CY}_3 \times \mathbb{C}$. These theories can be nicely understood via dimensional reduction from $4d$ theories that arise on the worldvolume of D3-branes probing the CY$_3$ factor of the Calabi-Yau 4-fold. They have non-chirally enhanced SUSY: $2d$ $(2,2)$, $(4,4)$ or $(8,8)$ for $4d$ $\mathcal{N}=1$, 2 and 4 parent theories, respectively.\footnote{Thinking more broadly, there can be $2d$ $(2,2)$ theories that are not obtained by dimensionally reducing consistent $4d$ $\mathcal{N}=1$ theories. For example, anomalies are determined differently in $2d$ and $4d$, and therefore one may obtain consistent $(2,2)$ theories by dimensionally reducing anomalous $4d$ theories. This is not going to be the case for the theories we obtain in this section.} 

\bigskip

\subsection{$2d$ $(2,2)$ Theories from Dimensional Reduction \label{section_dimensional_reduction}}

Let us briefly review how $4d$ $\mathcal{N}=1$ theories can be dimensionally reduced to $2d$ $(2,2)$ theories, which in turn can be expressed in $2d$ $(0,2)$ language. To do so, we consider all fields are independent of $x^2$ and $x^3$ and decompose the representations of the $4d$ Lorentz group into those of the $2d$ one. The $2d$ theory is described in terms of $(2,2)$ superspace $(y^\alpha,\theta^+,\theta^-,\bar{\theta}^+,\bar{\theta}^-)$, where $(y^0,y^1)=(x^0,x^1)$. We can further express the $(2,2)$ theory in $(0,2)$ language, which uses the $(0,2)$ superspace  $(y^\alpha,\theta^+,\bar{\theta}^+)$.

$4d$ $\mathcal{N}=1$ SUSY has vector multiplets $\mathcal{V}_i$ and a chiral multiplet $\mathcal{X}_{ij}$, where we have included subindices to stress the quiver nature of the theories that we are considering. Under dimensional reduction, they become $2d$ $(2,2)$ vector and chiral multiplets, respectively. Finally, $2d$ $(2,2)$ multiplets can be expressed in terms of $2d$ $(0,2)$ multiplets. In summary, we obtain the following reduction from $4d$ $\mathcal{N}=1$ to $2d$ $(0,2)$ multiplets:

\smallskip

\begin{itemize}
\item \underline{$4d$ $\mathcal{N}=1$ vector $\mathcal{V}_i$} $\rightarrow$ $2d$ $(0,2)$ vector $V_i$ + $2d$ $(0,2)$ adjoint chiral $\Phi_{ii}$
\item \underline{$4d$ $\mathcal{N}=1$ chiral $\mathcal{X}_{ij}$} $\rightarrow$ $2d$ $(0,2)$ chiral $X_{ij}$ + $2d$ $(0,2)$ Fermi $\Lambda_{ij}$
\end{itemize}  

\smallskip

\begin{figure}[h]
\begin{center}
\resizebox{0.9\hsize}{!}{
\includegraphics[trim=0cm 0cm 0cm 0cm,totalheight=10 cm]{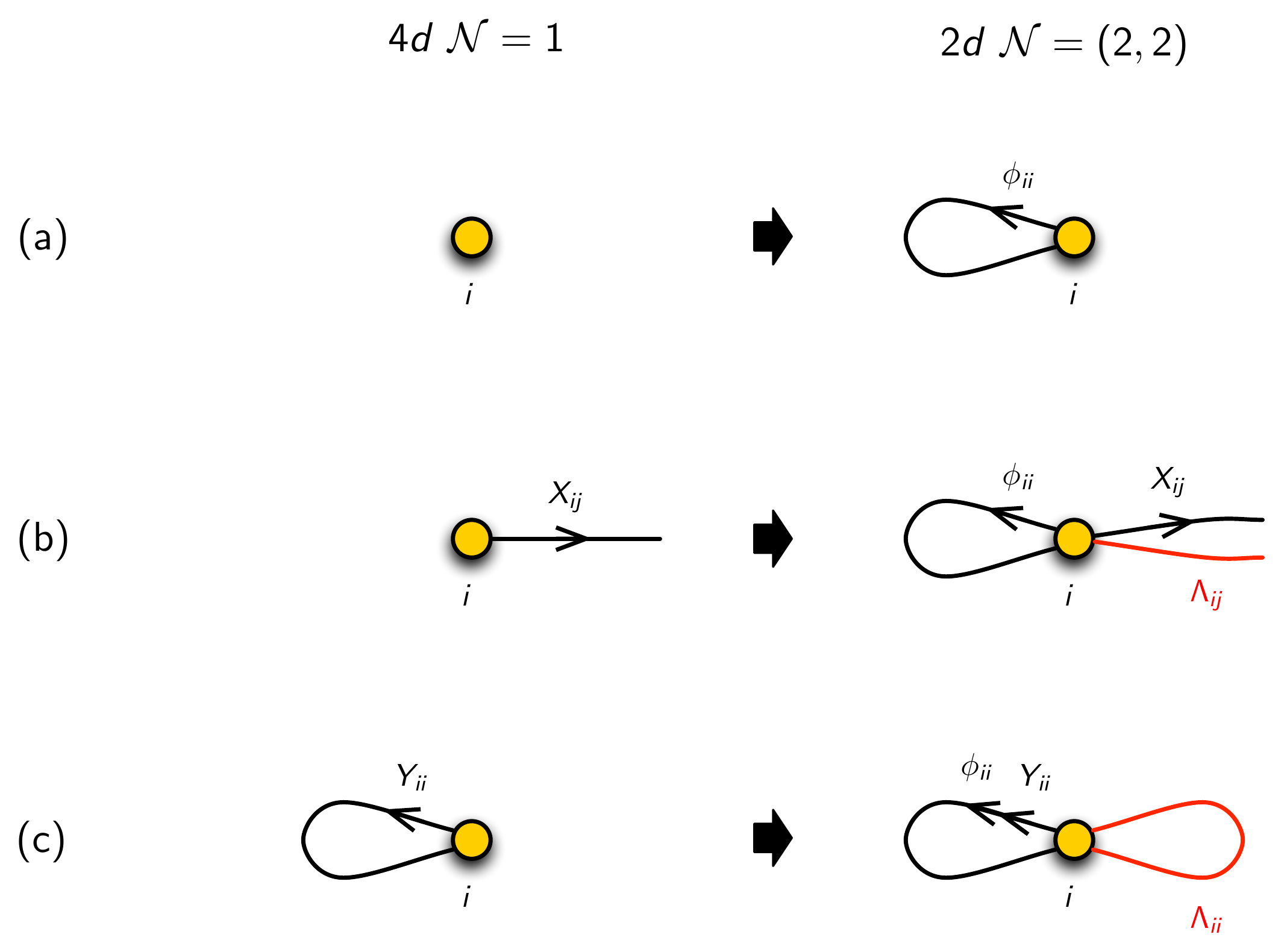}
}  
\caption{Dimensional reduction of $4d$ $\mathcal{N}=1$ quivers down to $2d$ $(2,2)$, expressed in $(0,2)$ language. (a) A 4d $\mathcal{N}=1$ vector multiplet reduces to a $(0,2)$ vector multiplet and a $(0,2)$ chiral multiplet in the adjoint representation. (b) A $4d$ $\mathcal{N}=1$ bifundamental chiral multiplet reduces to a $(0,2)$ chiral multiplet and a $(0,2)$ Fermi multiplet, both in the bifundamental representation. (c) Similarly, a $4d$ $\mathcal{N}=1$ adjoint  chiral multiplet reduces to a $(0,2)$ chiral multiplet and a $(0,2)$ Fermi multiplet, both in the adjoint representation.
\label{fdimreddic}}
 \end{center}
 \end{figure}

\noindent
\fref{fdimreddic} schematically shows how $4d$ $\mathcal{N}=1$ multiplets map to the $2d$ $(0,2)$ multiplets in terms of the quiver diagram.

The $J$-terms of the $2d$ theory follow from the $4d$ F-terms
 \bea
J_{ji} = \frac{\partial W}{\partial X_{ij}} ~,
\label{J_dim_red}
\eea
where $W$ is the $4d$ superpotential. The $J$-terms and $W$ are here understood as functions of the $2d$ $(0,2)$ chiral multiplets that descend from the $4d$ chiral multiplets.
 
The $E$-terms come from the gauge interaction of the $4d$ theory, and are given by
\beq
E_{ij} = \Phi_{ii} X_{ij} - X_{ij} \Phi_{jj}~.
\label{E_dim_red}
\eeq
The condition $\sum \mathrm{tr}(J_{ji} E_{ij})=0$ follows from gauge invariance of the superpotential $W$ \cite{GarciaCompean:1998kh}. 
As explained in section \sref{section_constructing_(0,2)_theories}, there is no invariant distinction between $J$- and $E$-terms. However, dimensionally reduced theories naturally distinguish between them due to the $4d$ parent theory. 

\bigskip

\subsection{Dimensional Reduction: from $4d$ to $2d$ Periodic Quivers \label{section_lift_periodic_quiver}}

Here we are interested in $2d$ $(2,2)$ theories on the worldvolume of D1-branes over toric Calabi-Yau 4-folds of the form $\mathrm{CY}_3 \times \mathbb{C}$. These theories can be obtained by dimensional reduction of $4d$ $\mathcal{N}=1$ theories on the worldvolume of D3-branes probing the toric CY$_3$ factor of the Calabi-Yau 4-fold.\footnote{The theories for D3-branes are generically $\mathcal{N}=1$ but, of course, there can be examples with enhanced SUSY.} Let us first discuss these 4d theories, which are given by {\it brane tilings}. These are bipartite graphs on a 2-torus where gauge groups, chiral fields and superpotential terms map to faces, edges and nodes in the graph. We refer the reader to \cite{Franco:2005rj,Kennaway:2007tq,Yamazaki:2008bt} for thorough discussions on brane tilings.

\paragraph{Brane Tilings.}
Brane tilings \cite{Hanany:2005ve,Franco:2005rj} have simplified immensely the connection between $4d$ $\mathcal{N}=1$ quiver gauge theories and their corresponding Calabi-Yau geometry. For example, GLSM fields in the toric description of the CY$_3$ admit a combinatorial implementation as perfect matchings in the brane tiling picture. A perfect matching $p_\alpha$ is a collection of edges in the tiling such that every node is the endpoint of exactly one edge in $p_\alpha$.

Brane tilings can be graph dualized into periodic quivers on $T^2$. Being equivalent to brane tilings, periodic quivers fully specify the corresponding gauge theories. In particular, every plaquette in the periodic quiver corresponds to a term in the superpotential. Periodic quivers nicely capture the global symmetries of the gauge theory, which at least contain a $U(1)^2 \times U(1)_R$ subgroup coming from the isometries of the toric CY$_3$. $U(1)_R$ is the R-symmetry whereas $U(1)\times U(1)$ is a mesonic flavor symmetry. Each fundamental cycle of the 2-torus corresponds to a $U(1)$ factor of the global flavor symmetry. 

The purpose of this section is to develop a systematic method for constructing the periodic quiver on $T^3$ corresponding to a $2d$ $(2,2)$ theory that comes from dimensional reduction of a $4d$ brane tiling. The periodic quiver on $T^3$ encodes $J$- and $E$-terms as minimal plaquettes.

\paragraph{Vertical direction and $T^3$.}
The third cycle in the periodic quivers on $T^3$ for the $2d$ theories corresponding to $\mathrm{CY}_3 \times \mathbb{C}$ is parameterized by the adjoint chiral multiplets arising from dimensional reduction of the $4d$ vector multiplets. This new cycle corresponds to an extra $U(1)$ mesonic global symmetry, which precisely comes from the additional $U(1)$ isometry associated with the $\mathbb{C}$ factor of the $\text{CY}_3 \times \mathbb{C}$. From now on, we refer to this third cycle as the \textit{vertical direction} in the picture of the $T^3$ periodic quiver.

Since mesonic symmetries map to the three fundamental directions of $T^3$, determining the periodic quiver for the $2d$ theory translates into the problem of assigning charges to fields. Once the $4d$ multiplets split into $2d$ $(0,2)$ multiplets, one needs to establish how each chiral multiplet is charged under the additional $U(1)$ flavor symmetry associated with the vertical direction. These charges determine how the original periodic quiver on $T^2$ gets `lifted' to $T^3$. The charges under the remaining $U(1)^2$ flavor symmetries are inherited from the $4d$ theory. The choice of a $U(1)$ subgroup of the flavor symmetry is determined up to an overall $GL(3,\mathbb{Z})$ symmetry of $T^3$ but, as we explain below, it is possible to come up with a natural prescription for performing the lift.

\paragraph{Vertical Shifts.}
Recall that every Fermi field $\Lambda_{ij}$ in the $2d$ theory descends from a $4d$ chiral field $\mathcal{X}_{ij}$ and gives rise to the $J$- and $E$-terms in \eref{J_dim_red} and \eref{E_dim_red}.
Assuming that $\Lambda_{ij}$ has a well-defined vertical shift, we note that $J_{ji}$ and $E_{ij}$ should have equal and opposite shifts along the vertical direction. Here vertical shifts for chiral fields are measured between the tail and the head of arrows or paths of arrows. For Fermi fields, we use the same prescription, with the orientation determined by the corresponding $4d$ chiral field. By convention, we may assign a $(+1)$ vertical shift to all $\Phi_{ii}$'s. It then follows that every term in $W$ must have a $(-1)$ vertical shift. Due to this, the oriented loops in the quiver, which correspond to the $4d$ superpotential, only close once the vertical periodicity is taken into account.
The vertical shifts of $X_{ij}$, if any, do not affect this property, as summarized in \tref{tref1}. 

\begin{table}[H] 
\begin{center}
\begin{tabular}{c|c} 
Object &
\multicolumn{1}{c}{Vertical shift} \\
\hline
$X_{ij}$ & $s_{ij}$ \\
$E_{ij}$ & $s_{ij}+1$ \\
$J_{ji}$ & $-s_{ij}-1$ \\
\hline
$W$ & $-1$
\label{table_vertical_shift}
\end{tabular}
\caption{Vertical shifts of chiral fields and their products in the periodic quiver on $T^3$.\label{tref1}}
\end{center}
\end{table}

The question now is how to determine the vertical shifts $s_{ij}$ of the chiral fields $X_{ij}$ such that all terms in $W$ acquire a $(-1)$ shift. This problem has a beautiful combinatorial answer in terms of perfect matchings of the parent brane tiling. Recall that a perfect matching is a collection of edges in the tiling that touches every node exactly once. Equivalently, it corresponds to a collection of chiral fields that contains precisely one field per superpotential term.
Accordingly, an efficient way of achieving the desired result is picking a perfect matching of the brane tiling and assigning a $(-1)$ vertical shift to all the chiral fields contained in it (regarded as $2d$ chiral fields), while leaving the remaining chiral fields without a vertical shift. This procedure automatically gives all terms in $W$ a $(-1)$ vertical shift. The freedom in choosing this perfect matching corresponds to the freedom in choosing a point in the CY$_3$ toric diagram. When the selected perfect matching receives the $(-1)$ shift, an additional point in the toric diagram is generated on top of it, which precisely corresponds to the $\mathbb{C}$ factor in the $\mathrm{CY}_3 \times \mathbb{C}$ geometry.

\begin{figure}[ht!!]
\begin{center}
\resizebox{1\hsize}{!}{
\includegraphics[trim=0cm 0cm 0cm 0cm,totalheight=10 cm]{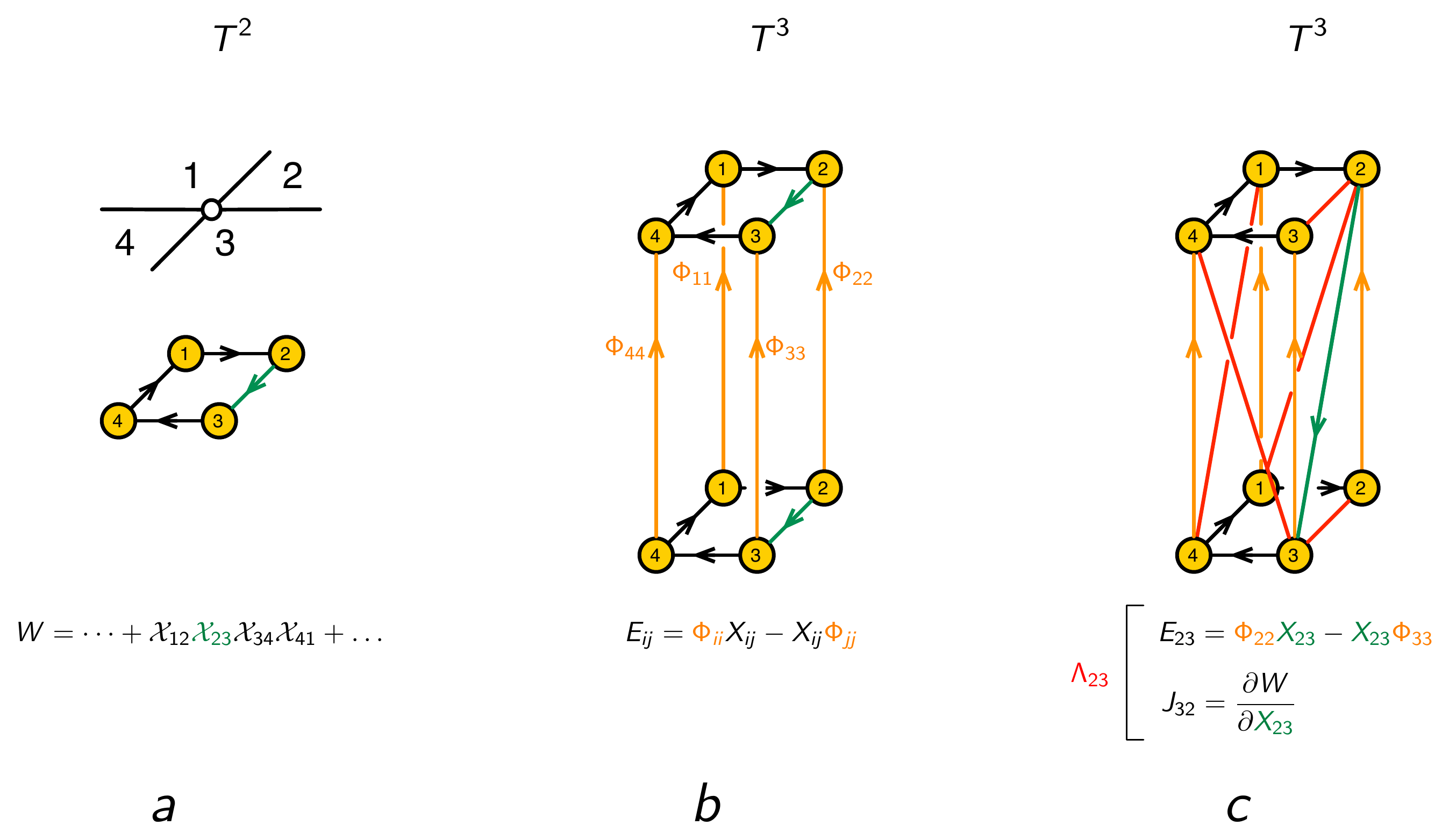}
}  
\caption{(a) A node in the brane tiling corresponds to a plaquette in the $T^2$ periodic quiver and a superpotential term in the $4d$ $\mathcal{N}=1$ parent theory. A perfect matching picks a single $4d$ chiral field in the plaquette, here shown in green. (b) An intermediate step in which we introduce two copies of the $T^2$ periodic quiver and connect their nodes by the $2d$ adjoint chiral multiplets $\Phi_{ii}$ coming from the $4d$ vector multiplets. (c) Finally, the $2d$ chiral field $X_{ij}$ and Fermi field $\Lambda_{ij}$ arising from a given $4d$ chiral field $\mathcal{X}_{ij}$ receive vertical shifts $s_{ij}$ and $(s_{ij}+1)$, respectively, with $s_{ij}=-1$ if $\mathcal{X}_{ij}$ is in the perfect matching and 0 otherwise.
\label{dimredschem}}
 \end{center}
 \end{figure}

Finally, we note that the $2d$ chiral field $X_{ij}$ and Fermi field $\Lambda_{ij}$ arising from a given $4d$ chiral field $\mathcal{X}_{ij}$ receive vertical shifts $s_{ij}$ and $(s_{ij}+1)$ respectively, and hence get split in the periodic quiver on $T^3$. \fref{dimredschem} shows the basics of the lifting algorithm for the periodic quiver. Section \sref{section_examples_CY3xC} is going to illustrate this construction with several examples. Before closing this section, let us emphasize that these periodic quivers can alternatively be derived from those of orbifolds by higgsing, following the ideas that are going to be discussed in section \sref{section_higgsing}.

\bigskip

\subsubsection{Thoughts Regarding Compactifications on Magnetized Tori}

We have just explained how the dimensional reduction from the $4d$ $\mathcal{N}=1$ gauge theory on D3-branes probing a toric $\mathrm{CY}_3$ to the $2d$ $(2,2)$ theory on D1-branes over toric $\mathrm{CY}_3 \times \mathbb{C}$ has a beautiful implementation as a lift of the periodic quiver from $T^2$ to $T^3$. Conversely, going to the 4d parent theory simply corresponds to a projection of the periodic quiver from $T^3$ to $T^2$ along the vertical direction. 

There are more general ways of going from $4d$ to $2d$, preserving less SUSY. In particular, if the $4d$ $\mathcal{N}=1$ theory has a $U(1)$ global (non-R) symmetry, it is possible to obtain a $2d$ $(0,2)$ theory by compactification on a 2-torus with background magnetic and $D$ fields for it \cite{Kutasov:2013ffl}.\footnote{Notice that the $2d$ $(0,2)$ theory obtained this way may break SUSY once quantum effects are taken into account.} In the presence of several $U(1)$ global symmetries, there are multiple possibilities for such a reduction.

Remarkably, the $4d$ toric theories under consideration have at least two $U(1)$ global flavor symmetries that, moreover, translate into the two fundamental directions of the corresponding periodic quivers on $T^2$. It is natural to wonder whether the $(0,2)$ reductions may be captured by a generalized lift of the periodic quiver that non-trivially takes into account the two directions of $T^2$. Conversely, it would be interesting to explore whether a given periodic quiver on $T^2$ can be obtained via different projections from periodic quivers on $T^3$ corresponding to $(2,2)$ or $(0,2)$ theories. We leave these questions for future investigation.

\bigskip

\subsection{General Analysis of the Classical Mesonic Moduli Space}

\label{section_general_moduli_space_dimensionally_reduced}

It is possible to present a concise discussion of the application of the forward algorithm to all the dimensionally reduced theories under consideration. We are going to see that, for this class of theories, the classical mesonic moduli space admits a simple combinatorial description. 

The $J$-terms are given by \eref{J_dim_red}, which we reproduce here for convenience
\bea
J_{ji} = \frac{\partial W}{\partial X_{ij}} ~.
\eea
It can be seen that the $J$-terms are equal to the F-terms of the $4d$ theory. Furthermore, they exclusively depend on variables that are in one-to-one correspondence with the $4d$ chiral fields. Solving these equations subject to D-terms thus gives rise to the $\mathrm{CY}_3$ factor of the moduli space. Moreover, this piece of the problem is exactly equivalent to finding the mesonic moduli space of the 4d toric theory. In this case, the $c-1$ GLSM fields are in one-to-one correspondence with perfect matchings of the corresponding brane tiling \cite{Franco:2005rj}. As introduced in section \sref{section_geometry_from_quivers}, we have 
\begin{align}
X_m = \prod_{\alpha=1}^{c-1} p_\alpha^{P_{m\alpha}} \,,
\end{align}
where $p_\alpha$, $\alpha=1,\ldots,c-1$ correspond to perfect matchings of the parent brane tiling.

Let us now consider the $E$-terms. Since we are interested in the abelian theory, \eref{J_dim_red} reduces to 
\beq
E_{ij} = \Phi_{ii}- \Phi_{jj}~.
\label{E_dim_red2}
\eeq
In other words, all the adjoint chiral fields $\Phi_{ii}$ must be equal. This is simply taken care of by introducing an additional GLSM field $s$ such that
\begin{align}
\Phi_{ii} = s ~,
\end{align}
for all gauge nodes $i$. This GLSM field gives rise to the additional point in the toric diagram representing the $\mathbb{C}$ factor of the geometry.

\bigskip

\subsection{Examples \label{sec:examplesOfCY3xC}}

\label{section_examples_CY3xC}

This section presents various explicit examples of $2d$ $(0,2)$ theories obtained by dimensional reduction of 4d $\mathcal{N}=1$ theories corresponding to brane tilings. We determine their classical mesonic moduli spaces using the forward algorithm, recovering the expected toric $\mathrm{CY}_3 \times \mathbb{C}$ geometry. The general structure of the corresponding periodic quivers and Calabi-Yau geometries that have been discussed in sections \sref{section_lift_periodic_quiver} and \sref{section_general_moduli_space_dimensionally_reduced} are recovered. 

\bigskip

\subsubsection{$\mathbb{C}^4/\mathbb{Z}_{3}$ $(1,1,1,0)$}

\begin{figure}[h]
\begin{center}
\resizebox{0.9\hsize}{!}{
\includegraphics[trim=0cm 0cm 0cm 0cm,totalheight=10 cm]{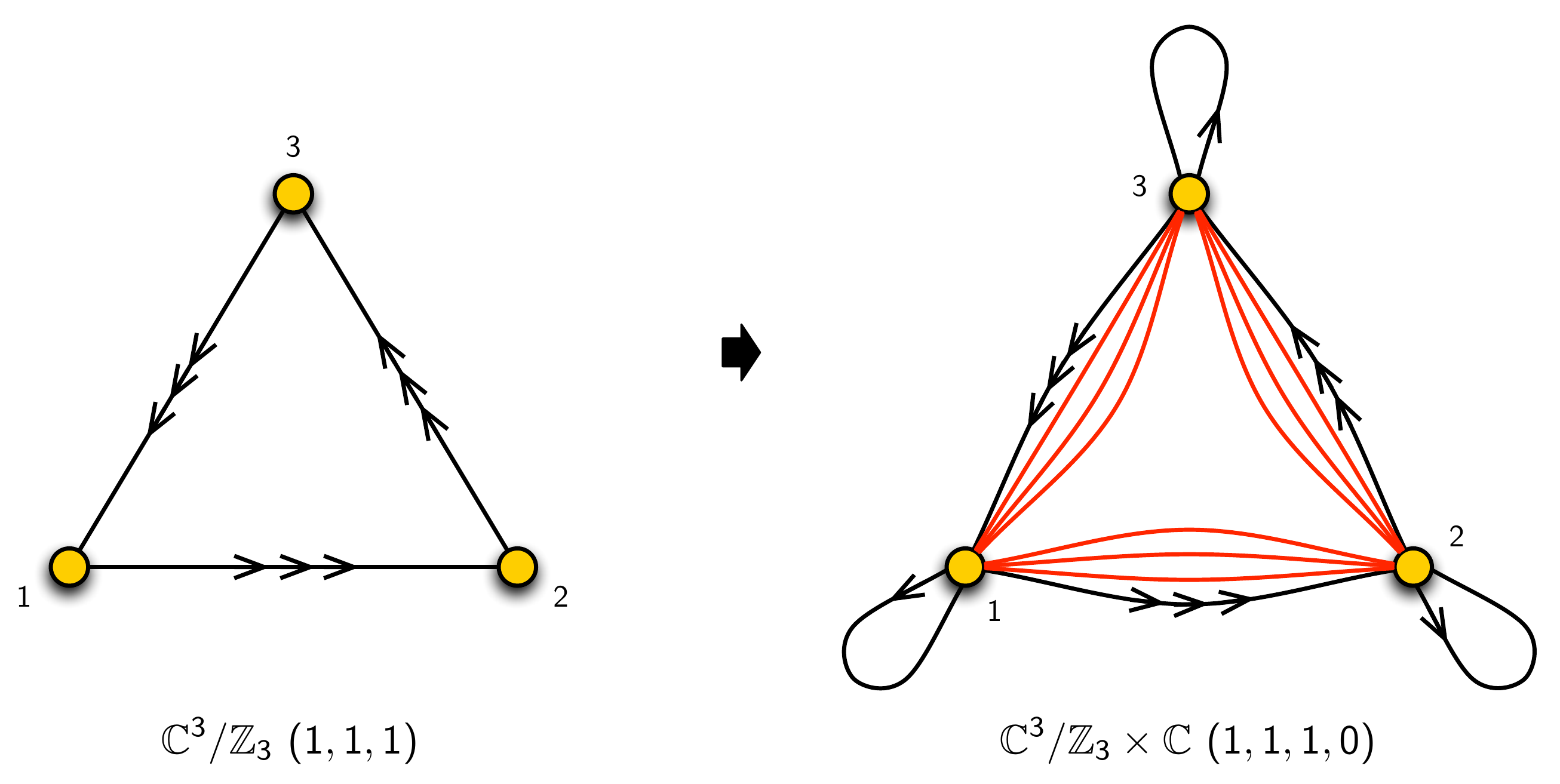}
}  
\caption{
Dimensional reduction from the quiver of the $4d$ $\mathcal{N}=1$ gauge theory corresponding to $\mathbb{C}^3/\mathbb{Z}_3$ with action $(1,1,1)$ to the quiver of the $2d$ $(2,2)$ theory corresponding to $\mathcal{C}^4/\mathbb{Z}_4$ with action $(1,1,1,0)$.
\label{fz30111}}
 \end{center}
 \end{figure}

Our first example of a dimensionally reduced theory corresponds to the $\mathbb{C}^4/\mathbb{Z}_{3}$ $(1,1,1,0)$ orbifold. Since this is an orbifold, it can be studied with the tools from both section \sref{sec:orbifold} and section \sref{section_lift_periodic_quiver}. The left of \fref{fz30111} shows the quiver for the $4d$ parent theory corresponding to $\mathbb{C}^3/\mathbb{Z}_{3}$ $(1,1,1)$. The $4d$ superpotential is given by 
\beal{es103e1b}
W= && X_{12} \cdot Y_{23} \cdot Z_{31} + Y_{12} \cdot Z_{23} \cdot X_{31} + Z_{12} \cdot X_{23} \cdot Y_{31} 
\nn\\
&&  - X_{12} \cdot Z_{23} \cdot Y_{31} - Y_{12} \cdot X_{23} \cdot Z_{31} - Z_{12} \cdot Y_{23} \cdot X_{31}~.~
\eea
Here and henceforth, overall traces over gauge indices are understood in superpotentials. Note that for simplicity, the font distinction between $4d$ and $2d$ chiral multiplets has been dropped. The theory has an $SU(3)$ global symmetry under which the three bifundamental fields with the same gauge charges transform in triplets. 
 
Dimensional reduction produces a $2d$ $(2,2)$ theory whose quiver is also depicted in \fref{fz30111}. The $J$- and $E$-term equations are 
 
\beq
\begin{array}{rcccccccc}
& & \ \ \ \ \ \ \ \ \ \ \ \ \ J & & & & \ \ \ \ \ \ \ \ \ \ \ \ \ E & & 
\\ 
  \Lambda_{12}^{1} : & \ \ \ & Y_{23}\cdot Z_{31} - Z_{23}\cdot Y_{31}& = & 0 & \ \ \ \ &   \Phi_{11}\cdot X_{12} -X_{12}\cdot \Phi_{22}  & = & 0\\
 \Lambda_{23}^{1} : & \ \ \ &  Y_{31}\cdot Z_{12} - Z_{31}\cdot Y_{12} & = & 0& \ \ \ \ & \Phi_{22}\cdot X_{23} - X_{23}\cdot \Phi_{33} & = & 0  \\ 
  \Lambda_{31}^{1} : & \ \ \ & Y_{12}\cdot Z_{23} - Z_{12}\cdot Y_{23}& = & 0 & \ \ \ \ & \Phi_{33}\cdot X_{31} - X_{31}\cdot \Phi_{11}   & = & 0  \\
  \Lambda_{12}^{2} : & \ \ \ &  Z_{23}\cdot X_{31} - X_{23}\cdot Z_{31} & = & 0& \ \ \ \ & \Phi_{11}\cdot Y_{12} - Y_{12}\cdot \Phi_{22}& = & 0 \\
  \Lambda_{23}^{2} : & \ \ \ & Z_{31}\cdot X_{12} -  X_{31}\cdot Z_{12} & = & 0 & \ \ \ \ & \Phi_{22}\cdot Y_{23} - Y_{23}\cdot \Phi_{33}& = & 0 \\
  \Lambda_{31}^{2} : & \ \ \ & Z_{12}\cdot X_{23} - X_{12}\cdot Z_{23} & = & 0 & \ \ \ \ & \Phi_{33}\cdot Y_{31} - Y_{31}\cdot \Phi_{11}& = & 0 \\
 \Lambda_{12}^{3} : & \ \ \ & X_{23}\cdot Y_{31} - Y_{23}\cdot X_{31}& = & 0& \ \ \ \ &  \Phi_{11}\cdot Z_{12} - Z_{12}\cdot \Phi_{22}  & = & 0  \\
 \Lambda_{23}^{3} : & \ \ \ &  X_{31}\cdot Y_{12} - Y_{31}\cdot X_{12}& = & 0& \ \ \ \ & \Phi_{22}\cdot Z_{23} - Z_{23}\cdot \Phi_{33} & = & 0 \\ 
 \Lambda_{31}^{3} : & \ \ \ &  X_{12}\cdot Y_{23} - Y_{12}\cdot X_{23}& = & 0  & \ \ \ \ &  \Phi_{33}\cdot Z_{31} - Z_{31}\cdot \Phi_{11} & = & 0\\
\end{array}
\label{es103e1}
\eeq

\medskip

\noindent Here, we use the notation for orbifold theories from section \sref{sec:orbifold}. In view of the general discussion on dimensional reduction, the adjoint chiral fields $\Phi_{ii}$ $(i=1,2,3)$ are precisely the fields originating from $4d$ vector multiplets. 

 \begin{figure}[H]
\begin{center}
\resizebox{0.4\hsize}{!}{
\includegraphics[trim=0cm 0cm 0cm 0cm,totalheight=10 cm]{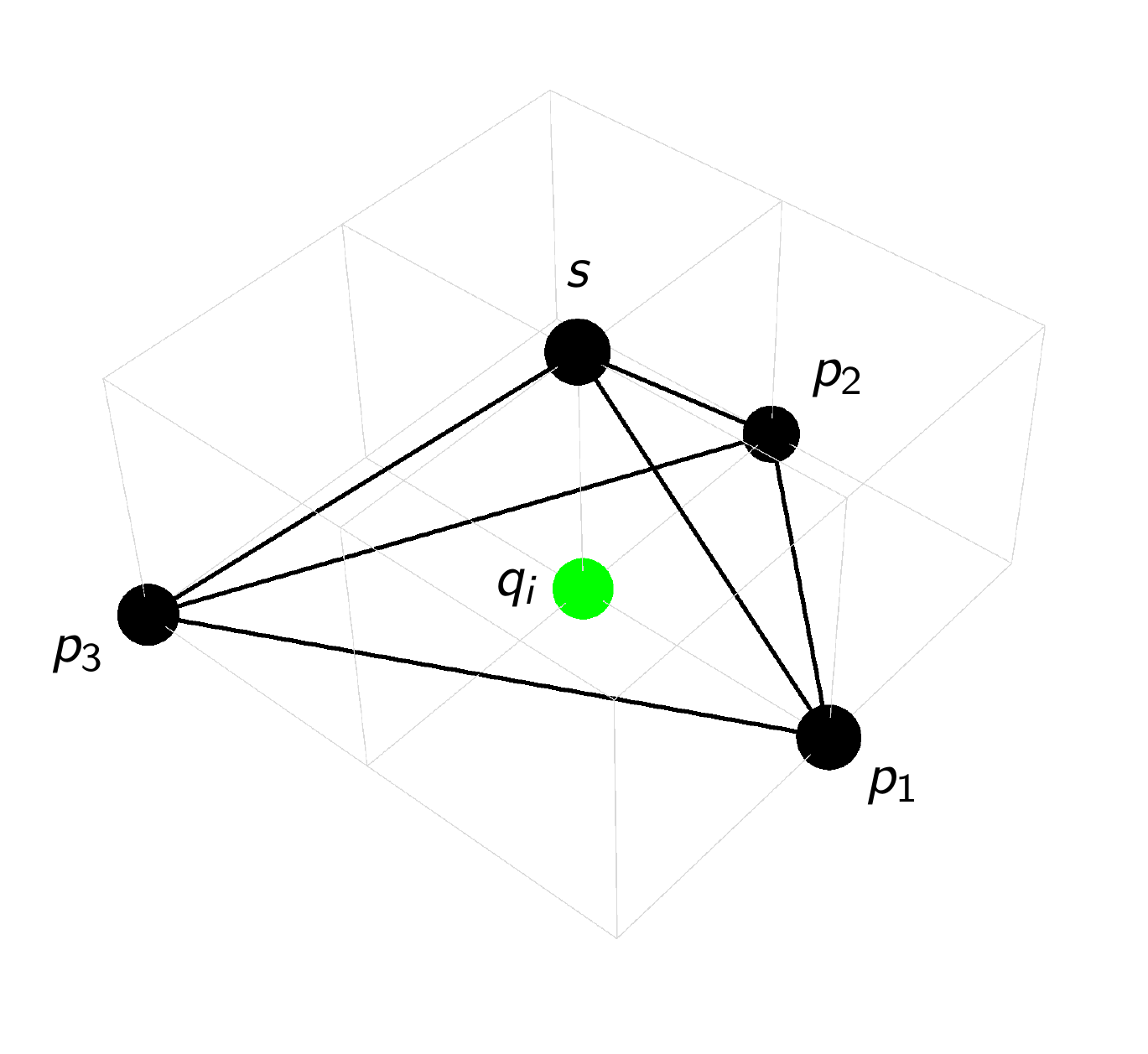}
}  
\vspace{-.8cm}\caption{
Toric diagram for the $\mathbb{C}^4/\mathbb{Z}_3$ orbifold with action $(1,1,1,0)$ obtained as the mesonic moduli space of the dimensionally reduced $\mathbb{C}^3/\mathbb{Z}_3$ with action $(1,1,1)$ gauge theory.
\label{ftoricz30111}}
 \end{center}
 \end{figure}

We now determine the classical mesonic moduli space of the gauge theory using the forward algorithm. For brevity, we only quote some of the matrices involved in the algorithm. Vanishing $J$- and $E$-terms can be reduced and summarized by the $K$-matrix as follows,
\beal{es103e3}
K=
\left(
\begin{array}{c|cccccccccccc}
\; & X_{12} & X_{23} & X_{31}& Y_{12}& Y_{23}&Y_{31}& Z_{12}& Z_{23}& Z_{31} & \Phi_{11} & \Phi_{22} & \Phi_{33} 
\\
\hline
X_{12}  & 1 & 0 & 0 & 0 & -1 & -1 & 0 & -1 & -1 & 0 & 0 & 0\\
X_{23}  & 0 & 1 & 0 & 0 & 1 & 0 & 0 & 1 & 0 & 0 & 0 & 0\\
X_{31}  & 0 & 0 & 1 & 0 & 0 & 1 & 0 & 0 & 1 & 0 & 0 & 0\\
Y_{12}  & 0 & 0 & 0 & 1 & 1 & 1 & 0 & 0 & 0 & 0 & 0 & 0\\
X_{12} & 0 & 0 & 0 & 0 & 0 & 0 & 1 & 1 & 1  & 0 & 0 & 0 \\
\Phi_{11} & 0 & 0 & 0 & 0 & 0 & 0 & 0 & 0 & 0 &  1 & 1 & 1 \\
\end{array}
\right).~
\eea
The $P$-matrix becomes
\beal{es103e4}
P=
\left(
\begin{array}{c|ccccccc}
\; & p_1 & p_3 & p_4 & q_1 & q_2 & q_3 & s\\
\hline
X_{12} & 1 & 0 & 0 & 1 & 0 & 0  & 0\\
X_{23} & 1 & 0 & 0 & 0 & 1 & 0  & 0\\
X_{31} & 1 & 0 & 0 & 0 & 0 & 1  & 0\\
Y_{12} & 0 & 1 & 0 & 1 & 0 & 0  & 0\\
Y_{23} & 0 & 1 & 0 & 0 & 1 & 0  & 0\\
Y_{31} & 0 & 1 & 0 & 0 & 0 & 1  & 0\\
Z_{12} & 0 & 0 & 1 & 1 & 0 & 0  & 0\\
Z_{23} & 0 & 0 & 1 & 0 & 1 & 0  & 0\\
Z_{31} & 0 & 0 & 1 & 0 & 0 & 1  & 0\\
\Phi_{11} & 0 & 0 & 0 & 0 & 0 & 0  & 1\\
\Phi_{22} & 0 & 0 & 0 & 0 & 0 & 0  & 1\\
\Phi_{33} & 0 & 0 & 0 & 0 & 0 & 0  & 1\\
\end{array}
\right)
~.~
\eea

   \begin{figure}[ht!]
\begin{center}
\resizebox{0.7\hsize}{!}{
\includegraphics[trim=0cm 0cm 0cm 0cm,totalheight=10 cm]{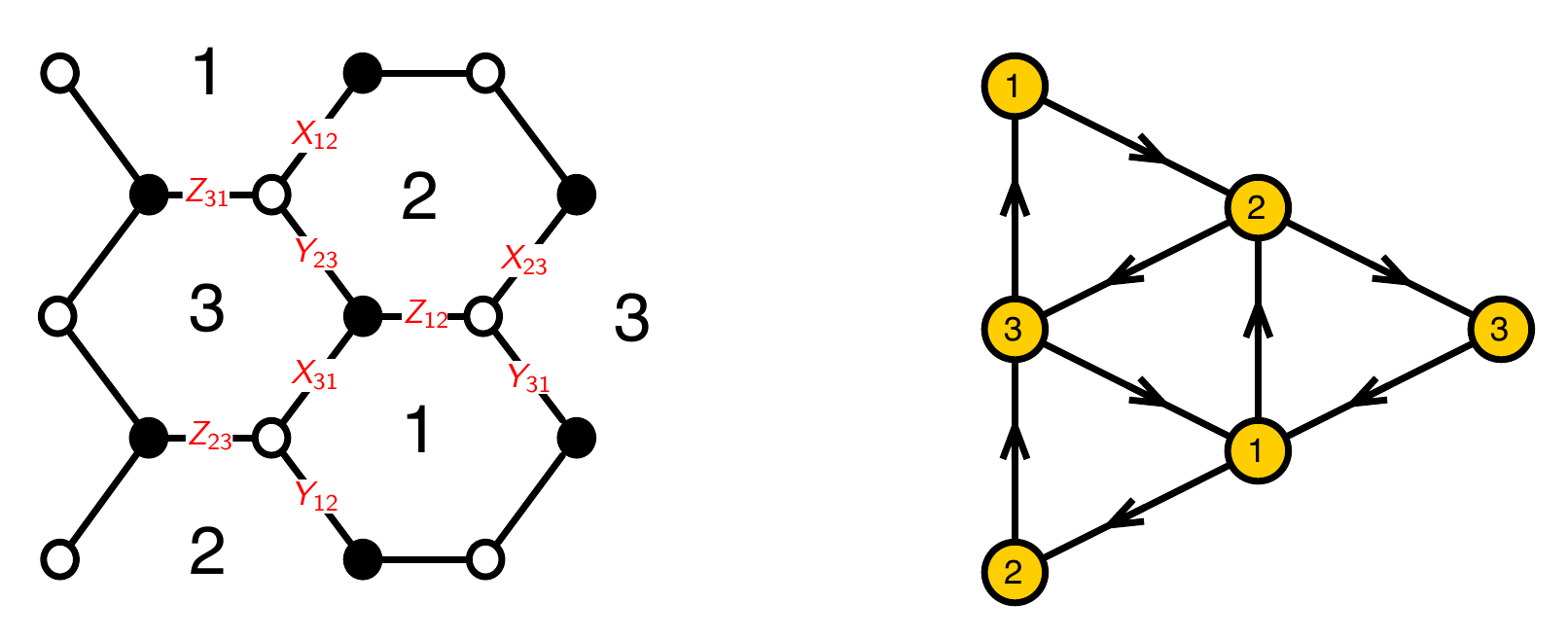}
}  
\caption{
Brane tiling for $\mathbb{C}^3/\mathbb{Z}_{3}$ $(1,1,1)$ encoding the corresponding $4d$ $\mathcal{N}=1$ gauge theory and its dual periodic quiver on $T^2$.
\label{fzc3z32}}
 \end{center}
 \end{figure}

 \begin{figure}[h]
\begin{center}
\resizebox{0.8\hsize}{!}{
\includegraphics[trim=0cm 0cm 0cm 0cm,totalheight=10 cm]{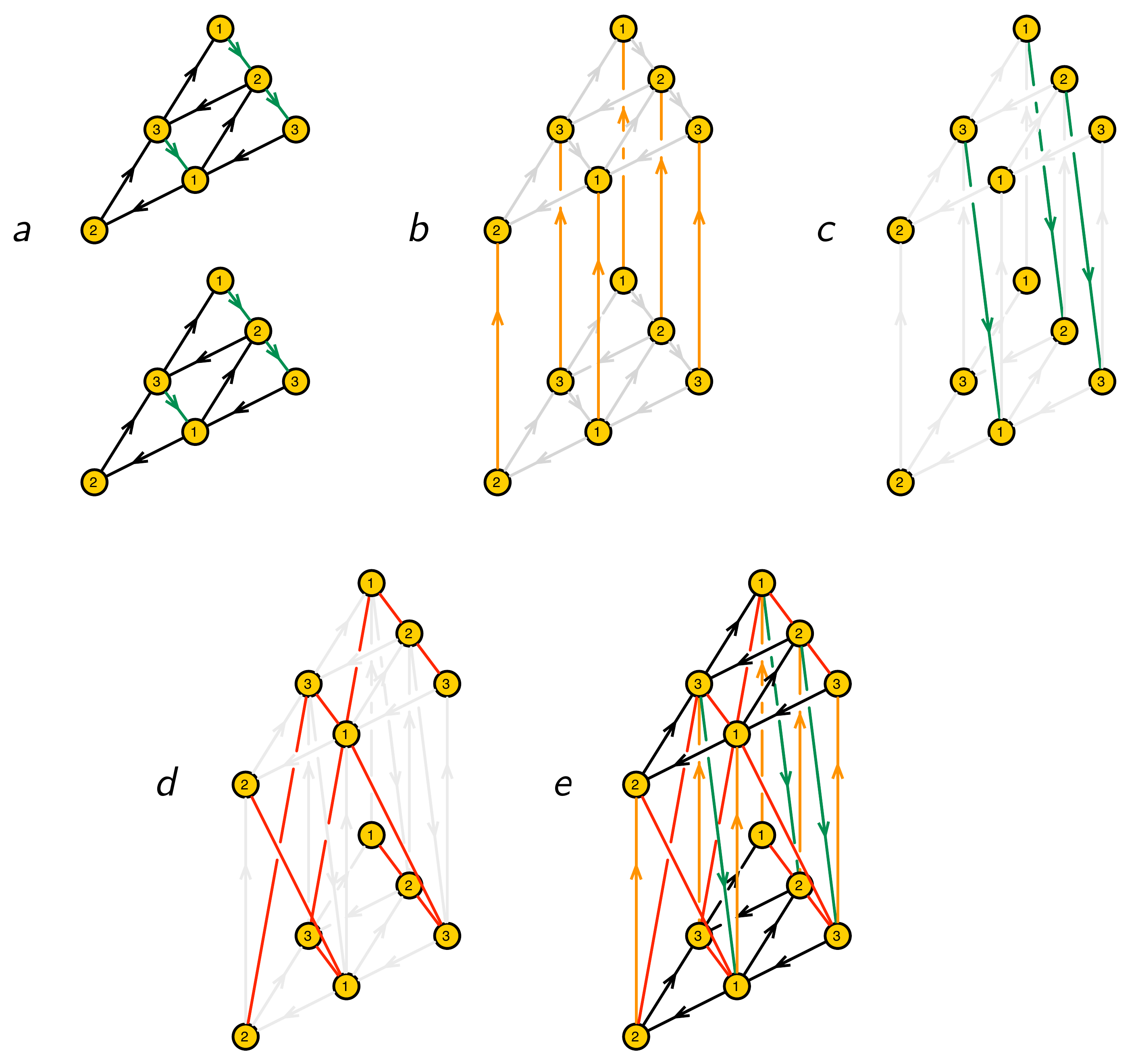}
}  
\caption{Lift of the periodic quiver from $T^2$ to $T^3$. (a) Two copies of the $4d$ periodic quiver with the chiral fields for the perfect matching $p_1$ shown in green. (b) The adjoints $\Phi_{ii}$ connect these planes with a vertical shift (+1). (c) The $2d$ chiral fields descending from those in $p_1$ receive a $(-1)$ vertical shift. (d) The Fermi fields coming from $4d$ chiral fields in $p_1$ have no vertical shift, while all others receive a $(-1)$ shift. (e) The full periodic quiver on $T^3$ for $\mathbb{C}^4/\mathbb{Z}_{3}$ $(1,1,1,0)$. 
\label{fzc3z3}}
 \end{center}
 \end{figure}

Next, we use the previous matrices to determine $Q_{JE}$ and the incidence matrix to find $Q_D$, which in turn we use to compute the toric diagram matrix
\beal{es103e5}
G=\left(
\begin{array}{ccccccc}
p_1 & p_2 & p_3 & q_1 & q_2 & q_3 & s \\
\hline
1 & 1 & 1 & 1 & 1 & 1 &  1 \\
 1 & 0 & -1 & 0 & 0 & 0 &  0\\
 0 & 1 & -1 & 0 & 0 & 0 &  0 \\
 0 & 0 & 0 & 0 & 0 & 0 &  1 \\
\end{array}
\right)
~,~
\eea
The corresponding toric diagram is shown in \fref{ftoricz30111}. It corresponds to the $\mathbb{C}^4/\mathbb{Z}_3$ orbifold with action $(1,1,1,0)$, as expected.
 
Let us now use the lifting algorithm introduced in section \sref{section_lift_periodic_quiver} to generate the periodic quiver on $T^3$ encoding the $2d$ $(2,2)$ theory. The starting point is the brane tiling and the dual periodic quiver for the $4d$ theory associated with $\mathbb{C}^3/\mathbb{Z}_{3}$ $(1,1,1)$, which are shown in \fref{fzc3z32}. The periodic quiver on $T^3$ is presented in \fref{fzc3z3}.
 
\bigskip
 
\subsubsection{$\mathcal{C} \times \mathbb{C}$}

\begin{figure}[ht!]
\begin{center}
\resizebox{1\hsize}{!}{
\includegraphics[trim=0cm 0cm 0cm 0cm,totalheight=10 cm]{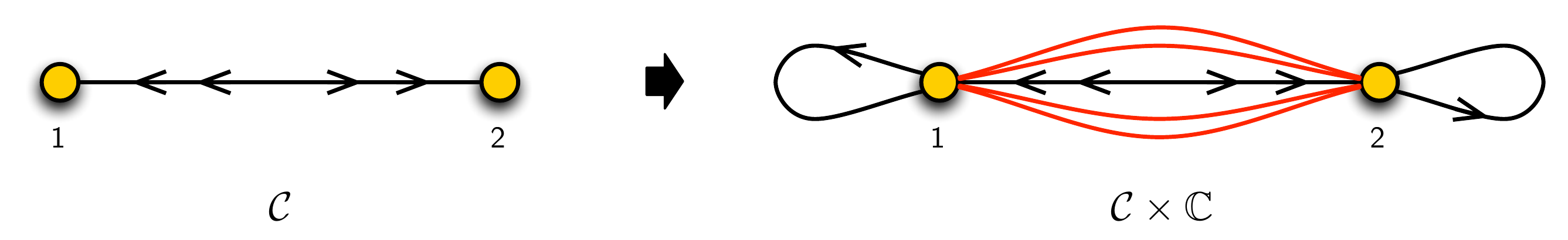}
}  
\caption{Dimensional reduction from the quiver of the $4d$ $\mathcal{N}=1$ gauge theory corresponding to the conifold $\mathcal{C}$ to the quiver of the $2d$ $(2,2)$ theory corresponding to $\mathcal{C}\times \mathbb{C}$.
\label{fzconfioldaaa}}
 \end{center}
 \end{figure}

Let us now investigate the theory for $\mathcal{C}\times \mathbb{C}$, where $\mathcal{C}$ refers to the conifold \cite{Klebanov:1998hh}. The 4d quiver for the conifold theory is given on the left of \fref{fzconfioldaaa}. The corresponding superpotential is 
\beal{es200a1c}
W = X_{12} \cdot Y_{21} \cdot Y_{12} \cdot X_{21} - X_{12} \cdot X_{21}  \cdot Y_{12} \cdot Y_{21}  ~.~
\eea
The theory has an $SU(2) \times SU(2)$ global mesonic symmetry.
 
The dimensionally reduced quiver is shown on the right of \fref{fzconfioldaaa}. The $J$- and $E$-terms are given by
\beq
\begin{array}{rclccclcc}
& & \ \ \ \ \ \ \ \ \ \ \ \ \ \ \ \ \ \ \ \ J & & & & \ \ \ \ \ \ \ \ \ \ \ \ \ \ E & & \\
 \Lambda_{12}^{1} : & \ \ \ & X_{21}\cdot X_{12}\cdot Y_{21} - Y_{21}\cdot X_{12}\cdot X_{21}& = & 0  & \ \ \ \ & \Phi_{11}\cdot Y_{12} - Y_{12}\cdot \Phi_{22} & = & 0  \\
 \Lambda_{21}^{1} : & \ \ \ & X_{12}\cdot Y_{21}\cdot Y_{12} - Y_{12}\cdot Y_{21}\cdot X_{12}& = & 0  & \ \ \ \ & \Phi_{22}\cdot X_{21} - X_{21}\cdot \Phi_{11} & = & 0  \\
 \Lambda_{12}^{2} : & \ \ \ & Y_{21}\cdot Y_{12}\cdot X_{21} -X_{21}\cdot Y_{12}\cdot Y_{21} & = & 0  & \ \ \ \ & \Phi_{11}\cdot X_{12} - X_{12}\cdot \Phi_{22} & = & 0 \\
 \Lambda_{21}^{2} : & \ \ \ &  Y_{12}\cdot X_{21}\cdot X_{12} -X_{12}\cdot X_{21}\cdot Y_{12}& = & 0  & \ \ \ \ & \Phi_{22}\cdot Y_{21} - Y_{21}\cdot \Phi_{11}& = & 0
\end{array}
\label{es200a1-new}
\eeq

\begin{figure}[H]
\begin{center}
\resizebox{0.3\hsize}{!}{
\includegraphics[trim=0cm 0cm 0cm 0cm,totalheight=10 cm]{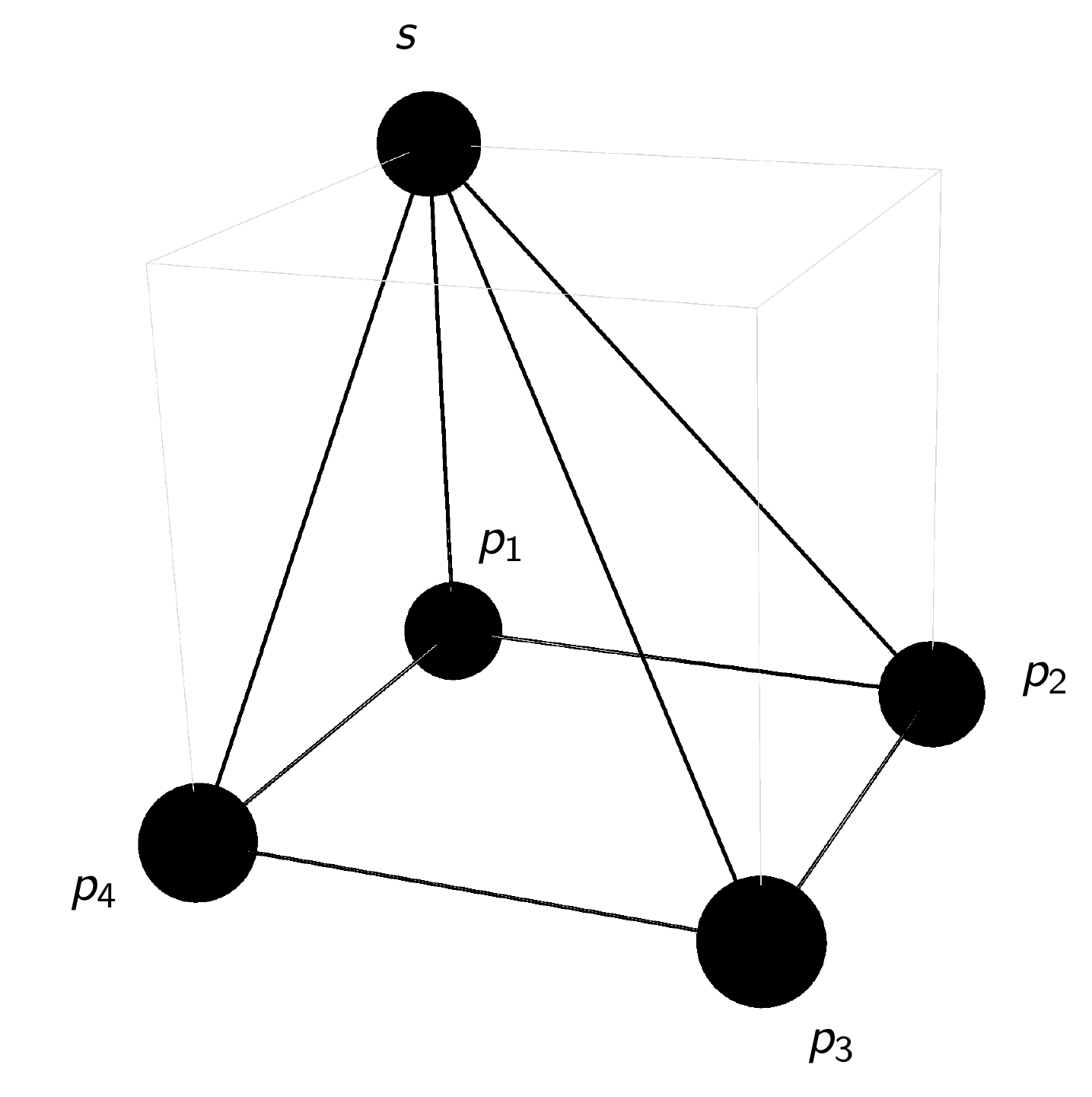}
}  
\caption{Toric diagram for $\mathcal{C} \times \mathbb{C}$ obtained as the mesonic moduli space of the dimensionally reduced conifold gauge theory.
\label{fzconfiold}}
 \end{center}
 \end{figure}

Proceeding with the forward algorithm, we obtain the $K$-matrix
\beal{es200a2}
K=
\left(
\begin{array}{c|cccccc}
\; &X_{21} & X_{12} & Y_{21} & Y_{12} & \Phi_{11} & \Phi_{22}\\
\hline
X_{21} & 1 & 0 & 0 & 0 & 0 & 0 \\
X_{12} & 0 & 1 & 0 & 0 & 0 & 0 \\
Y_{21} & 0 & 0 & 1 & 0 & 0 & 0 \\
Y_{12} & 0 & 0 & 0 & 1 & 0 & 0 \\
\Phi_{22} & 0 & 0 & 0 & 0 & 1 & 1 \\
\end{array}
\right)
~,~
\eea
the $P$-matrix
\beal{es200a3}
P=
\left(
\begin{array}{c|ccccc}
\; & p_1 & p_2 & p_3 & p_4 & s \\
\hline
X_{21} & 1 & 0 & 0 & 0 & 0 \\
X_{12} & 0 & 1 & 0 & 0 & 0 \\
Y_{21} & 0 & 0 & 1 & 0 & 0 \\
Y_{12} & 0 & 0 & 0 & 1 & 0 \\
\Phi_{11} & 0 & 0 & 0 & 0 & 1 \\
\Phi_{22} & 0 & 0 & 0 & 0 & 1 \\
\end{array}
\right)
~,~
\eea
and the toric diagram matrix 
\beal{es200a4}
G=
\left(
\begin{array}{ccccc}
p_1 & p_2 & p_3 & p_4 & s \\
\hline
 1 & 1 & 1 & 1 & 1 \\
 1 & 1 & 1 & 1 & 0 \\
 0 & 1 & 1 & 0 & 0 \\
 0 & 0 & 1 & 1 & 0 \\
\end{array}
\right)
~.~
\eea
The corresponding toric diagram is shown in \fref{fzconfiold} and it indeed corresponds to $\mathcal{C} \times \mathbb{C}$.

\begin{figure}[ht!]
\begin{center}
\resizebox{0.4\hsize}{!}{
\includegraphics[trim=0cm 0cm 0cm 0cm,totalheight=5 cm]{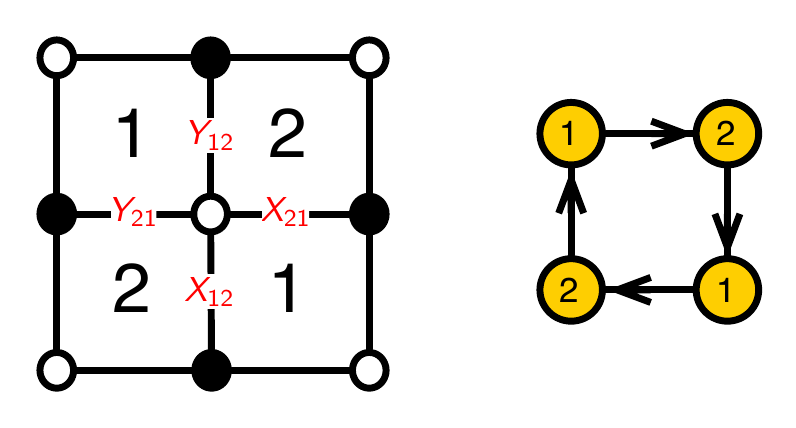}
}  
\caption{Brane tiling for $\mathcal{C}$ encoding the corresponding $4d$ $\mathcal{N}=1$ gauge theory and its dual periodic quiver on $T^2$.
\label{fzconifold2}}
 \end{center}
 \end{figure}

\begin{figure}[ht!]
\begin{center}
\resizebox{0.8\hsize}{!}{
\includegraphics[trim=0cm 0cm 0cm 0cm,totalheight=10 cm]{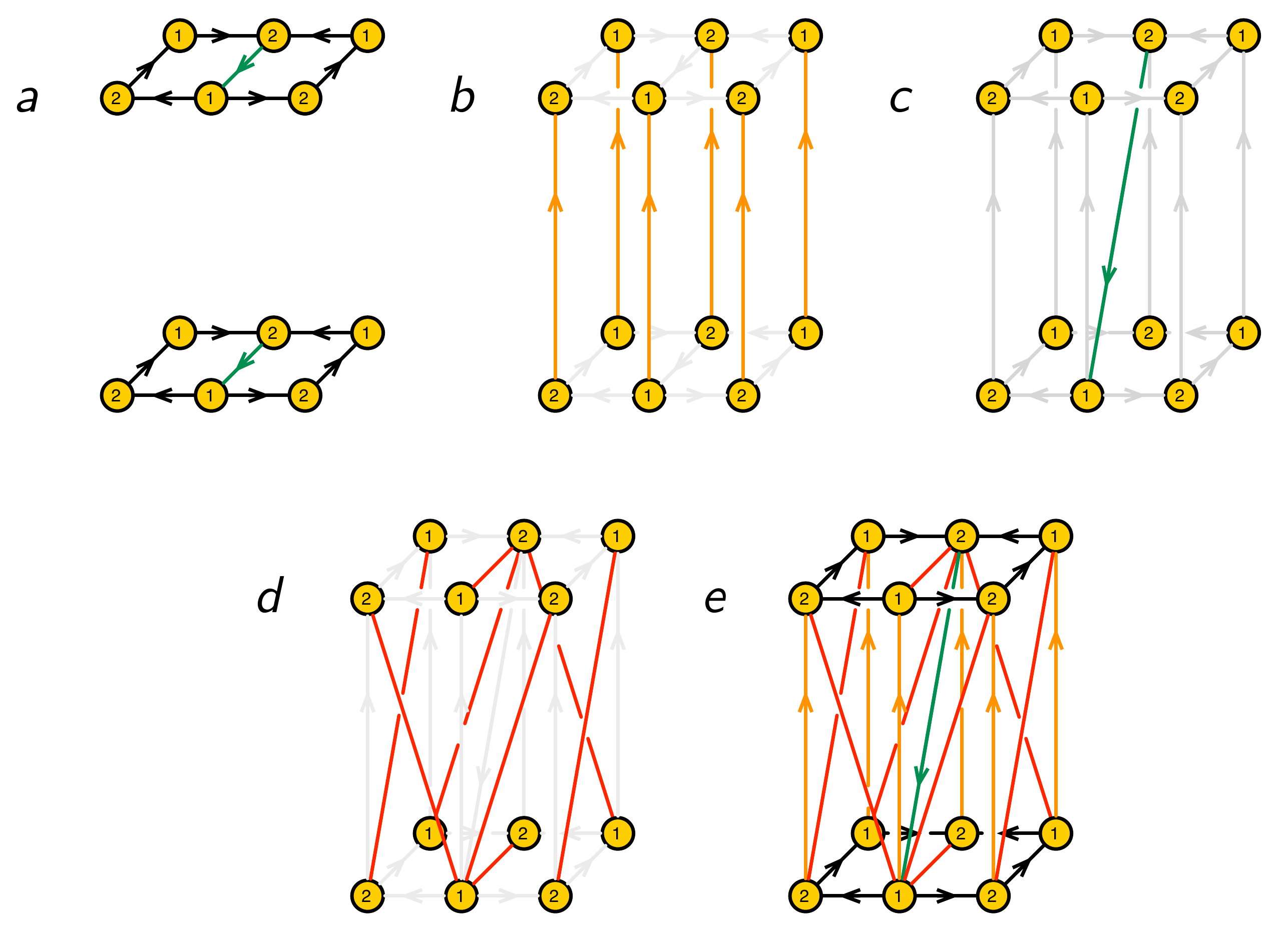}
}  
\caption{Lift of the periodic quiver from $T^2$ to $T^3$. (a) Two copies of the $4d$ periodic quiver with the chiral fields for the perfect matching $p_1$ shown in green. (b) The adjoints $\Phi_{ii}$ connect these planes with a vertical shift (+1). (c) The $2d$ chiral fields descending from those in $p_1$ receive a $(-1)$ vertical shift. (d) The Fermi fields coming from $4d$ chiral fields in $p_1$ have no vertical shift, while all others receive a $(-1)$ shift. (e) The full periodic quiver on $T^3$ for $\mathcal{C} \times\mathbb{C}$.
\label{fzconifold}}
 \end{center}
 \end{figure}

Let us now construct the periodic quiver for the $2d$ theory. \fref{fzconifold2} shows the brane tiling and the dual periodic quiver for the parent conifold theory. The lift to a periodic quiver on $T^3$ is presented in \fref{fzconifold}.

\bigskip 
 
\subsubsection{$\text{SPP}\times \mathbb{C}$}

\begin{figure}[H]
\begin{center}
\resizebox{0.9\hsize}{!}{
\includegraphics[trim=0cm 0cm 0cm 0cm,totalheight=10 cm]{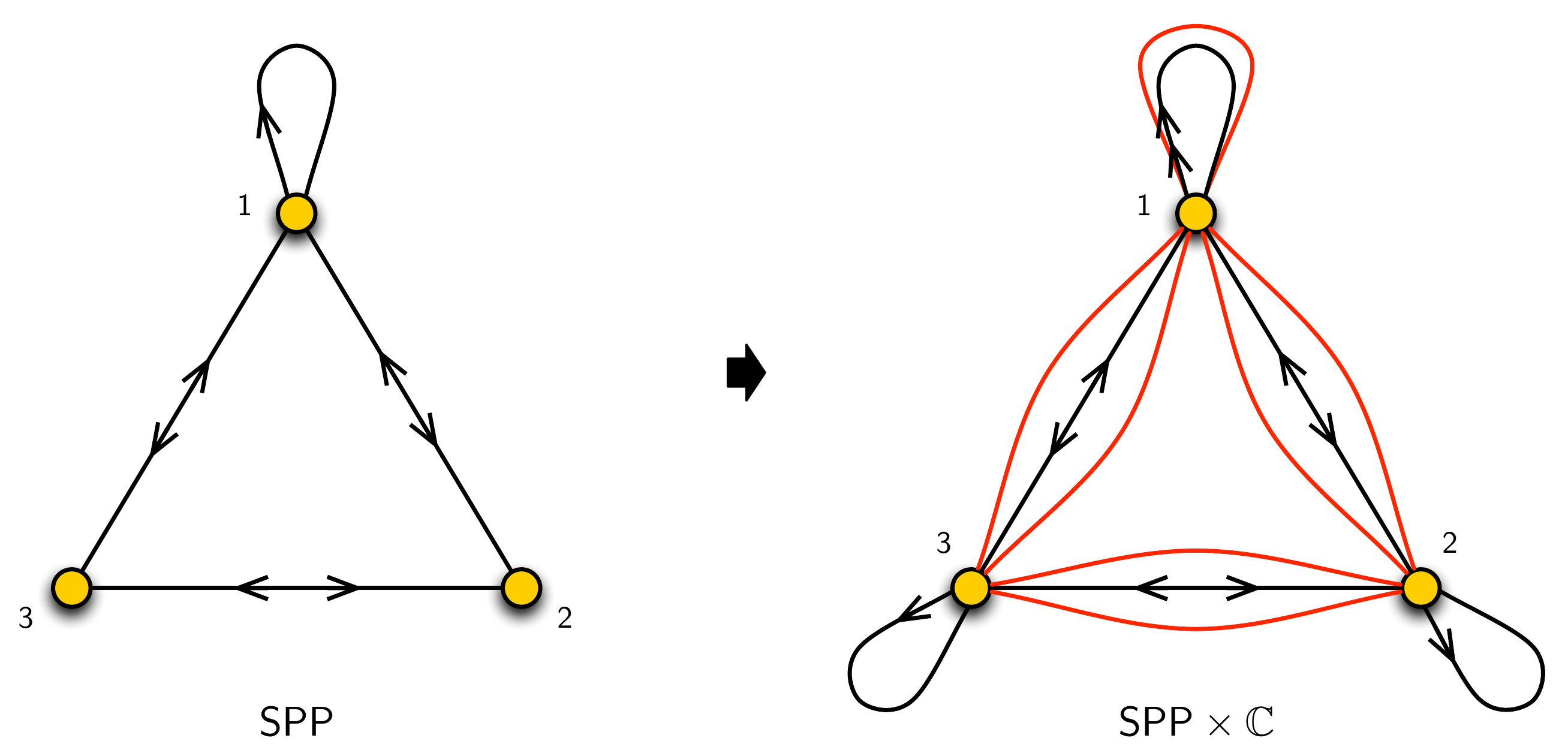}
}  
\caption{Dimensional reduction from the quiver of the $4d$ $\mathcal{N}=1$ gauge theory corresponding to SPP to the quiver of the $2d$ $(0,2)$ theory corresponding to $\text{SPP}\times \mathbb{C}$.
\label{fzspp}}
 \end{center}
 \end{figure}

The last example of this section is $\text{SPP}\times \mathbb{C}$, where SPP indicates the complex cone over the suspended pinch point \cite{Morrison:1998cs}. The $4d$ quiver for the SPP theory is shown on the left of \fref{fzspp} and the superpotential is
\beal{es200a20c}
W =  X_{13} \cdot X_{31}  \cdot X_{11} + X_{12}\cdot X_{23}\cdot X_{32}\cdot X_{21} - X_{12}\cdot X_{21}\cdot X_{11} - X_{13}\cdot X_{32}\cdot X_{23}\cdot X_{31}  ~.~
\nn\\
\eea
This theory has an $SU(2)$ global symmetry.

The dimensionally reduced quiver is shown on the right of \fref{fzspp}. The $J$- and $E$-terms are given by
 
\beq
\begin{array}{rcrccclcc}
& & J  \ \ \ \ \ \ \ \ \ \ \ \ \ & & & & \ \ \ \ \ \ \ \ \ \ \ \ \ \ \ \ \ \ \ E & & \\
\Lambda_{11 } : & \ \ \ & X_{13} \cdot X_{31} - X_{12} \cdot X_{21}& = & 0 & \ \ \ \ &\Phi_{11} \cdot X_{11} - X_{11} \cdot \Phi_{11}&= & 0 \\
\Lambda_{21} : & \ \ \ & X_{12} \cdot X_{23} \cdot X_{32} - X_{11} \cdot X_{12} & = & 0 & \ \ \ \ &\Phi_{22} \cdot X_{21} - X_{21} \cdot \Phi_{11}& = & 0 \\
\Lambda_{12} : & \ \ \ & X_{21} \cdot X_{11} - X_{23} \cdot X_{32} \cdot X_{21}& = & 0 & \ \ \ \ &\Phi_{11} \cdot X_{12} - X_{12} \cdot \Phi_{22}& = & 0 \\
\Lambda_{31} : & \ \ \ & X_{13} \cdot X_{32} \cdot X_{23}- X_{11} \cdot X_{13}& = & 0 & \ \ \ \ & \Phi_{33} \cdot X_{31} - X_{31} \cdot \Phi_{11}& = & 0 \\
\Lambda_{13} : & \ \ \ & X_{31} \cdot X_{11} - X_{32} \cdot X_{23} \cdot X_{31}& = & 0 & \ \ \ \ &\Phi_{11} \cdot X_{13} - X_{13} \cdot \Phi_{33}& = & 0 \\
\Lambda_{32} : & \ \ \ & X_{21} \cdot X_{12} \cdot X_{23} -X_{23} \cdot X_{31} \cdot X_{13}  & = & 0 & \ \ \ \ &\Phi_{33} \cdot X_{32} - X_{32} \cdot \Phi_{22} & = & 0 \\
\Lambda_{23} : & \ \ \ & X_{32} \cdot X_{21} \cdot X_{12} - X_{31} \cdot X_{13} \cdot X_{32}& = & 0 & \ \ \ \ & \Phi_{22} \cdot X_{23} - X_{23} \cdot \Phi_{33}& = & 0 
\end{array}
\label{es200a20}
\eeq

\begin{figure}[H]
\begin{center}
\resizebox{0.4\hsize}{!}{
\includegraphics[trim=0cm 0cm 0cm 0cm,totalheight=10 cm]{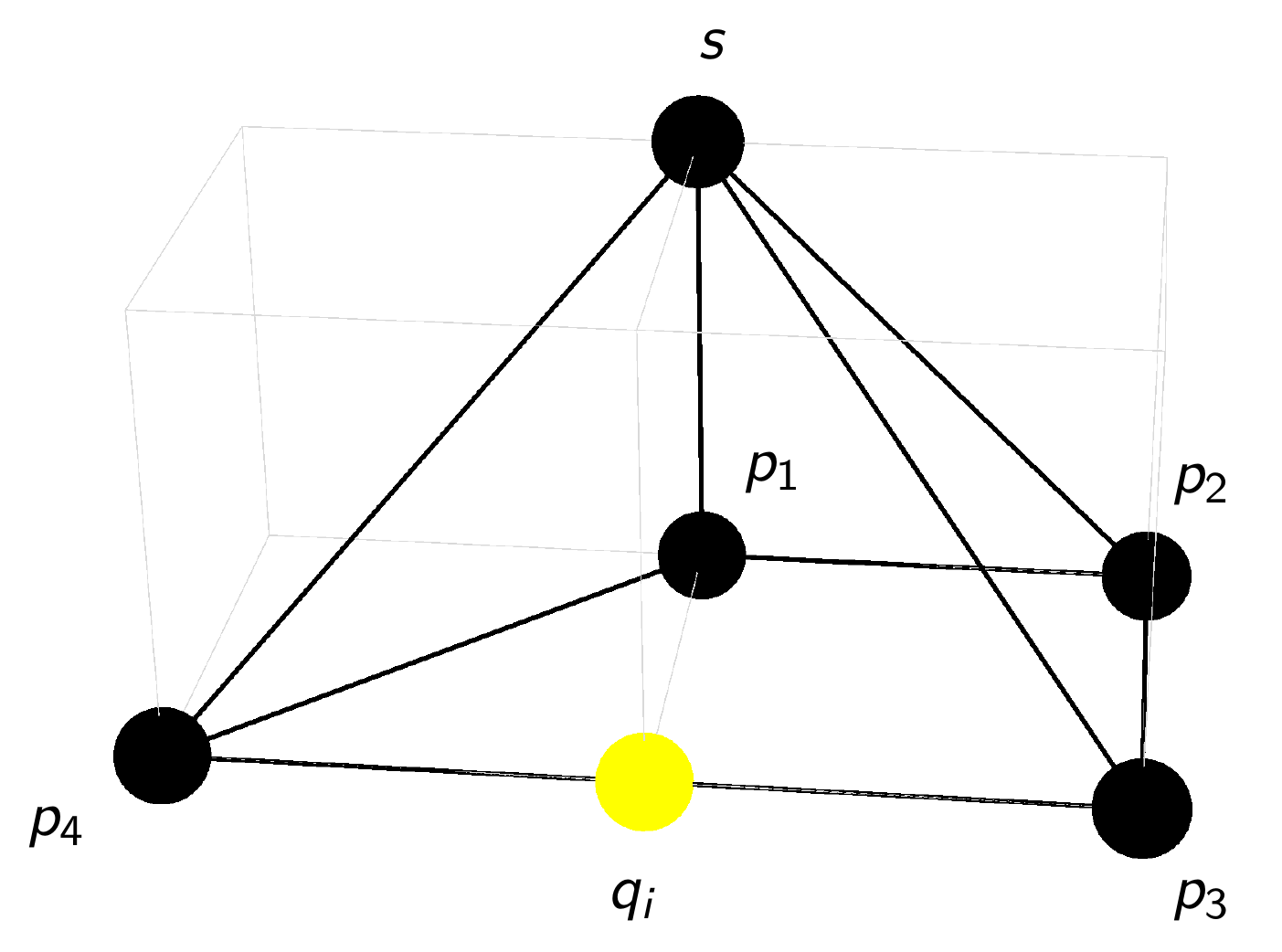}
}  
\caption{Toric diagram for $\text{SPP} \times \mathbb{C}$ obtained as the mesonic moduli space of the dimensionally reduced SPP gauge theory.
\label{fzsppc}}
 \end{center}
 \end{figure}

Using the forward algorithm to find the classical mesonic moduli space of this theory, we obtain the $K$-matrix
\beal{es200a21}
K=
\left(
\begin{array}{c|cccccccccc}
\; & X_{23}&  X_{11}& X_{32}& X_{13}& X_{21}& X_{12}& X_{31} & \Phi_{11}& \Phi_{22}& \Phi_{33}
\\
\hline
X_{23}  & 1 & 1 & 0 & 0 & 0 & 0 & 0 & 0 & 0 & 0\\
X_{32} & 0 & 1 & 1 & 0 & 0 & 0 & 0 & 0 & 0 & 0 \\
X_{13}  & 0 & 0 & 0 & 1 & 0 & 0 & -1 & 0 & 0 & 0\\
X_{21}  & 0 & 0 & 0 & 0 & 1 & 0 & 1 & 0 & 0 & 0\\
X_{12} & 0 & 0 & 0 & 0 & 0 & 1 & 1 & 0 & 0 & 0  \\
X_{22} & 0 & 0 & 0 & 0 & 0 & 0 & 0 & 1 & 1 & 1 \\
\end{array}
\right)  
~,~
\eea
the $P$-matrix
\beal{es200a22}
P=
\left(
\begin{array}{c|ccccccc}
\; & p_1 & p_2 & p_3 & p_4 & q_1 & q_2 & s \\
\hline
X_{23} & 1 & 0 & 0 & 0 & 0 & 0 & 0 \\
X_{11} & 1 & 1 & 0 & 0 & 0 & 0 & 0 \\
X_{32} & 0 & 1 & 0 & 0 & 0 & 0 & 0 \\
X_{13} & 0 & 0 & 1 & 0 & 1 & 0 & 0 \\
X_{21} & 0 & 0 & 1 & 0 & 0 & 1 & 0 \\
X_{12} & 0 & 0 & 0 & 1 & 1 & 0 & 0 \\
X_{31} & 0 & 0 & 0 & 1 & 0 & 1 & 0 \\
\Phi_{11} & 0 & 0 & 0 & 0 & 0 & 0 & 1 \\
\Phi_{22} & 0 & 0 & 0 & 0 & 0 & 0 & 1 \\
\Phi_{33} & 0 & 0 & 0 & 0 & 0 & 0 & 1 \\
\end{array}
\right)
\eea
and the toric diagram matrix
\beal{es200a23}
G=
\left(
\begin{array}{ccccccc}
 p_1 & p_2 & p_3 & p_4 & q_1 & q_2 & s \\
\hline
 1 & 1 & 1 & 1 & 1 & 1 & 1 \\
 1 & 1 & 1 & 1 & 1 & 1 & 0 \\
 0 & 1 & 1 & -1 & 0 & 0 & 0 \\
 0 & 0 & 1 & 1 & 1 & 1 & 0 \\
 \end{array}
\right)~.~
\eea
The corresponding toric diagram is shown in \fref{fzsppc} and it indeed is the one for $\text{SPP}\times \mathbb{C}$.

\begin{figure}[ht!]
\begin{center}
\resizebox{0.9\hsize}{!}{
\includegraphics[trim=0cm 0cm 0cm 0cm,totalheight=10 cm]{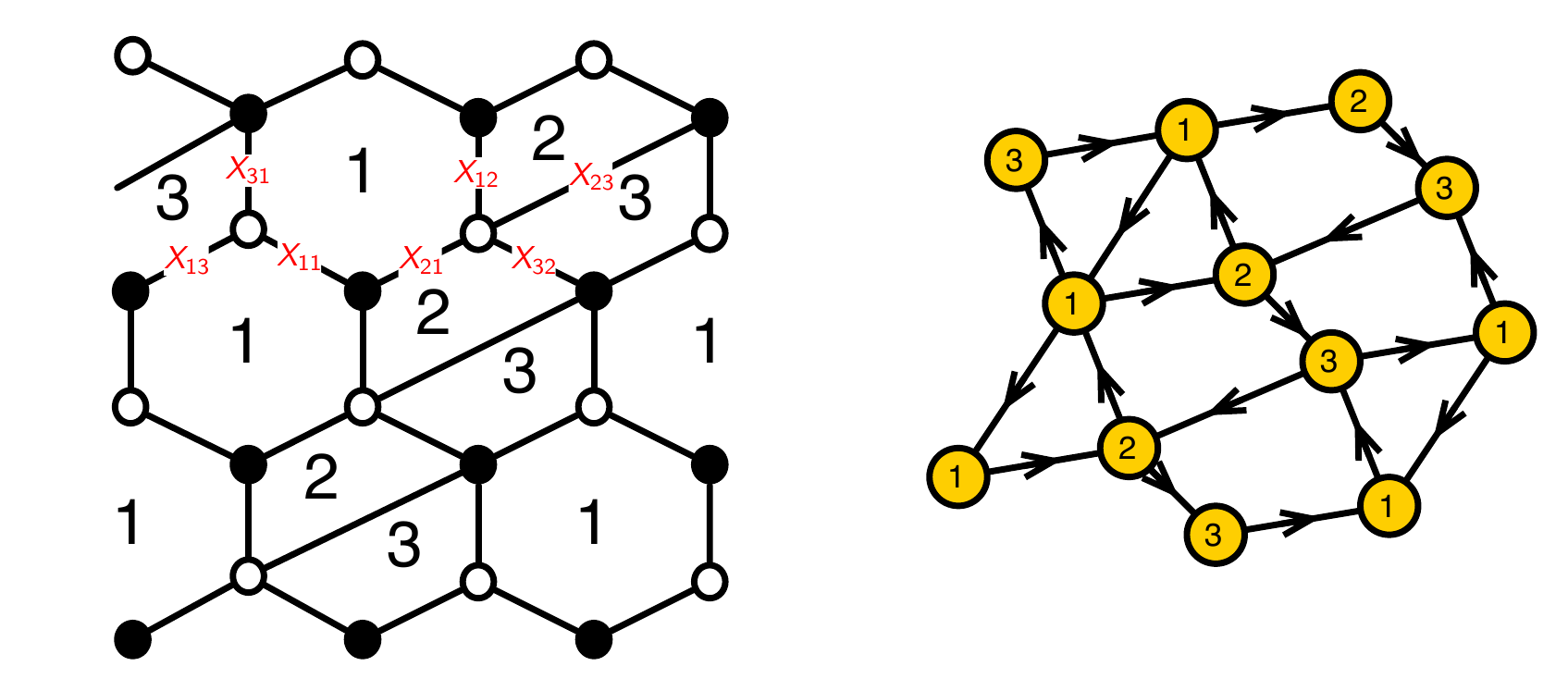}
}  
\caption{Brane tiling for SPP encoding the corresponding $4d$ $\mathcal{N}=1$ gauge theory and its dual periodic quiver on $T^2$.
\label{fzspp2}}
 \end{center}
 \end{figure}

\begin{figure}[ht!]
\begin{center}
\resizebox{0.8\hsize}{!}{
\includegraphics[trim=0cm 0cm 0cm 0cm,totalheight=10 cm]{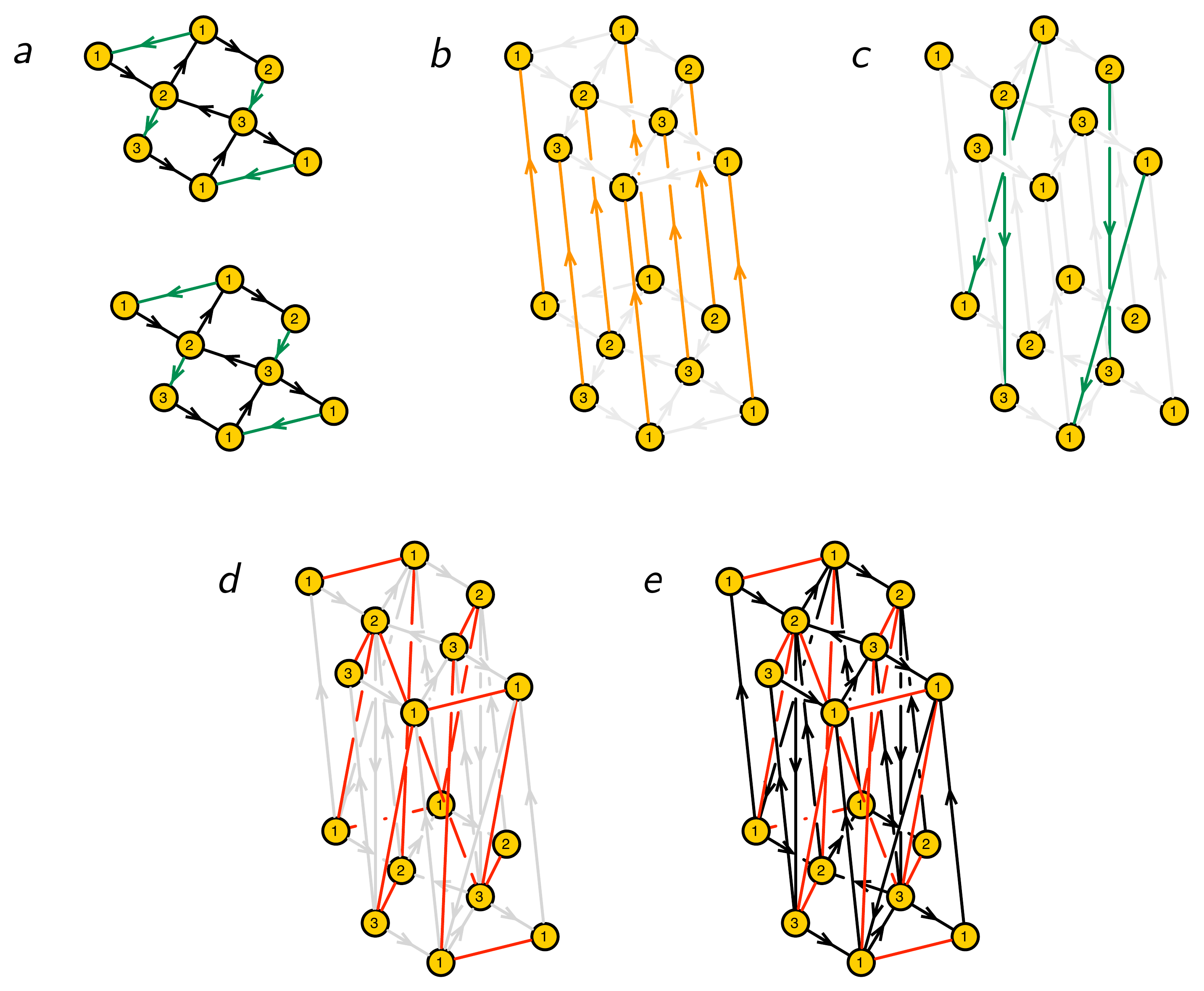}
}  
\caption{Lift of the periodic quiver from $T^2$ to $T^3$. (a) Two copies of the $4d$ periodic quiver with the chiral fields for the perfect matching $p_1$ shown in green. (b) The adjoints $\Phi_{ii}$ connect these planes with a vertical shift (+1). (c) The $2d$ chiral fields descending from those in $p_1$ receive a $(-1)$ vertical shift. (d) The Fermi fields coming from $4d$ chiral fields in $p_1$ have no vertical shift, while all others receive a $(-1)$ shift. (e) The full periodic quiver on $T^3$ for $\text{SPP}\times\mathbb{C}$ .
\label{fzspp3}}
 \end{center}
 \end{figure}

To construct the periodic quiver for the $2d$ theory, we start from the brane tiling and its dual periodic quiver for the $4d$ theory, which are shown in \fref{fzspp2}. 
The lift to the periodic quiver on $T^3$ is presented in \fref{fzspp3}.

\bigskip

\section{Partial Resolution and Higgsing}

\label{section_higgsing}

In this section, we study how to connect theories for different singularities via partial resolution. In terms of the gauge theory, partial resolution translates into higgsing, namely into RG flows triggered by turning on non-zero VEVs for the scalar component of certain chiral multiplets.

\bigskip

\subsection{Higgsing}

\begin{figure}[h]
\begin{center}
\includegraphics[width=11cm]{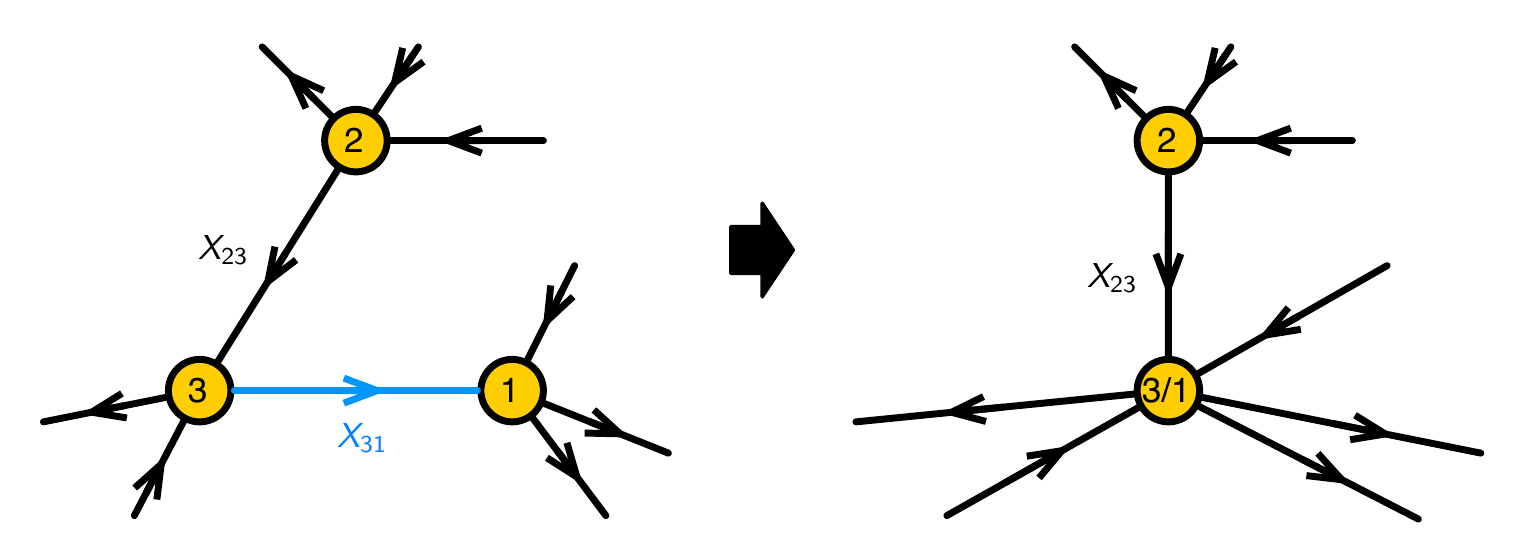}
\caption{Effect of bifundamental higgsing on the quiver. The chiral field with a non-zero VEV, in this case $X_{31}$, disappears from the quiver and the two nodes connected by it are condensed into a single one.
\label{higgsing0}}
 \end{center}
 \end{figure}

Let us consider the effect of turning a non-zero VEV for the scalar component of a bifundamental chiral field $X_{ij}$. In the abelian theory, this follows from turning on FI terms of equal magnitude and opposite signs for nodes $i$ and $j$. As a result, the gauge groups associated with nodes $i$ and $j$ are higgsed to the diagonal subgroup, while the anti-diagonal combination becomes massive. In terms of the quiver, the $X_{ij}$ arrow is removed and nodes $i$ and $j$ are condensed into a single one, as schematically shown in \fref{higgsing0}. At the same time, we replace $X_{ij}$ by its VEV, which for simplicity can be taken to be $1$, in all $J$- and $E$-terms.

\bigskip

\paragraph{Massive Fields.}

As usual, a possible additional outcome of higgsing is the generation of masses for some of the matter fields. For $2d$ $(0,2)$ theories, such massive fields correspond to Fermi-chiral pairs. Massive pairs arise when either a $J$- or $E$-term develops a linear term. For concreteness, let us consider the case of an $J$-term with a linear term. The case of a linear $E$-term is identical, due to the symmetry under the exchange of $J$- and $E$-terms. This situation arises when, before turning on a VEV for $X_{ij}$, the original $J$-term for $\Lambda_{ki}$ takes the general form given on the left of
\beq
J_{ik}=X_{ij}X_{jk}-f_{ik}(X) \ \ \longrightarrow \ \ X_{jk}-f_{ik}(X),
\label{higgsed_J}
\eeq
where $f_{ij}(X)$ indicates a product of scalar fields associated with an oriented path of chiral fields in the quiver connecting nodes $i$ and $k$. The right hand side of \eref{higgsed_J} shows $J_{ik}$ after the VEV. The massive pair in this case consists of the Fermi field $\Lambda_{ki}$ (the one for which the $J$-term becomes linear) and the chiral field $X_{jk}$ (the one in the linear term). Notice that after higgsing identifies nodes $i$ and $j$, $\Lambda_{ki}$ and $X_{jk}$ end up connecting the same pair of nodes, transforming in conjugate representations. It is straightforward to see that all the on-shell degrees of freedom in $\Lambda_{ki}$ and $X_{jk}$ become massive. Plugging \eref{higgsed_J} into \eref{V_JE}, we obtain a mass term for $\phi_{jk}$. Masses for the fermions $\psi_{+,jk}$ and $\lambda_{-,ki}$ arise from replacing \eref{higgsed_J} in \eref{VY}.

\begin{figure}[h]
\begin{center}
\resizebox{1\hsize}{!}{
\includegraphics[trim=0cm 0cm 0cm 0cm,totalheight=10 cm]{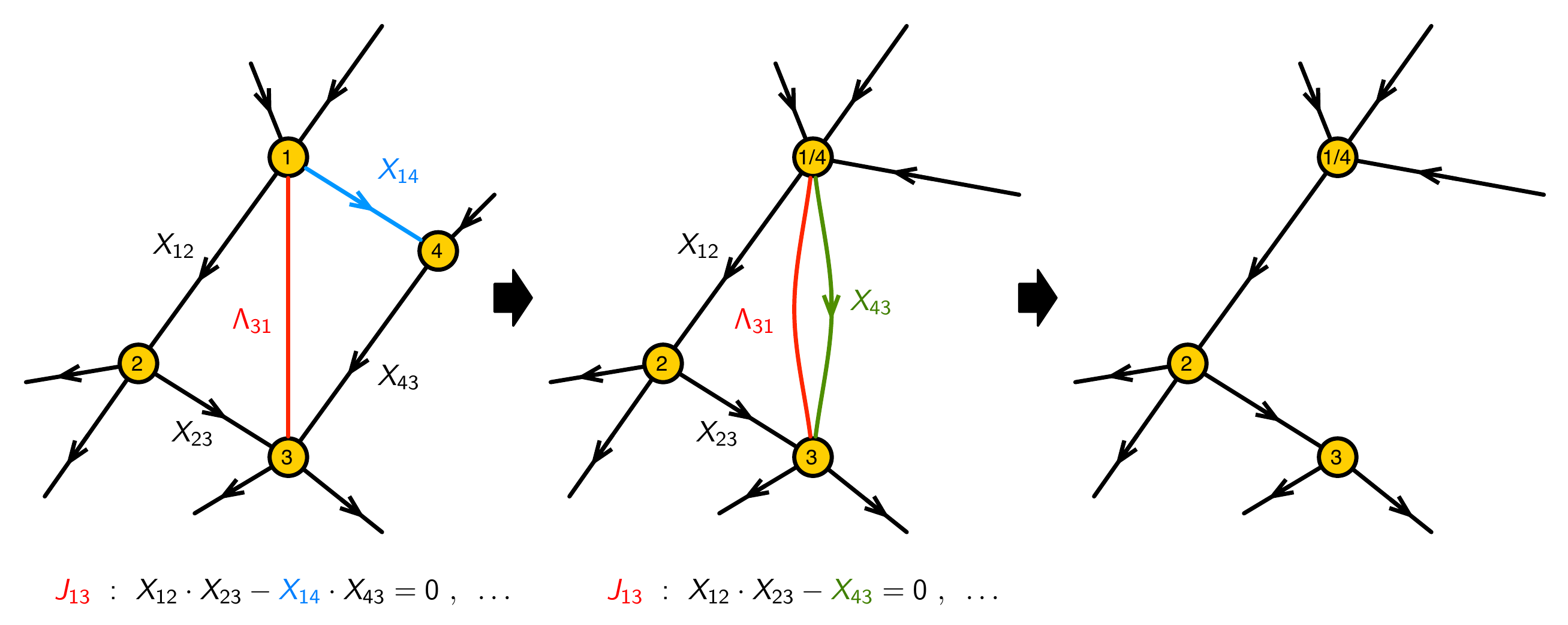}
}  
\caption{
An example showing the appearance of a Fermi-chiral massive pair when higgsing and how it is integrated out. Let us consider the original $J$-term for $\Lambda_{31}$ is $J_{13}=X_{12}\cdot X_{23} - X_{14} \cdot X_{43}$. Turning on a VEV for $X_{14}$ condenses nodes $1$ and $4$. In addition, we get $J_{13}=X_{12}\cdot X_{23} - X_{43}$, which results in a massive pair consisting of $\Lambda_{13}$ and $X_{43}$. When integrating out these fields, we set $J_{13}=0$ replacing $X_{43} \to X_{12}\cdot X_{23}$.
\label{fhiggsing}}
 \end{center}
 \end{figure}

At low energies, we can integrate out $\Lambda_{ki}$ and $X_{jk}$. When doing so, the terms $J_{ik}$ and $E_{ki}$ associated with the Fermi field $\Lambda_{ki}$ are removed from the Lagrangian. We explicitly set $J_{ik}$ to zero, using \eref{higgsed_J} to make the replacement $X_{jk}\to f_{ik}(X)$ wherever it appears in the Lagrangian. $E _{ik}$ also disappears with the Fermi field. The process of generating a massive pair by higgsing and integrating it out is schematically illustrated in \fref{fhiggsing}.

In terms of the periodic quiver, the Fermi and chiral fields in a massive pair not only connect the same pair of nodes but they overlap, hence giving rise to the plaquette associated with the linear $J$- or $E$-term.

 \bigskip

\subsection{Partial Resolution via Higgsing}

\label{section_partial_resolution_higgsing}

Different toric singularities can be connected by {\it partial resolution}, which corresponds to the removal of some points in the toric diagram. Partial resolution translates into higgsing in the gauge theory. From this viewpoint, the change in the singularity corresponds to the change in the mesonic moduli space after higgsing. This strategy for generating gauge theories associated with new geometries from known ones has been successfully exploited for the determination of $4d$ $\mathcal{N}=1$ gauge theories on D3-branes over toric Calabi-Yau 3-folds \cite{Beasley:1999uz,Feng:2000mi}. Based on this precedent, we extend its application to the $2d$ $(0,2)$ theories on D1-branes over toric Calabi-Yau 4-folds.

As explained earlier, every point in the toric diagram corresponds to one or various GLSM fields. In turn, GLSM fields are related to chiral fields in the quiver via the map \eref{map_X_p}, which is controlled by the $P$-matrix introduced in section \sref{sfa} as part of the forward algorithm. Giving a VEV to a chiral field corresponds to eliminating all the GLSM fields that contribute to it.\footnote{If these GLSM fields contain {\it all} the ones in another chiral field, the latter also gets a VEV. The additional VEV would be the result of relations coming from vanishing $J$- and $E$-terms.} A point in the toric diagram is removed once all the GLSM associated with it disappear.  The $P$-matrix thus gives us a systematic method for identifying the chiral field whose VEV implements any desired partial resolution.

\begin{figure}[ht!!]
\begin{center}
\resizebox{0.9\hsize}{!}{
\includegraphics[trim=0cm 0cm 0cm 0cm,totalheight=10 cm]{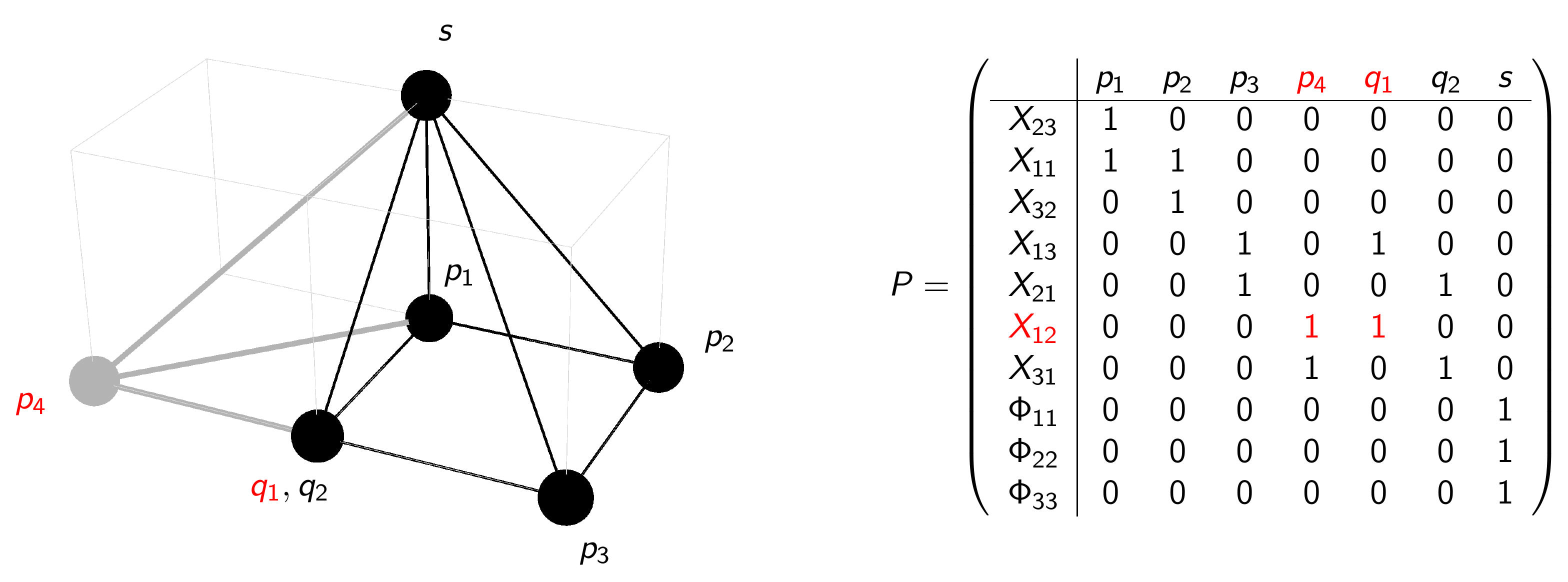}
}  
\caption{
Higgsing the $2d$ theory for $\text{SPP}\times\mathbb{C}$ to the theory for $\mathcal{C}\times\mathbb{C}$. According to the $P$-matrix \eref{es200a22}, which we reproduce here for convenience, giving a VEV to the chiral field $Z_{12}$ corresponds to removing the GLSM fields $p_4$ and $q_1$.
\label{fspphiggs}}
\end{center}
\end{figure}

The method we just outlined is illustrated in \fref{fspphiggs} for the partial resolution $\text{SPP}\times \mathbb{C} \to \mathcal{C}\times\mathbb{C}$. In this example, the point associated with $p_4$ can only be removed by giving a VEV to $X_{12}$ which, in turn, also requires the removal of $q_1$. Deleting $q_1$, however, does not result in the elimination of an additional point in the toric diagram due to the presence of $q_2$.

\begin{figure}[hht!!]
\begin{center}
\resizebox{0.5\hsize}{!}{
\includegraphics[trim=0cm 0cm 0cm 0cm,totalheight=10 cm]{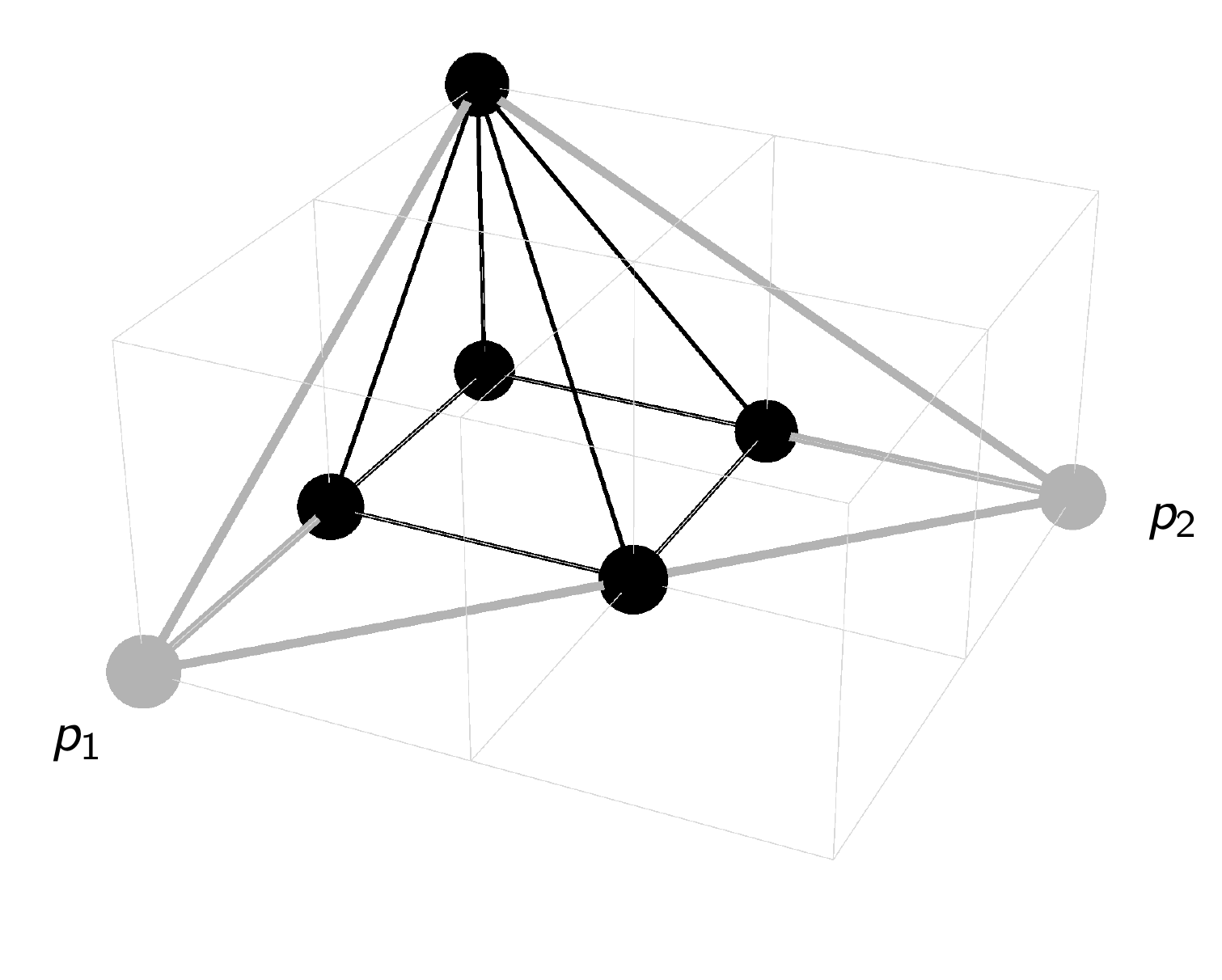}
}  
\caption{The black toric diagram of $\mathcal{C}\times \mathbb{C}$ can be embedded into the toric diagram of $\mathbb{C}^4/\mathbb{Z}_2 \times \mathbb{Z}_2$, shown in grey. Consequently, its gauge theory can be obtained via higgsing from that for the orbifold.
\label{ftoricembed}}
 \end{center}
 \end{figure}

In summary, partial resolution corresponds to removing points in the toric diagram that, in turn, translates into the Higgs mechanism in the gauge theory. This provides us with a {\it systematic algorithm} for constructing gauge theories associated with {\it arbitrary} toric singularities: they can be obtained by higgsing from theories whose toric diagram contains the one we are interested in. Of course such an approach would not be of much use if determining the gauge theory for the original singularity were difficult. However, there is a standard choice for the starting point. Any toric diagram can be embedded into the one for a $\mathbb{C}^4/\mathbb{Z}_{n_1}\times \mathbb{Z}_{n_2} \times \mathbb{Z}_{n_3}$ orbifold with action $(1,0,0,-1)(0,1,0,-1)(0,0,1,-1)$ for sufficiently large $n_1$, $n_2$ and $n_3$. In this case the toric diagram is a tetrahedron of length $n_1$, $n_2$ and $n_3$ along the three axes. The gauge theories for these orbifolds can be straightforwardly constructed using the ideas in section \sref{sec:orbifold}. \fref{ftoricembed} shows an example of a toric diagram embedded into the one of such an orbifold.

\bigskip

\subsection{Vanishing Trace Condition}

As reviewed in section \sref{stheory}, the $J$- and $E$-terms must satisfy 
\beal{v-trace}
\sum_{a} \text{Tr}(E_{a}J_{a})=0~,~
\eea
where the sum runs over Fermi fields. We call this requirement the vanishing trace condition. Recall that for the theories under consideration the $J$- and $E$-terms take the toric form
\beal{toric-EJ}
E_a = E_a^{(+)} - E_a^{(-)} \,,
\quad 
J_a = J_a^{(+)} - J_a^{(-)} \,, 
\eea
where $E_a^{(\pm)}$ and $J_a^{(\pm)}$ are products of 
matrices. The contribution of a Fermi field to the trace above has the following general form
\beal{ev1}
\text{Tr}(E_{a}J_{a}) = \text{Tr}(E_a^{(+)}J_a^{(+)}) + 
\text{Tr}(E_a^{(-)}J_a^{(-)})
-\text{Tr}(E_a^{(+)}J_a^{(-)})
-\text{Tr}(E_a^{(-)}J_a^{(+)})
~.~
\eea
The vanishing trace condition holds because there is a pairing of terms in \eref{v-trace} of the form
\beal{esv2}
E_{a}^{m} J_{a}^{n} - E_{b}^{k} J_{b}^{l} = 0 ~,~
\eea
where $m,n,k.l=(\pm)$ such that the product is $mnkl = (-)$. For conciseness, we have left out the trace.

Let us assume \eref{v-trace} is true for a theory with all the terms in the trace being paired as in \eref{esv2}.
When a chiral field $X$ gets a non-zero VEV, the following scenarios apply to the vanishing trace condition in \eref{v-trace}:
\begin{itemize}
\item \textit{No massive fields:} The simplest situation occurs when turning on VEV $\langle X \rangle = m$, does not produce masses for other matter fields. In this case, \eref{v-trace} takes the form
\beal{esv3}
E_{a}^{m} J_{a}^{n} - E_{b}^{k} J_{b}^{l}  \Big |_{\langle X \rangle = m}
= m \left( \tilde{E}_{a}^{m} \tilde{J}_{a}^{n} - \tilde{E}_{b}^{k} \tilde{J}_{b}^{l} \right) ~.~
\eea
This contribution vanishes provided that \eref{esv2} is true prior to giving the VEV. Accordingly, the vanishing trace condition holds when chiral fields receive VEVs, 
\beal{esv4}
E_{a}^{m} J_{a}^{n} - E_{b}^{k} J_{b}^{l} = 0
~~\Rightarrow~~
\tilde{E}_{a}^{m} \tilde{J}_{a}^{n} - \tilde{E}_{b}^{k} \tilde{J}_{b}^{l} = 0 ~.~
\eea

\item \textit{Massive fields:} As we discussed earlier, $\langle X \rangle = m$ can sometimes result in a Fermi-chiral massive pair. This is reflected in a linear $J$- or $E$-term. For concreteness, let us consider the case in which
\beal{esv5}
E_c \Big|_{\langle X \rangle = m}= m \tilde{E}_c^{(+)} - E_c^{(-)} ~,~
\quad 
J_c \Big|_{\langle X \rangle = m}= J_c^{(+)} - J_c^{(-)} ~,~ 
\eea
where $ \tilde{E}_c^{(+)}$ corresponds to a single chiral field that, following the discussion in the previous section, becomes massive. 
When integrating it out, we impose
\beal{esv6}
 \tilde{E}_c^{(+)} = \frac{1}{m} E_c^{(-)} ~,~
\eea
which implies
\beal{esv7}
E_c \Big|_{\langle X \rangle = m} = 0 ~.~
\eea
As a result, the contribution to the trace condition \eref{v-trace} associated with the massive Fermi field vanishes,
\beal{esv8}
\Tr (E_c J_c) \Big|_{\langle X \rangle = m}  = 0 ~.~
\eea
We thus have
\beal{esv9}
\sum_{a} \text{Tr}(E_{a}J_{a}) \Big|_{\langle X \rangle = m} = \sum_{a\neq k } \text{Tr}(E_{a}J_{a})  \Big|_{\langle X \rangle = m}~.~
\eea
Assuming that the vanishing trace condition is satisfied in the original theory, \eref{esv4} and \eref{esv9} imply that it continues to hold after higgsing, i.e.  
\beal{esv10}
\sum_{a} \text{Tr}(E_{a}J_{a}) = 0 
~~\Rightarrow~~
\sum_{a\neq k } \text{Tr}(E_{a}J_{a})  \Big|_{\langle X \rangle = m} = 0 ~,~
\eea 
The argument holds even if there are multiple massive fields that are integrated out.

\end{itemize}
The vanishing trace condition has been shown to hold for abelian orbifolds of $\mathbb{C}^4$ in \cite{GarciaCompean:1998kh}. Since, as we explained in the previous section, the gauge theory for any toric Calabi-Yau 4-fold can be obtained from that of an abelian orbifold via higgsing, the arguments we have just presented imply that the vanishing trace condition continues to hold for them.

\bigskip

\subsection{Cancellation of Non-Abelian Anomalies in General Toric Theories}

It is possible to use the previous ideas regarding higgsing to show that all gauge theories on D1-branes probing toric CY 4-folds are free of non-abelian anomalies. To do so, we exploit once more the fact that any such theory can be obtained as a partial resolution of an appropriate abelian orbifold of $\mathbb{C}^4$. Since, as discussed in section \sref{sec:orbifold}, all orbifold theories are free of non-abelian anomalies, it is sufficient to show that higgsing preserves the cancellation.

Consider higgsing a parent theory by an expectation value for the scalar in a bifundamental chiral field which, without loss of generality, we can call $X_{12}$. To start, let us assume that this VEV does not result in any massive field. Hence, the only anomaly cancellation conditions we should care about are the ones for nodes 1 and 2, since the other ones remain unaltered. In the parent theory, nodes 1 and 2 satisfy \eref{simple_anomaly_cancellation}, i.e.
\beq
\begin{array}{lcl}
n^\chi_1-n^F_1 & =& 2 \\
n^\chi_2-n^F_2 & =& 2 ~.
\end{array}
\eeq
Giving a VEV to $X_{12}$ higgses nodes 1 and 2 into a single one, which we call 1/2. $X_{12}$ is removed from the theory, which results in $n^\chi_i$ decreasing by one for $i=1,2$. In addition, the Fermi fields originally charged under nodes 1 and 2 remain unaffected. We thus conclude that the combined node 1/2 is free of non-abelian anomalies, since
\beq
\left.
\begin{array}{lcl}
n^\chi_{1/2}&=&n^\chi_1+n^F_1 -2 \\
n^F_{1/2}&=&n^F_1 + n^F_2
\end{array} \right\} \ \ \Rightarrow \ \ n^\chi_{1/2} - n^F_{1/2} =2
\eeq
Let us now consider what happens if massive fields are generated while higgsing. In this case, both the anomaly cancellation conditions of node 1/2 and other nodes can be affected. However, every massive pair consists of a chiral field and a Fermi field stretching between the same pair of nodes. The net contribution of such a pair of fields to the non-abelian anomaly is zero, so the theory remains free of non-abelian anomalies after integrating them out.

\bigskip

\section{Beyond Orbifolds and $\mathrm{CY}_3 \times \mathbb{C}$}

\label{section_beyond_orbifolds}

We now present explicit examples of the connection of different singularities by partial resolution and higgsing. As a warm-up, we first consider some $\mathrm{CY}_3 \times \mathbb{C}$ theories. For them, partial resolution closely resembles the one for the underlying CY 3-folds. We then use partial resolution to generate the first known examples of gauge theories on D1-branes probing singularities that are neither abelian orbifolds of $\mathbb{C}^4$ nor of the form $\mathrm{CY}_3 \times \mathbb{C}$.

\bigskip

\subsection{$\mathbb{C}^4/\mathbb{Z}_2\times \mathbb{Z}_2 \, (1,1,0,0)(1,0,1,0) \to \mathrm{SPP}\times \mathbb{C} \to \mathcal{C}\times \mathbb{C}$}

Let us construct some of the dimensionally reduced theories considered in section \sref{sec:examplesOfCY3xC} by partial resolution. Specifically, we are going to  consider the partial resolution sequence $\mathbb{C}^4/\mathbb{Z}_2\times \mathbb{Z}_2 \, (1,1,0,0)(1,0,1,0) \to \mathrm{SPP}\times \mathbb{C} \to \mathcal{C}\times \mathbb{C}$. All the necessary information regarding $\mathbb{C}^4/\mathbb{Z}_2\times \mathbb{Z}_2 \, (1,1,0,0) (1,0,1,0)$, which is also a dimensionally reduced theory,  is given in appendix~\ref{sc4z2z2a}. In order to illustrate how partial resolution is implemented, our presentation of these first theories is going to be rather detailed. Our treatment of subsequent examples will be considerably shorter.

Let us first consider the resolution from $\mathbb{C}^4/\mathbb{Z}_2\times \mathbb{Z}_2 \, (1,1,0,0)(1,0,1,0)$ to $\mathrm{SPP}\times \mathbb{C}$. The reduction in the volume of the toric diagram implies that the gauge theory looses a gauge group. Thus, we conclude it is necessary to turn on a VEV for a bifundamental chiral field. Since all bifundamental chiral fields are equivalent in this orbifold theory, we can pick any of them for higgsing. We can arrive at the same conclusion using the systematic approach introduced in section \sref{section_partial_resolution_higgsing}  based on the $P$-matrix, which for this theory is given by \eref{es115e4}. We see that every chiral bifundamental involves a multiplicity 1 GLSM field $p_i$, which correspond to the point in the toric diagram that is removed when higgsing, and two additional GLSM fields coming from different points in the toric diagram with multiplicity 2, which hence are not removed. Without loss of generality, let us give a VEV to $X_{14}$. The $J$- and $E$-terms from \eref{es115e1} become
\beq
\begin{array}{rcrcccrcc}
& & J \ \ \ \ \ \ \ \ \ \ \ \ \ \ \ & & & & E \ \ \ \ \ \ \ \ \ \ \ \ \ & & \\
\Lambda_{12} : & \ \ \ & X_{23} \cdot Y_{31} - Y_{24} \cdot X_{41}& = & 0 & \ \ \ \ &D_{11} \cdot Z_{12} - Z_{12} \cdot D_{22}&= & 0 \\ 
\textcolor{red}{\Lambda_{21} }: & \ \ \ & Y_{42} - Y_{13} \cdot X_{32}& = & 0 & \ \ \ \ & D_{22} \cdot Z_{21} - Z_{21} \cdot D_{11}&= & 0 \\ 
\Lambda_{13} : & \ \ \ & Z_{34} \cdot X_{41} - X_{32} \cdot Z_{21}& = & 0 & \ \ \ \ &D_{11} \cdot Y_{13} - Y_{13} \cdot D_{33}&= & 0 \\ 
\textcolor{red}{\Lambda_{31}} : & \ \ \ &  Z_{12} \cdot X_{23} - Z_{43}& = & 0 & \ \ \ \ &D_{33} \cdot Y_{31} - Y_{31} \cdot D_{11}&= & 0 \\ 
\textcolor{red}{\Lambda_{14}} : & \ \ \ & Y_{42} \cdot Z_{21} - Z_{43} \cdot Y_{31}& = & 0 & \ \ \ \ &D_{11}  - D_{44}&= & 0 \\ 
\Lambda_{41} : & \ \ \ & Y_{13} \cdot Z_{34} - Z_{12} \cdot Y_{24}& = & 0 & \ \ \ \ &D_{44} \cdot X_{41} - X_{41} \cdot D_{11}&= & 0 \\ 
\Lambda_{23} : & \ \ \ & Y_{31} \cdot Z_{12} - Z_{34} \cdot Y_{42}& = & 0 & \ \ \ \ &D_{22} \cdot X_{23} - X_{23} \cdot D_{33}&= & 0 \\ 
\Lambda_{32} : & \ \ \ & Y_{24} \cdot Z_{43} - Z_{21} \cdot Y_{13}& = & 0 & \ \ \ \ &D_{33} \cdot X_{32} - X_{32} \cdot D_{22}&= & 0 \\ 
\Lambda_{24} : & \ \ \ & Z_{43} \cdot X_{32} -X_{41} \cdot Z_{12}& = & 0 & \ \ \ \ &D_{22} \cdot Y_{24} - Y_{24} \cdot D_{44}&= & 0 \\ 
\textcolor{red}{\Lambda_{42}} : & \ \ \ & Z_{21}  -X_{23} \cdot Z_{34}& = & 0 & \ \ \ \ &D_{44} \cdot Y_{42} - Y_{42} \cdot D_{22}&= & 0 \\ 
\Lambda_{34} : & \ \ \ & X_{41} \cdot Y_{13} - Y_{42} \cdot X_{23}& = & 0 & \ \ \ \ &D_{33} \cdot Z_{34} - Z_{34} \cdot D_{44}&= & 0 \\
\textcolor{red}{\Lambda_{43}} : & \ \ \ & X_{32} \cdot Y_{24} - Y_{31} & = & 0 & \ \ \ \ &D_{44} \cdot Z_{43} - Z_{43} \cdot D_{33}&= & 0
\end{array}
\label{es115e1a}
\eeq
where we indicate in red the Fermi fields developing linear $J$- or $E$-terms. The corresponding Fermi-chiral massive pairs are
\beal{es110a3a}
\{\Lambda_{21}, Y_{42}\} ~,~
\{\Lambda_{31}, Z_{43}\} ~,~
\{\Lambda_{14}, (D_{11}-D_{44})/2\} ~,~
\{\Lambda_{42}, Z_{21}\} ~,~
\{\Lambda_{43}, Y_{31}\} ~.~
\eea
Notice that while $(D_{11}-D_{44})/2$ becomes massive, the orthogonal combination $\Phi_{11}=(D_{11}+D_{44})/2$ remains massless. After identifying nodes 1 and 4 and integrating out the massive fields, the quiver diagram becomes the one for $\mathrm{SPP}\times \mathbb{C}$, which was given in \fref{fzspp}. Imposing the relations
\beq
\begin{array}{c}
Y_{42} = Y_{13}\cdot X_{32} \ \ \ \ Z_{43} =  Z_{12}\cdot X_{23} \ \ \ \ Z_{21} = X_{23}\cdot Z_{34} \\
Y_{31} = X_{32}\cdot Y_{24} \ \ \ \  D_{44} = D_{11} 
\end{array}
\label{es110a4another}
\eeq
and relabeling node $4\to1$, the $J$- and $E$-terms for the surviving Fermi fields become
\beq
\begin{array}{rcrcccrcc}
& & J  \ \ \ \ \ \ \ \ \ \ \ \ \ \ \ & & & & E \ \ \ \ \ \ \ \ \ \ \ \ \ & & \\
\Lambda_{12} : & \ \ \ & X_{23} \cdot X_{32}\cdot Y_{21} - Y_{21} \cdot X_{11}& = & 0 & \ \ \ \ &\Phi_{11} \cdot Z_{12} - Z_{12} \cdot D_{22}&= & 0 \\ 
\Lambda_{13} : & \ \ \ & Z_{31} \cdot X_{11} - X_{32} \cdot X_{23}\cdot Z_{31}& = & 0 & \ \ \ \ &\Phi_{11} \cdot Y_{13} - Y_{13} \cdot D_{33}&= & 0 \\ 
\Lambda_{11} : & \ \ \ & Y_{13} \cdot Z_{31} - Z_{12} \cdot Y_{21}& = & 0 & \ \ \ \ & \Phi_{11} \cdot X_{11} - X_{11} \cdot \Phi_{11}&= & 0 \\ 
\Lambda_{23} : & \ \ \ & X_{32}\cdot Y_{21} \cdot Z_{12} - Z_{31} \cdot Y_{13}\cdot X_{32}& = & 0 & \ \ \ \ &D_{22} \cdot X_{23} - X_{23} \cdot D_{33}&= & 0 \\ 
\Lambda_{32} : & \ \ \ & Y_{21} \cdot Z_{12}\cdot X_{23} - X_{23}\cdot Z_{31} \cdot Y_{13}& = & 0 & \ \ \ \ &D_{33} \cdot X_{32} - X_{32} \cdot D_{22}&= & 0 \\ 
\Lambda_{21} : & \ \ \ & Z_{12}\cdot X_{23} \cdot X_{32} -X_{11} \cdot Z_{12}& = & 0 & \ \ \ \ &D_{22} \cdot Y_{21} - Y_{21} \cdot \Phi_{11}&= & 0 \\ 
\Lambda_{31} : & \ \ \ & X_{11} \cdot Y_{13} - Y_{13}\cdot X_{32} \cdot X_{23}& = & 0 & \ \ \ \ &D_{33} \cdot Z_{31} - Z_{31} \cdot \Phi_{11}&= & 0
\end{array}
\label{es115e1a2}
\eeq
which precisely agree with \eref{es200a20} after relabeling of fields.

Let us continue and perform the partial resolution from $\mathrm{SPP}\times \mathbb{C}$ to $\mathcal{C}\times \mathbb{C}$. Translating the $P$-matrix in \eref{es200a22} to the notation in \eref{es115e1a2}, we conclude that this particular partial resolution can be achieved by turning on a VEV for $Z_{12}$, $Y_{21}$, $Y_{13}$ or $Z_{31}$. Giving a VEV to either $X_{23}$ or $X_{32}$ instead would generate a resolution to $\mathbb{C}^4/\mathbb{Z}_2 \, (1,1,0,0)$. Consider giving a VEV to $Y_{13}$. The $J$- and $E$-terms in \eref{es115e1a2} become
\beq
\begin{array}{rcrcccrcc}
& & J  \ \ \ \ \ \ \ \ \ \ \ \ \ \ \ & & & & E \ \ \ \ \ \ \ \ \ \ \ \ \ & & \\
\Lambda_{12} : & \ \ \ & X_{23} \cdot X_{32}\cdot Y_{21} - Y_{21} \cdot X_{11}& = & 0 & \ \ \ \ &\Phi_{11} \cdot Z_{12} - Z_{12} \cdot D_{22}&= & 0 \\ 
\textcolor{red}{\Lambda_{13}} : & \ \ \ & Z_{31} \cdot X_{11} - X_{32} \cdot X_{23}\cdot Z_{31}& = & 0 & \ \ \ \ &\Phi_{11} - D_{33}&= & 0 \\ 
\textcolor{red}{\Lambda_{11}} : & \ \ \ & Z_{31} - Z_{12} \cdot Y_{21}& = & 0 & \ \ \ \ & \Phi_{11} \cdot X_{11} - X_{11} \cdot \Phi_{11}&= & 0 \\ 
\Lambda_{23} : & \ \ \ & X_{32}\cdot Y_{21} \cdot Z_{12} - Z_{31} \cdot X_{32}& = & 0 & \ \ \ \ &D_{22} \cdot X_{23} - X_{23} \cdot D_{33}&= & 0 \\ 
\Lambda_{32} : & \ \ \ & Y_{21} \cdot Z_{12}\cdot X_{23} - X_{23}\cdot Z_{31} & = & 0 & \ \ \ \ &D_{33} \cdot X_{32} - X_{32} \cdot D_{22}&= & 0 \\ 
\Lambda_{21} : & \ \ \ & Z_{12}\cdot X_{23} \cdot X_{32} -X_{11} \cdot Z_{12}& = & 0 & \ \ \ \ &D_{22} \cdot Y_{21} - Y_{21} \cdot \Phi_{11}&= & 0 \\ 
\textcolor{red}{\Lambda_{31}} : & \ \ \ & X_{11} - X_{32} \cdot X_{23}& = & 0 & \ \ \ \ &D_{33} \cdot Z_{31} - Z_{31} \cdot \Phi_{11}&= & 0
\end{array}
\label{es115e1a3}
\eeq
The massive pairs are now
\beq
\{\Lambda_{13},(\Phi_{11} - D_{33})/2\} ~,~
\{\Lambda_{11}, Z_{31}\} ~,~
\{\Lambda_{31}, X_{11} \} ~.~
\eeq
while the linear combination $\tilde{\Phi}_{11}=(\Phi_{11} + D_{33})/2$ is massless. Integrating out the massive fields and identifying nodes $3$ and $1$, we arrive at the quiver in \fref{fzconfioldaaa} with
\beq
\begin{array}{rclccclcc}
& & \ \ \ \ \ \ \ \ \ \ \ \ \ \ \ \ \ \ \ J  & & & & \ \ \ \ \ \ \ \ \ \ \ \ \ E & & \\
\Lambda_{12}^1 : & \ \ \ & X_{21} \cdot X_{12}\cdot Y_{21} - Y_{21} \cdot X_{12} \cdot X_{21}& = & 0 & \ \ \ \ &\tilde{\Phi}_{11} \cdot Z_{12} - Z_{12} \cdot D_{22}&= & 0 \\ 
\Lambda_{21}^1 : & \ \ \ & X_{12}\cdot Y_{21} \cdot Z_{12} - Z_{12} \cdot Y_{21} \cdot X_{12}& = & 0 & \ \ \ \ &D_{22} \cdot X_{21} - X_{21} \cdot \tilde{\Phi}_{11}&= & 0 \\ 
\Lambda_{12}^2 : & \ \ \ & Y_{21} \cdot Z_{12}\cdot X_{21} - X_{21}\cdot Z_{12} \cdot Y_{21} & = & 0 & \ \ \ \ &\tilde{\Phi}_{11} \cdot X_{12} - X_{12} \cdot D_{22}&= & 0 \\ 
\Lambda_{21}^2 : & \ \ \ & Z_{12}\cdot X_{21} \cdot X_{12} -X_{12} \cdot X_{21} \cdot Z_{12}& = & 0 & \ \ \ \ &D_{22} \cdot Y_{21} - Y_{21} \cdot \tilde{\Phi}_{11}&= & 0 \\ 
\end{array}
\label{es115e1a4}
\eeq
which is in agreement with \eref{es200a1-new}.

\bigskip

\subsection{$D_3$ \label{sd3}}

\begin{figure}[ht!]
\begin{center}
\includegraphics[trim=0cm 0cm 0cm 0cm,width=9.5 cm]{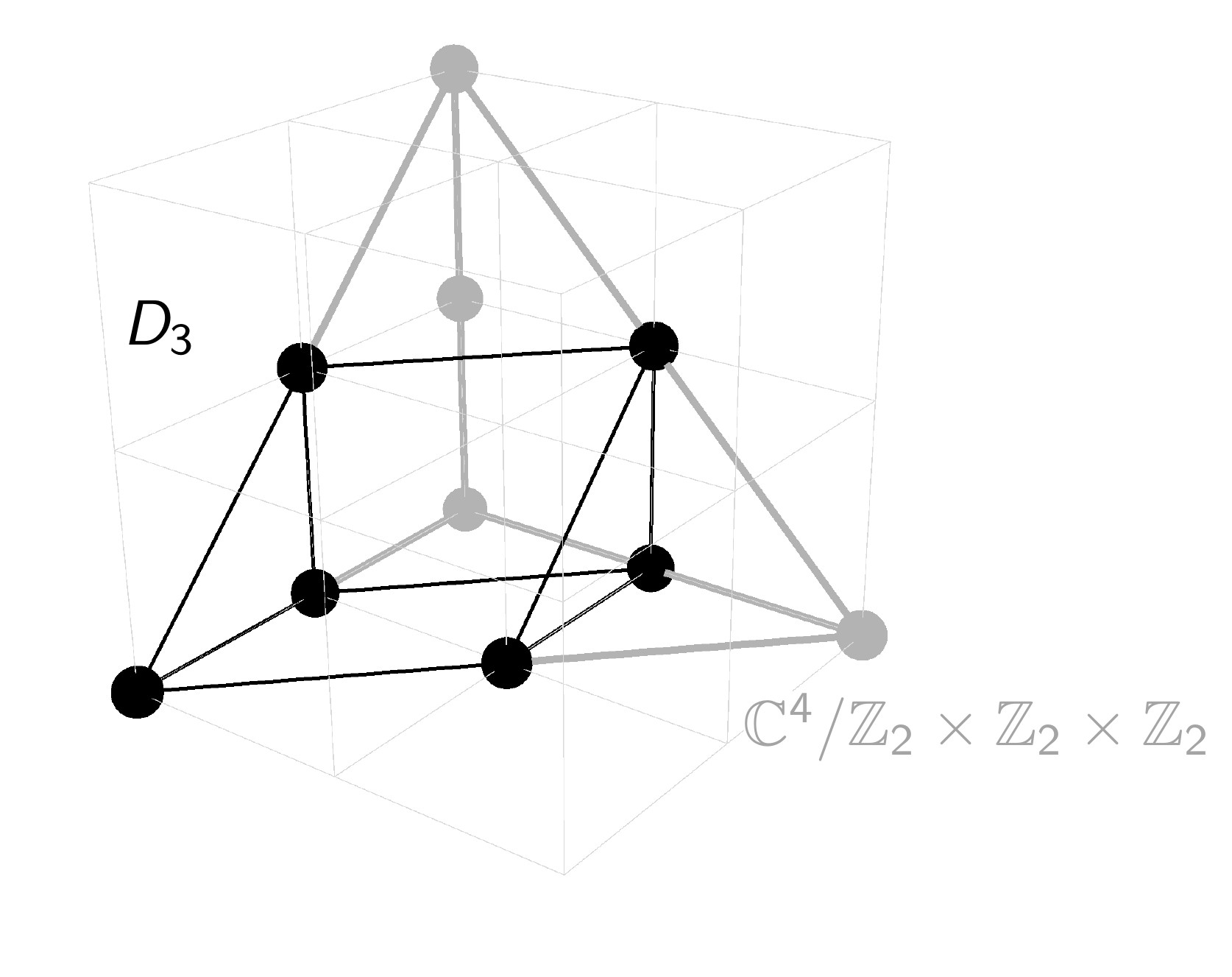}
\vspace{-.7cm}\caption{
The toric diagram for $D_3$ (shown in black) can be embedded into the toric diagram of $\mathbb{C}^4/\mathbb{Z}_2 \times\mathbb{Z}_2\times \mathbb{Z}_2 \, (1,0,0,1)(0,1,0,1)(0,0,1,1)$ (shown in grey) and hence can be obtained from it by partial resolution.
\label{fd3inc4}}
 \end{center}
 \end{figure}

Having developed some familiarity with the implementation of partial resolution as higgsing, we illustrate in this and the coming sections how to use it for constructing gauge theories for general toric singularities that are neither abelian orbifolds of $\mathbb{C}^4$ nor $\mathrm{CY}_3 \times \mathbb{C}$. The first geometry we consider is the so-called $D_3$ singularity \cite{Franco:2008um}, whose toric diagram is shown in black in \fref{fd3inc4}. From this figure, we also conclude that $D_3$ can be obtained from the $\mathbb{C}^4/\mathbb{Z}_2 \times\mathbb{Z}_2\times \mathbb{Z}_2 \, (1,0,0,1)(0,1,0,1)(0,0,1,1)$  orbifold by partial resolution.

\begin{figure}[h]
\begin{center}
\resizebox{0.5\hsize}{!}{
\includegraphics[trim=0cm 0cm 0cm 0cm,totalheight=10 cm]{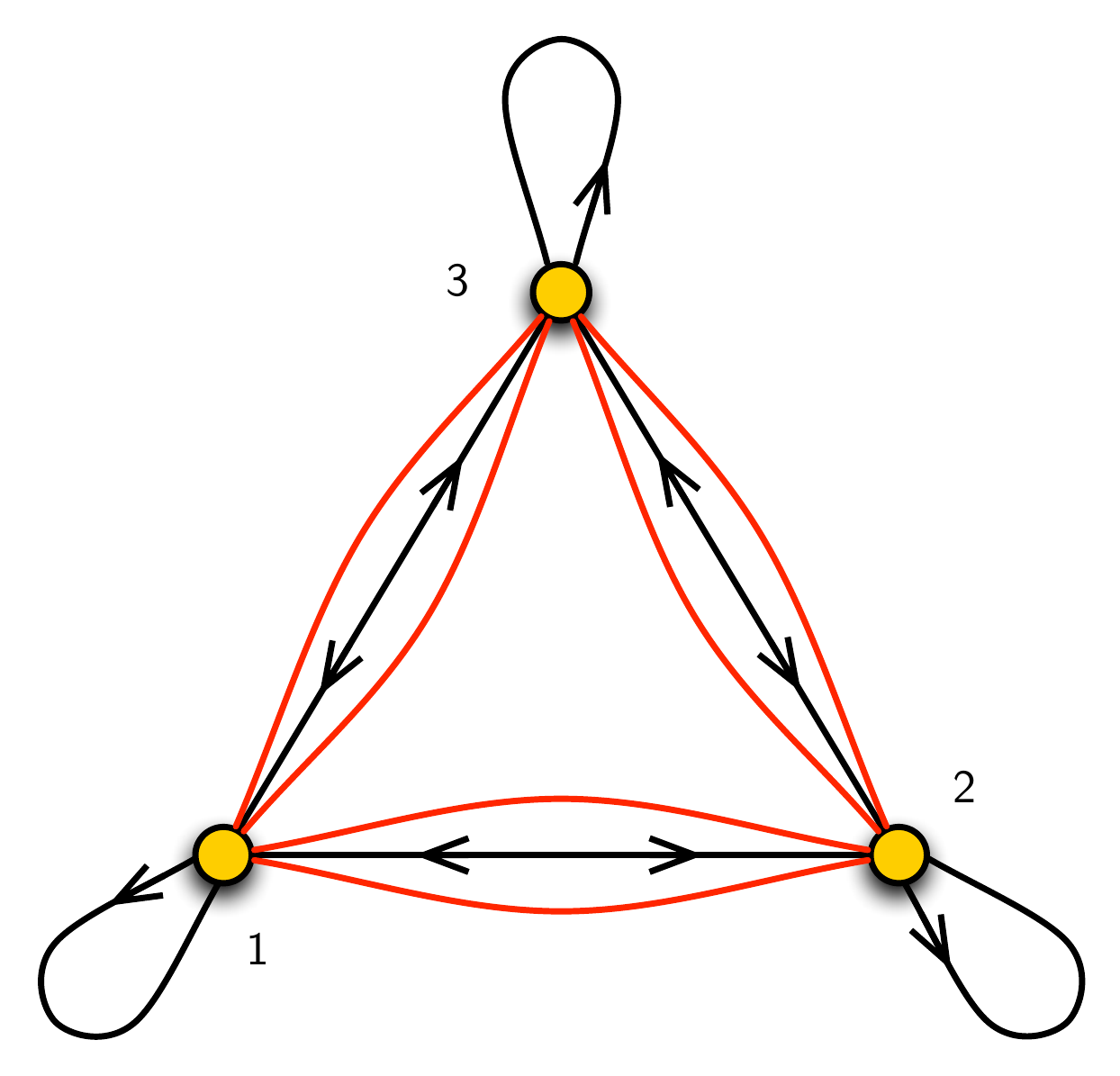}
}  
\caption{Quiver diagram for $D_3$.
\label{fquiverd3}}
 \end{center}
 \end{figure}

All necessary information regarding the $\mathbb{C}^4/\mathbb{Z}_2 \times\mathbb{Z}_2\times \mathbb{Z}_2 \, (1,0,0,1)(0,1,0,1)(0,0,1,1)$ theory is collected in appendix \ref{sc4z2z2z2}. The corresponding $P$-matrix \eref{es117aaa2} indicates that the desired partial resolution can be achieved, for example, by giving VEVs to $Y_{12}$, $X_{37}$, $Z_{42}$, $X_{48}$ and $Y_{56}$. In the notation of \fref{fc4z2z2z2toric}, the surviving GLSM fields are $p_4$, $q_1$, $r_1$, $u_1$, $v_1$ and $w_2$, which precisely agrees with \fref{fd3inc4}. After higgsing, integrating out massive fields and relabeling nodes, the final theory for the $D_3$ singularity is given by the quiver in \fref{fquiverd3}, with the following $J$- and $E$-terms

\beq
\begin{array}{rcrcccrcc}
& &  J \ \ \ \ \ \ \ \ \ \ \ \ \ & & & &  E \ \ \ \ \ \ \ \ \ \ \ \ \ & & \\
\Lambda_{21} : & \ \ \ & X_{13}\cdot X_{31}\cdot Y_{12}-Y_{12}\cdot X_{22 }& = & 0 & \ \ \ \ & D_{23}\cdot Z_{32}\cdot Z_{21}-Z_{21}\cdot D_{11}&= & 0 \\
 \Lambda_{12} : & \ \ \ & Z_{21}\cdot X_{13}\cdot X_{31} -X_{22}\cdot Z_{21}& = & 0 & \ \ \ \ &D_{11}\cdot Y_{12} - Y_{12}\cdot D_{23}\cdot Z_{32}&= & 0 \\
 \Lambda_{31}: & \ \ \ & X_{13}\cdot Y_{33} - Y_{12}\cdot Z_{21}\cdot X_{13}& = & 0 & \ \ \ \ & X_{31}\cdot D_{11} - Z_{32}\cdot D_{23}\cdot X_{31}&= & 0 \\
  \Lambda_{13} : & \ \ \ & X_{31}\cdot Y_{12}\cdot Z_{21} -Y_{33}\cdot X_{31}& = & 0 & \ \ \ \ &D_{11}\cdot X_{13} - X_{13}\cdot Z_{32}\cdot D_{23}&= & 0 \\
 \Lambda_{23}^{1 } : & \ \ \ & Y_{33}\cdot Z_{32} - Z_{32}\cdot Z_{21}\cdot Y_{12}& = & 0 & \ \ \ \ & D_{23}\cdot X_{31}\cdot X_{13} -X_{22}\cdot D_{23} &= & 0 \\
 \Lambda_{23}^{2 } : & \ \ \ & Z_{32}\cdot X_{22} - X_{31}\cdot X_{13}\cdot Z_{32}& = & 0 & \ \ \ \ &D_{23}\cdot Y_{33} - Z_{21}\cdot Y_{12}\cdot D_{23} &= & 0
\end{array}
\label{es130a1}
\eeq

\bigskip

\begin{figure}[ht!]
\begin{center}
\resizebox{0.5\hsize}{!}{
\includegraphics[trim=0cm 0cm 0cm 0cm,totalheight=10 cm]{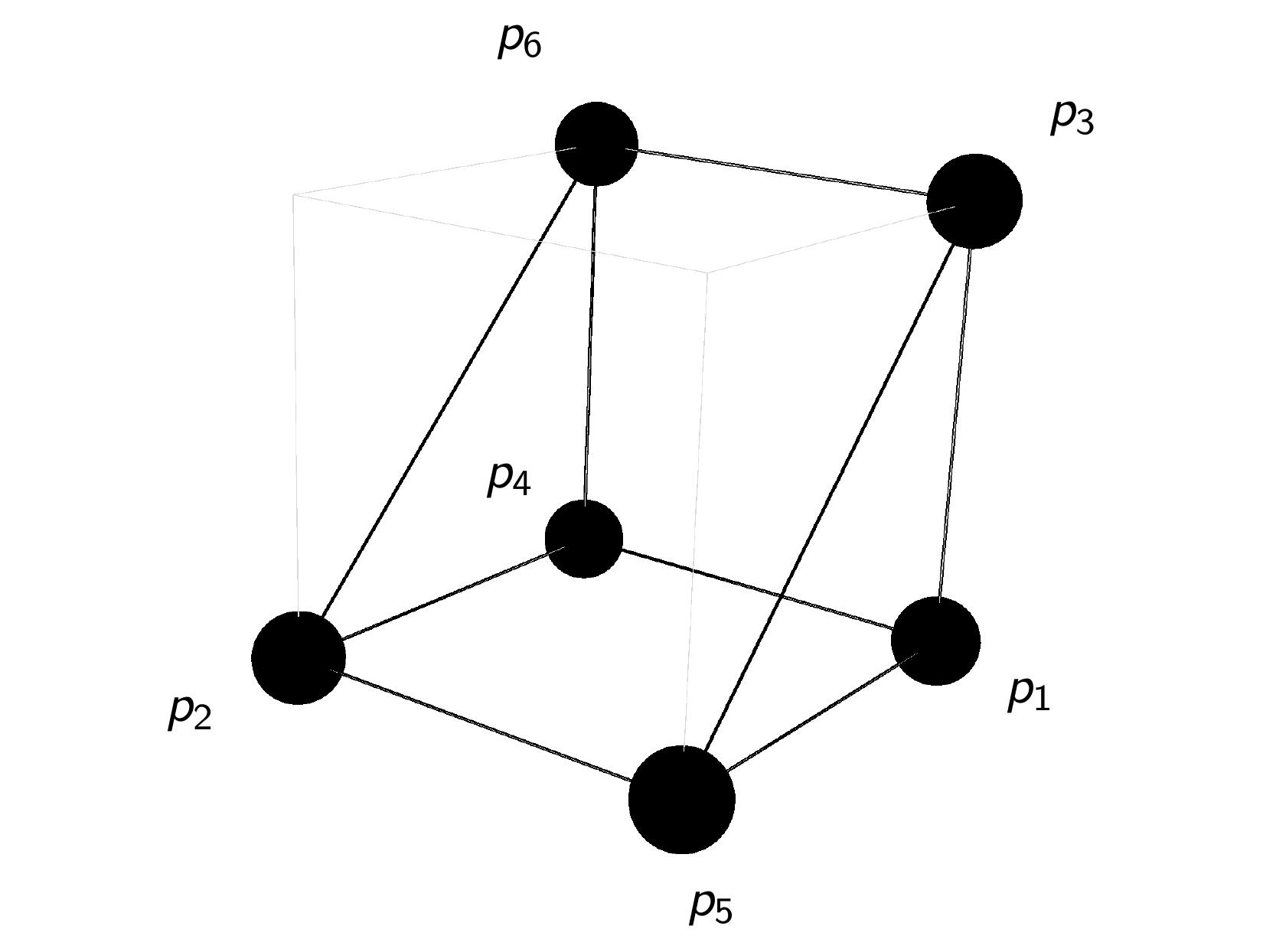}
}  
\caption{
The toric diagram associated with \eref{es130a6} is the one for $D_3$. This geometry has been obtained as the mesonic moduli space of the corresponding gauge theory.
\label{ftoricd3}}
 \end{center}
 \end{figure}

As an additional check of the theory we have just obtained, let us apply to it the forward algorithm and confirm that its mesonic moduli space indeed corresponds to the desired geometry. The quiver incidence matrix is 
\beal{es130a2}
d=
\left(
\begin{array}{l|ccccccccc}
\; &
D_{11}& D_{23}& X_{13}& X_{31}& X_{22}& Y_{33}& Y_{12}& Z_{21}& Z_{32}
\\
\hline
\encircle{1}& 0 & 0 & 0 & 0 & 0 & 1 & -1 & 1 & -1 \\
\encircle{2}& 0 & 0 & 0 & 1 & -1 & 0 & 0 & -1 & 1 \\
\encircle{3}& 0 & 0 & 0 & -1 & 1 & -1 & 1 & 0 & 0 \\
\end{array}
\right)
~,~
\eea
and the $K$- and $P$-matrices are given by
{\small \beq
K=
\left(
\begin{array}{c|cccccccccc}
\; & D_{11}& X_{22} & Y_{33} & D_{23} & Z_{32} & X_{13}& X_{31} & Y_{12}& Z_{21}
\\
\hline
D_{23} & 1 & 0 & 0 & 1 & 0 & 0 & 0   & 0 & 0  \\
Z_{32} & 1 & 0 & 0 & 0 & 1& 0  & 0 & 0 & 0  \\
X_{13} & 0 & 1 & 0 & 0 & 0 & 1 & 0  & 0 & 0  \\
X_{31} & 0 & 1 & 0 & 0 & 0 & 0 & 1  & 0 & 0  \\
Y_{12} & 0 & 0 & 1 & 0 & 0 & 0 & 0 & 1 & 0 \\
Z_{21} & 0 & 0 & 1 & 0 & 0 & 0 & 0  & 0 & 1  \\
\end{array}
\right)
\ \ ,\ \ 
P=
\left(
\begin{array}{c|cccccc}
\; & p_1 & p_2 & p_3 & p_4 & p_5 & p_6 
\\
\hline
D_{11} &1 & 0 & 0 & 1 & 0 & 0 \\
X_{22} & 0 & 1 & 0 & 0 & 1 & 0 \\
Y_{33} & 0 & 0 & 1 & 0 & 0 & 1 \\
D_{23} & 1 & 0 & 0 & 0 & 0 & 0 \\
Z_{32} & 0 & 0 & 0 & 1 & 0 & 0 \\
X_{13} & 0 & 1 & 0 & 0 & 0 & 0 \\
X_{31} & 0 & 0 & 0 & 0 & 1 & 0 \\
Y_{12} & 0 & 0 & 1 & 0 & 0 & 0 \\
Z_{21} & 0 & 0 & 0 & 0 & 0 & 1 \\
\end{array}
\right)
~.~ \\
\label{es130a3}
\eeq}
The GLSM charge matrices become
\beal{es130a5}
Q_{JE}=
\emptyset
\ \ , \ \
Q_D=
\left(
\begin{array}{cccccc}
 p_1 & p_2 & p_3 & p_4 & p_5 & p_6 
\\
\hline
 1 & 0 & -1 & -1 & 0 & 1 \\
 -1 & -1 & 0 & 1 & 1 & 0 \\
\end{array}
\right)
~.~
\eea
From them we obtain
\beal{es130a6}
G=
\left(
\begin{array}{cccccc}
 p_1 & p_2 & p_3 & p_4 & p_5 & p_6 
\\
\hline
 \ 1 \ & \ 1 \ & \ 1 \ & \ 1 \ & 1 & \ 1 \ \\
 0 & 0 & 0 & 1 & -1 & 1 \\
 0 & 1 & 0 & 0 & 1 & 0 \\
 0 & 0 & 1 & 0 & 0 & 1 \\
 \end{array}
\right)
~,~
\eea
which, following our expectations, corresponds to the toric diagram for $D_3$, as shown in \fref{ftoricd3}.

 \bigskip

\subsection{$Q^{1,1,1}$ \label{sq111}}

\begin{figure}[ht!]
\begin{center}
\includegraphics[trim=0cm 0cm 0cm 0cm,width=9.5 cm]{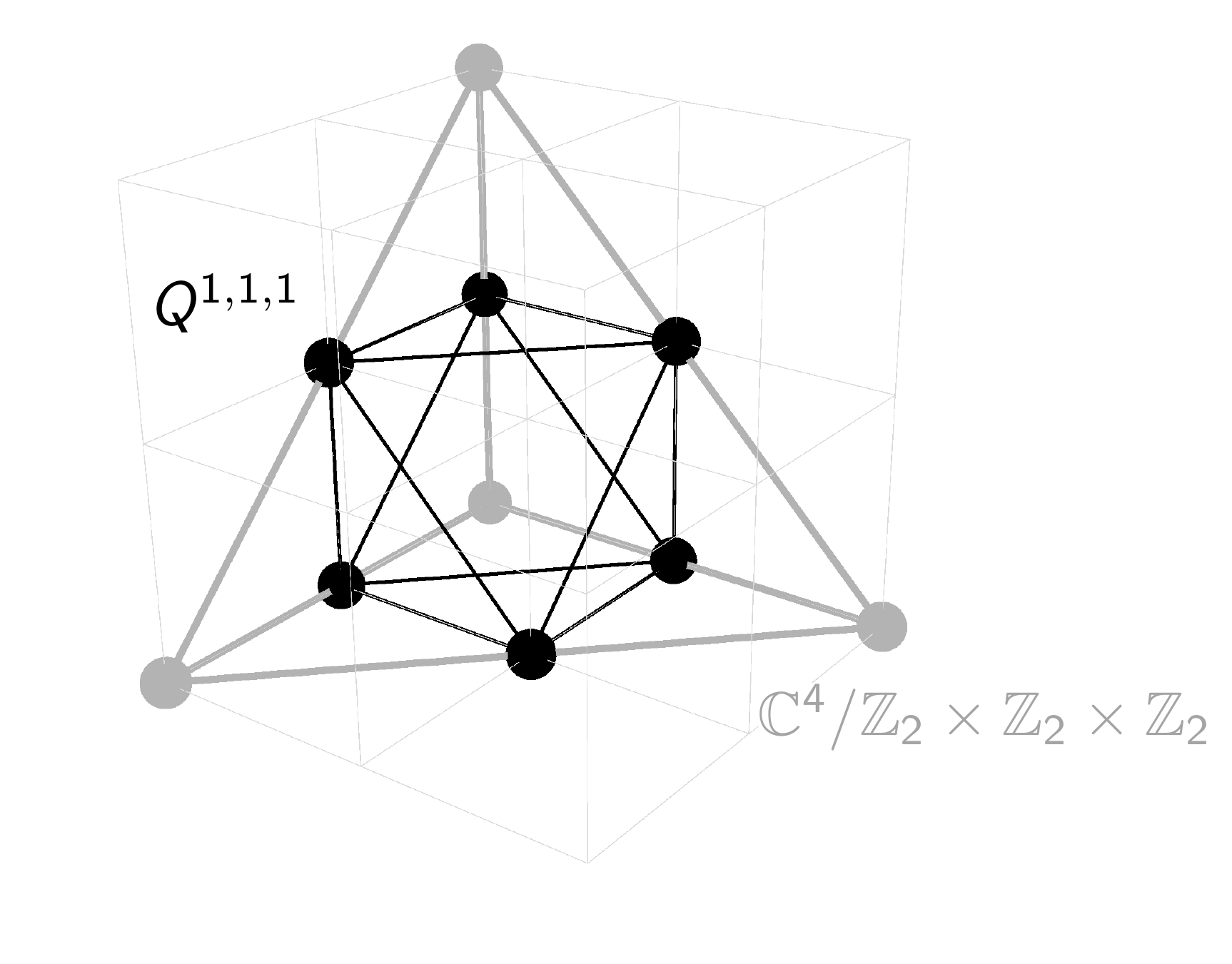}
\vspace{-.7cm}\caption{
 The toric diagram for $Q^{1,1,1}$ (shown in black) can be embedded into the toric diagram of $\mathbb{C}^4/\mathbb{Z}_2 \times\mathbb{Z}_2\times \mathbb{Z}_2 \, (1,0,0,1)(0,1,0,1)(0,0,1,1)$ (shown in grey) and hence can be obtained from it by partial resolution.
\label{fq111inc4}}
 \end{center}
 \end{figure}

We now construct the gauge theory on D1-branes probing the real cone over the 7d Sasaki-Einstein manifold $Q^{1,1,1}$, which is the homogeneous coset space
\beq
{SU(2) \times SU(2) \times SU(2)\over U(1) \times U(1)}
\eeq
and has a $U(1)_R \times SU(2)^3$ isometry \cite{D'Auria:1983vy, Nilsson:1984bj,Sorokin:1984ca,Sorokin:1985ap}. It can be written as a $U(1)$ fibration over $S^2\times S^2 \times S^2$. For brevity, we simply refer to the full cone geometry as $Q^{1,1,1}$. The toric diagram for $Q^{1,1,1}$ is shown in black in \fref{fq111inc4}, from which we also conclude that, like the theory in the section above, it can be obtained by partial resolution of $\mathbb{C}^4/\mathbb{Z}_{2}\times\mathbb{Z}_{2}\times\mathbb{Z}_{2}\, (1,0,0,1)(0,1,0,1)(0,0,1,1)$.

\begin{figure}[ht!]
\begin{center}
\resizebox{0.4\hsize}{!}{
\includegraphics[trim=0cm 0cm 0cm 0cm,totalheight=10 cm]{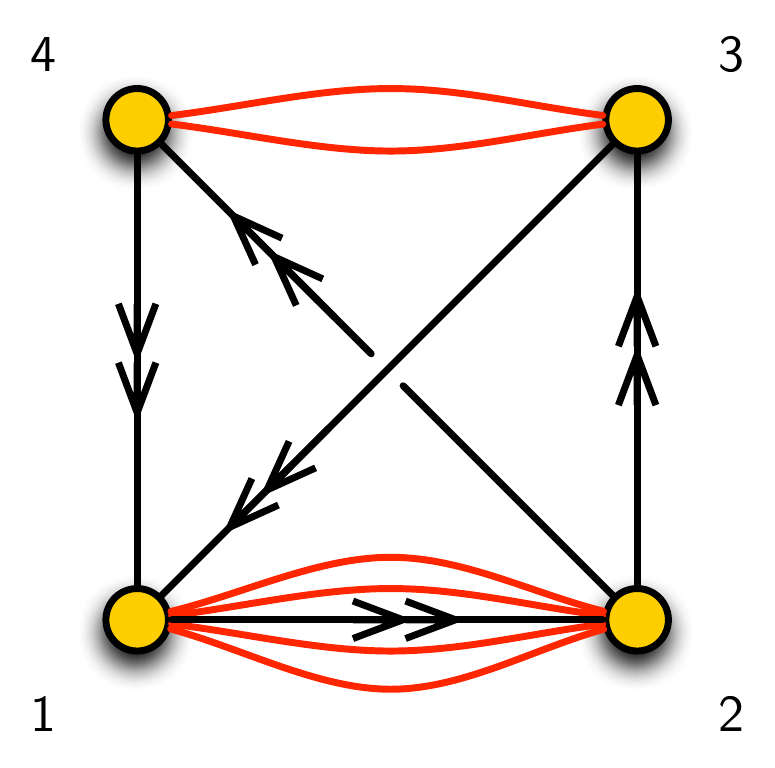}
}  
\caption{
Quiver diagram for $Q^{1,1,1}$.
\label{fquiverq111}}
 \end{center}
 \end{figure}

From the $P$-matrix for the orbifold given in \eref{es117aaa2}, we see that the desired partial resolution can be obtained by giving VEVs to $Z_{13}$, $X_{15}$, $D_{27}$ and $Y_{87} $. After higgsing, integrating out massive fields and relabeling nodes we obtain the theory for $Q^{1,1,1}$, which corresponds to the quiver in \fref{fquiverq111} with $J$- and $E$-terms
{\footnotesize
\beq
\begin{array}{rcrcccrcc}
& & J  \ \ \ \ \ \ \ \ \ \ \ \ \ \ \ \ \ \ \ \ \ & & & & E \ \ \ \ \ \ \ \ \ \ \ \ & & \\
\Lambda_{21}^{1} : & \ \ \ & D_{12}\cdot Z_{24}\cdot Y_{41}\cdot X_{12} - X_{12}\cdot Z_{23}\cdot D_{31}\cdot D_{12}& = & 0 & \ \ \ \ &X_{23}\cdot Y_{31} - X_{24}\cdot D_{41}&= & 0 \\
\Lambda_{21}^{2}: & \ \ \ & X_{12}\cdot X_{24}\cdot Y_{41}\cdot D_{12}-D_{12}\cdot Z_{23}\cdot Y_{31}\cdot X_{12}& = & 0 & \ \ \ \ &X_{23}\cdot D_{31} - Z_{24}\cdot D_{41}&= & 0 \\
\Lambda_{21}^{3} : & \ \ \ & D_{12}\cdot Z_{24}\cdot D_{41}\cdot X_{12} - X_{12}\cdot X_{23}\cdot D_{31}\cdot D_{12}& = & 0 & \ \ \ \ & X_{24}\cdot Y_{41} - Z_{23}\cdot Y_{31}&= & 0 \\
\Lambda_{21}^{4} : & \ \ \ & D_{12}\cdot X_{23}\cdot Y_{31}\cdot X_{12} - X_{12}\cdot X_{24}\cdot D_{41}\cdot D_{12}& = & 0 & \ \ \ \ &Z_{23}\cdot D_{31} - Z_{24}\cdot Y_{41}&= & 0 \\
\Lambda_{43} : & \ \ \ & D_{31}\cdot X_{12}\cdot X_{24} - Y_{31}\cdot X_{12}\cdot Z_{24}& = & 0 & \ \ \ \ &D_{41}\cdot D_{12}\cdot Z_{23} - Y_{41}\cdot D_{12}\cdot X_{23}&= & 0 \\
\Lambda_{34} : & \ \ \ & Y_{41}\cdot X_{12}\cdot X_{23}-D_{41}\cdot X_{12}\cdot Z_{23}& = & 0 & \ \ \ \ &D_{31}\cdot D_{12}\cdot X_{24} - Y_{31}\cdot D_{12}\cdot Z_{24}&= & 0 
\end{array} 
\label{es110a8}
\eeq}
 
\begin{figure}[ht!]
\begin{center}
\resizebox{0.35\hsize}{!}{
\includegraphics[trim=0cm 0cm 0cm 0cm,totalheight=10 cm]{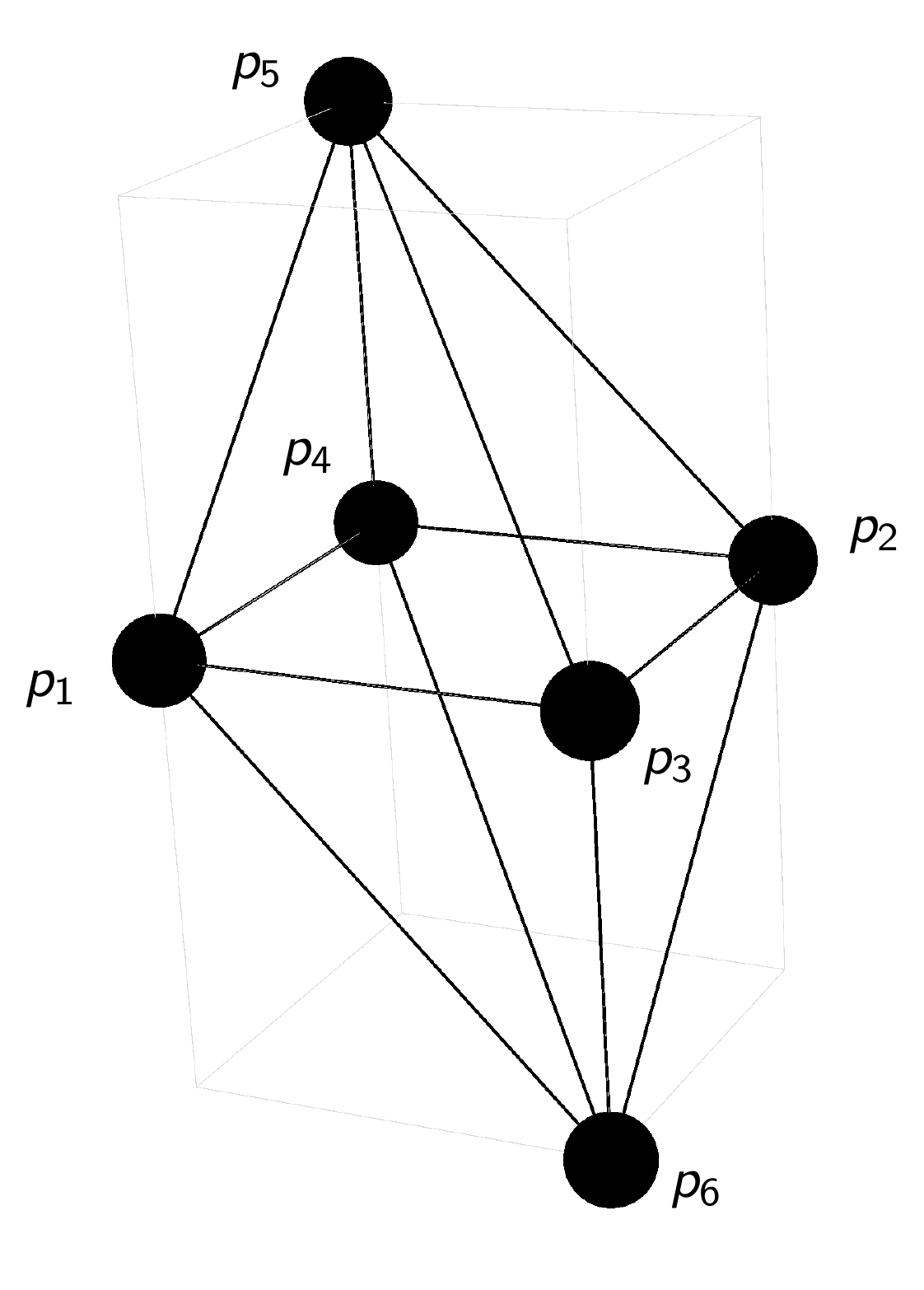}
}  
\caption{
After removing the extra GLSM fields $q_1$ and $q_2$, the toric diagram associated with \eref{es110a24} is the one for $Q^{1,1,1}$. This geometry has been obtained as the mesonic moduli space of the corresponding gauge theory.
\label{ftoricq111}}
 \end{center}
 \end{figure}

In the notation of \fref{fc4z2z2z2toric}, the surviving GLSM fields are $q_2$, $r_1$, $s_1$, $u_2$, $v_1$ and $w_1$, which give rise to \fref{fq111inc4}, as well as $e_4$ and $e_7$, which are two extra GLSM fields inherited from the parent orbifold. Let us confirm this is the case with the forward algorithm. The quiver incidence matrix is given by
 \beal{es110a20}
 d=
 \left(
\begin{array}{l|cccccccccc}
\; & D_{12} & X_{12} & Y_{31} & D_{31} & X_{24} & Z_{24} & X_{23} & Z_{23} & D_{41} & Y_{41}
\\
\hline
\encircle{1} & 1 & 1 & -1 & -1 & 0 & 0 & 0 & 0 & -1 & -1 \\
\encircle{2} & -1 & -1 & 0 & 0 & 1 & 1 & 1 & 1 & 0 & 0 \\
\encircle{3} & 0 & 0 & 1 & 1 & 0 & 0 & -1 & -1 & 0 & 0 \\
\encircle{4} & 0 & 0 & 0 & 0 & -1 & -1 & 0 & 0 & 1 & 1 \\
\end{array}
\right)
~.~
 \eea
The $K$- and $P$-matrices are
{\footnotesize
 \beq
 K=
 \left(
\begin{array}{l|cccccccccc}
\; & D_{12} & X_{12} & Y_{31} & D_{31} & D_{41} & Y_{41} & X_{24} & Z_{24} & X_{23} & Z_{23} 
\\
\hline
D_{12} &   1 & 0 & 0 & 0 & 0 & 0 & 0 & 0 & 0 & 0 \\
X_{12} & 0 & 1 & 0 & 0 & 0 & 0 & 0 & 0 & 0 & 0 \\
Y_{31} &  0 & 0 & 1 & 0 & 0 & 0 & 0 & -1 & -1 & -1  \\
D_{31} & 0 & 0 & 0 & 1 & 0 & 0 & 0 & 1 & 0 & 0  \\
D_{41} &0 & 0 & 0 & 0 & 1 & 0 & 0 & 0 & 1 & 0  \\
Y_{41} & 0 & 0 & 0 & 0 & 0 & 1 & 0 & 0 & 0 & 1  \\
X_{24} &  0 & 0 & 0 & 0 & 0 & 0 & 1 & 1 & 1 & 1  \\
\end{array}
\right)
\ \ \ 
 P=
 \left(
\begin{array}{c|cccccccc}
\; & p_1 & p_2 & p_3 & p_4 & p_5 & p_6 & q_1 & q_2 \\
\hline
 D_{12} & 1 & 0 & 0 & 0 & 0 & 0 & 0 & 0 \\
 X_{12} & 0 & 1 & 0 & 0 & 0 & 0 & 0 & 0 \\
 Y_{31} & 0 & 0 & 1 & 0 & 0 & 0 & 1 & 0 \\
D_{31} & 0 & 0 & 0 & 1 & 0 & 0 & 1 & 0 \\
D_{41} & 0 & 0 & 0 & 0 & 1 & 0 & 1 & 0 \\
Y_{41} & 0 & 0 & 0 & 0 & 0 & 1 & 1 & 0 \\
X_{24} & 0 & 0 & 1 & 0 & 0 & 0 & 0 & 1 \\
Z_{24} & 0 & 0 & 0 & 1 & 0 & 0 & 0 & 1 \\
X_{23} & 0 & 0 & 0 & 0 & 1 & 0 & 0 & 1 \\
Z_{23} & 0 & 0 & 0 & 0 & 0 & 1 & 0 & 1 \\
\end{array}
\right)
~.~
 \eeq}
Finally, the GLSM charge matrices become
 \beal{es110a23}
  Q_{JE}=
 \left(
\begin{array}{cccccccc}
p_1 & p_2 & p_3 & p_4 & p_5 & p_6 & q_1 & q_2 \\
\hline
 0 & 0 & -1 & -1 & -1 & -1 & 1 & 1 \\
\end{array}
\right) ~~,~~
Q_D=
\left(
\begin{array}{cccccccc}
p_1 & p_2 & p_3 & p_4 & p_5 & p_6 & q_1 & q_2 \\
\hline
 -1 & -1 & 0 & 0 & 0 & 0 & 0 & 1 \\
 0 & 0 & 1 & 1 & 0 & 0 & 0 & -1 \\
 0 & 0 & 0 & 0 & 1 & 1 & 0 & -1 \\
\end{array}
\right)
~.~
\nn\\
 \eea
This results in
 \beal{es110a24}
 G=
\left(
\begin{array}{cccccccc}
p_1 & p_2 & p_3 & p_4 & p_5 & p_6 & q_1 & q_2 \\
\hline
 1 & 1 & 1 & 1 & 1 & 1 & 2 & 2 \\
 0 & 1 & 1 & 0 & 0 & 1 & 1 & 1 \\
 0 & 1 & 0 & 1 & 1 & 0 & 1 & 1 \\
 0 & 0 & 0 & 0 & 1 & -1 & 0 & 0 \\
\end{array}
\right)
~.~
 \eea
Following our discussion in section \sref{sextrahs}, it is possible to identify $q_1$ and $q_2$ as possible extra GLSM fields and to verify this is the case using the Hilbert series. The toric diagram for the mesonic moduli space then corresponds to removing $q_1$ and $q_2$ and, as shown in \fref{ftoricq111}, is precisely the one for $Q^{1,1,1}$.

 \bigskip
 
\paragraph{Global Symmetry Enhancement.} 
The full matrix of complexified $U(1)$ charges for GLSM fields is given by
\beal{es399a5}
Q_t =
\left(\ba{c}
Q_{JE} \\ Q_D
\ea\right)
=
 \left(
\begin{array}{cccccccc}
p_1 & p_2 & p_3 & p_4 & p_5 & p_6 & q_1 & q_2 \\
\hline
 0 & 0 & -1 & -1 & -1 & -1 & 1 & 1 \\
  -1 & -1 & 0 & 0 & 0 & 0 & 0 & 1 \\
 0 & 0 & 1 & 1 & 0 & 0 & 0 & -1 \\
 0 & 0 & 0 & 0 & 1 & 1 & 0 & -1 \\
\end{array}
\right) ~.~
\eea
Each of the pairs $\{p_1,p_2\}$, $\{p_3,p_4\}$ and $\{p_5,p_6\}$ have the same complexified $U(1)$ charges and indeed transform as doublets of independent $SU(2)$ global symmetry factors. This indicates an enhancement of the global symmetry to the $U(1)_R \times SU(2)^3$ expected from the geometry. 

\bigskip
 
\subsection{The Higgsing Web}
 
\begin{figure}[ht!!]
\begin{center}
\includegraphics[trim=0cm 0cm 0cm 0cm,width=11 cm]{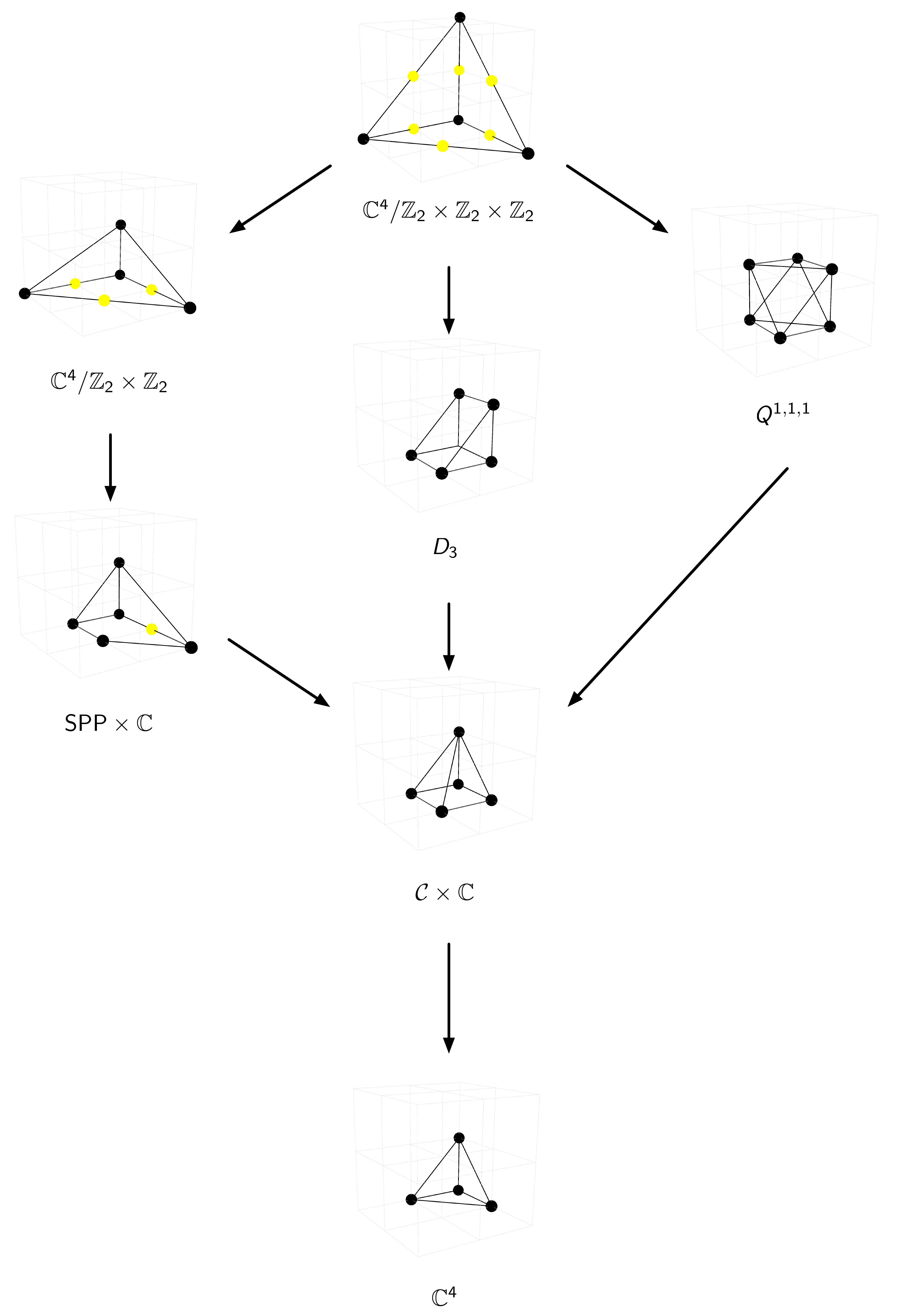}
\caption{A web showing several of the geometries for which we determined a $2d$ gauge theory and their connections through higgsing.
\label{fhiggstree}}
 \end{center}
 \end{figure}

\fref{fhiggstree} shows a map of several geometries whose corresponding gauge theories are considered in this work. The figure also indicates how we connected the theories by higgsing. Many of the gauge theories in this web can be obtained by two independent methods out of: orbifold techniques, dimensional reduction and partial resolution. This provides compelling support for the consistency of our results.
 
Of these theories, only $\mathbb{C}^4/\mathbb{Z}_{2}\times\mathbb{Z}_{2}\times\mathbb{Z}_{2}$ $(1,0,0,1)(0,1,0,1)(0,0,1,1)$ and $Q^{1,1,1}$ get extra GLSM fields in the forward algorithm.\footnote{It is in principle possible that ``dual" gauge theories exist for some of these geometries. It is also possible that some theories with extra GLSM fields have duals without them. We postpone this interesting question for future research.} Our systematic approach to partial resolution implies that it maintains or reduces the number of extra GLSM fields in a theory and that theories without extra GLSM fields can originate from those that have them. As illustrated with explicit examples in section \sref{section_orbifold_examples} and appendix \ref{sc4z2z2z2}, even standard orbifolds can sometimes contain extra GLSM fields. These two facts support our idea that extra GLSM fields do not imply any pathology of the corresponding gauge theories, but rather are a over-parameterization of the mesonic moduli space under the forward algorithm.
\\

 \bigskip

\section{Towards Brane Bricks}

\label{section_brane_bricks}

This final section provides an appetizer anticipating some of the results that will be presented in detail in the forthcoming works \cite{topub1,topub2}. Our goal is to construct a new type of brane configuration that serves as a more direct bridge connecting toric CY$_4$ singularities to the $2d$ $(0,2)$ quiver gauge theories that arise on the D1-branes probing them. Such a brane configuration explicitly encodes the $2d$ gauge theory with its defining quiver and $J$- and $E$-terms. It also bypasses the intricacies of the forward algorithm, which indeed becomes computationally demanding for moderately complicated geometries. This brane construction corresponds to a periodic tessellation of the 3-torus that we call {\it brane brick model}. 

Brane brick models are analogous to the brane tilings for $4d$ $\mathcal{N}=1$ quiver gauge theories on D3-branes over toric CY 3-folds \cite{Franco:2005rj},  to which we have referred throughout this work. It is thus instructive to review a few more facts about them before embarking on the construction of brane brick models.

\bigskip

\paragraph{Brane Tilings.}

Brane tilings, also known as dimers \cite{Franco:2005rj}, are bipartite periodic graphs on $T^2$ that encode a class of $4d$ $\mathcal{N}=1$ gauge theories. Faces, edges and nodes of the graph represent gauge groups, chiral fields and superpotential terms respectively.

One of the important features of brane tilings is that they allow a direct connection with the underlying CY$_3$ geometry. As mentioned in the above sections, solutions to F- and D-term constraints of the $4d$ supersymmetric gauge theory are encoded in terms of perfect matchings, providing an extremely efficient combinatorial alternative to the forward algorithm \cite{Franco:2005rj,Franco:2006gc}. Conversely, it is straightforward to construct brane tilings starting from geometry in terms of so-called zig-zag paths \cite{Hanany:2005ss, Feng:2005gw}.

\begin{figure}[h]
\begin{center}
\resizebox{1\hsize}{!}{
\includegraphics[trim=0cm 0cm 0cm 0cm,totalheight=10 cm]{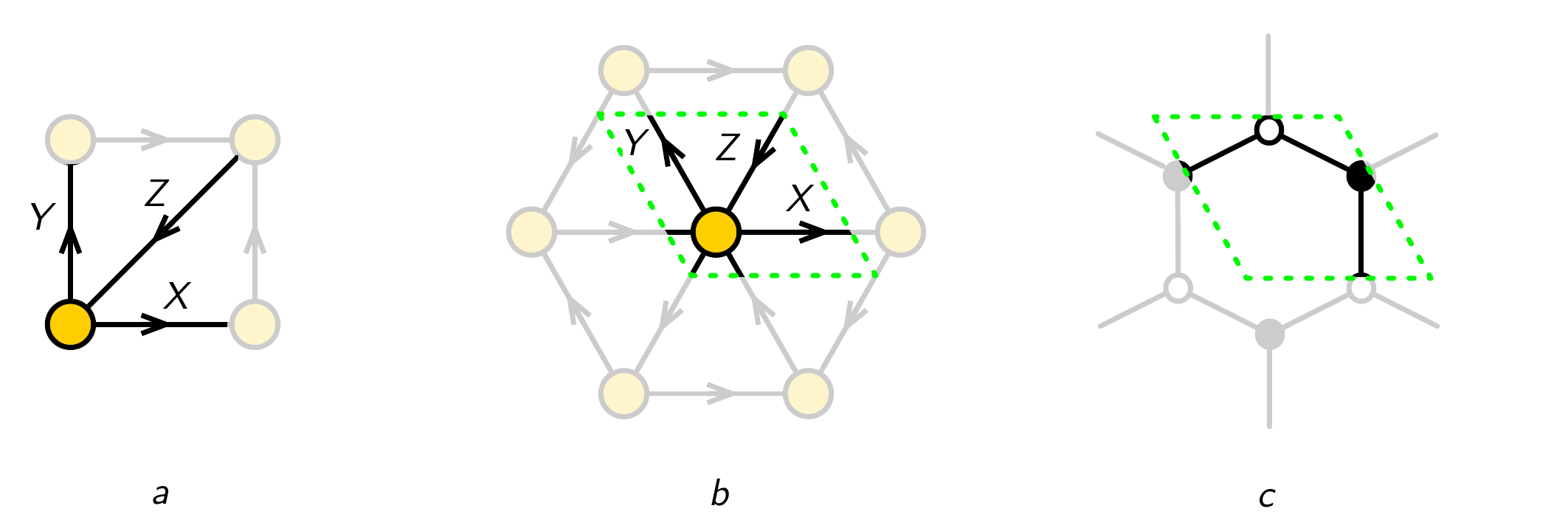}
}  
\caption{
\textit{Dualizing the $4d$ periodic quiver on $T^2$ into the brane tiling for the $\mathbb{C}^3$ example.}
(a) The periodic quiver can be fitted into a unit cell of $T^2$ which is taken to be a square here. (b) When the periodic quiver is drawn such that the symmetry of the quiver is manifest, the unit cell is not necessarily a square anymore.  (c) The dual graph of the periodic quiver is the brane tiling on $T^2$.}
	\label{fig:C3-tiling}
 \end{center}
 \end{figure}

From the point of view of string theory, a brane tilings represents a configuration consisting of an NS5- and D5-branes in Type IIB string theory. The NS5-brane extends in the 0123 directions and wraps a holomorphic curve embedded in the 4567 directions (with 5 and 7 compactified in a $T^2$). The D5-branes span 012357 and are suspended inside holes of the NS5-brane like soap bubbles. These configurations are connected to D3-branes over toric Calabi-Yau 3-folds by T-duality. It is also important to note that the brane tiling is the graph dual to the periodic quiver of the $4d$ $\mathcal{N}=1$ toric theory, as illustrated in \fref{fig:C3-tiling} for $\mathbb{C}^3$. 

\bigskip

\paragraph{Brane Bricks.}

\begin{figure}[h]
\begin{center}
\resizebox{1\hsize}{!}{
\includegraphics[trim=0cm 0cm 0cm 0cm,totalheight=10 cm]{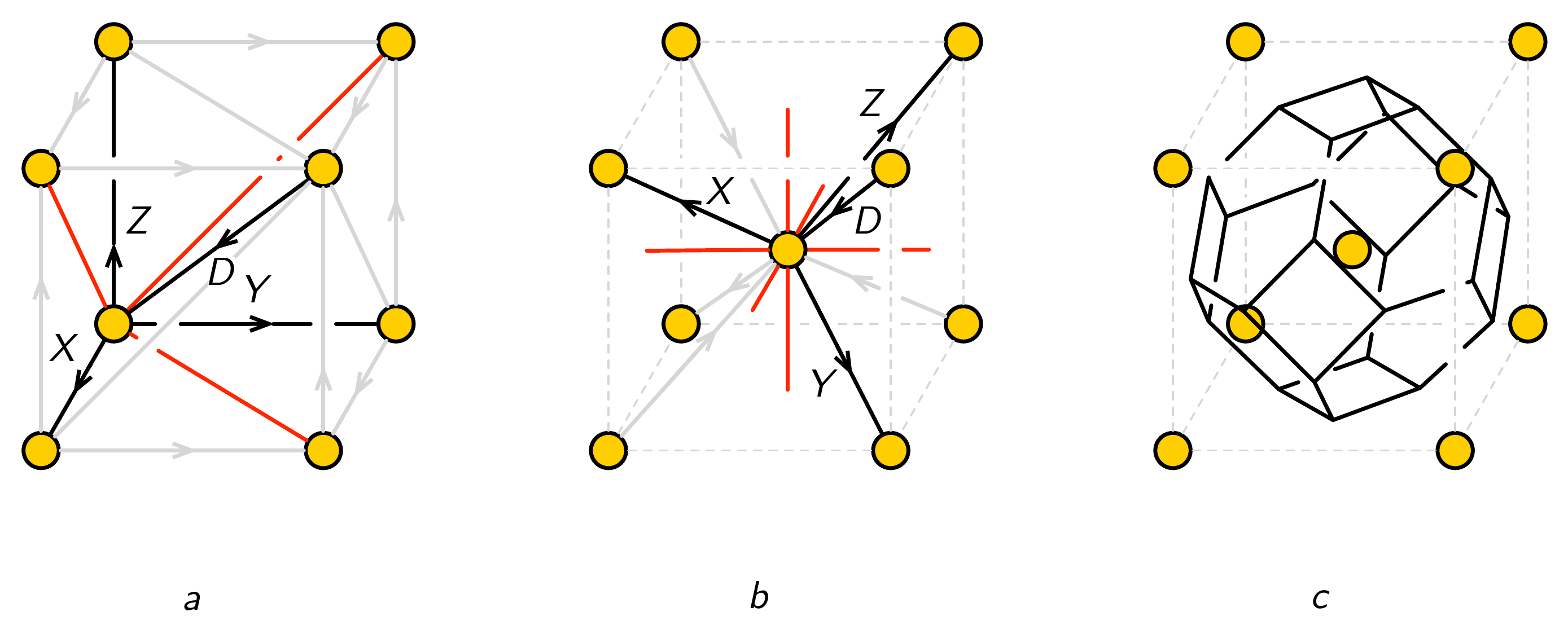}
}  
\caption{
\textit{Dualizing the $2d$ periodic quiver on $T^3$ into the brane tiling for the $\mathbb{C}^4$ example.}
(a) The periodic quiver of the $\mathbb{C}^4$ theory can be manifestly symmetrized to give (b) the body centered cubic (bcc) lattice. (c) The graph-dual of the bcc lattice is the bitruncated cubic honeycomb composed of truncated octahedra.
\label{fig:bcc2}}
 \end{center}
 \end{figure}

\begin{figure}[h]
\begin{center}
\includegraphics[trim=0cm 0cm 0cm 0cm,width=12 cm]{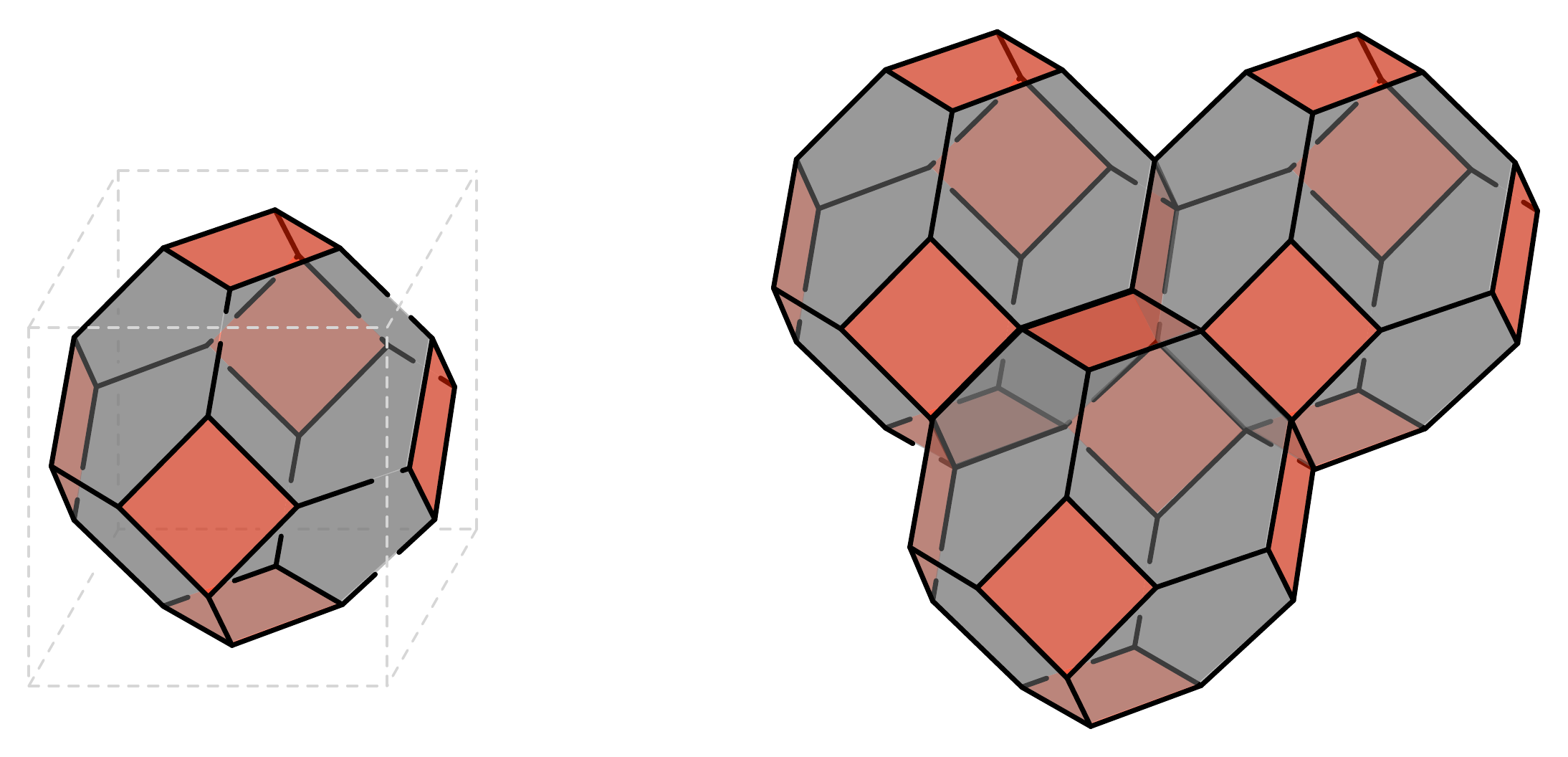}
\caption{
A single brane brick for the $\mathbb{C}^4$ theory and its tessellation on $T^3$.
\label{fbrick}}
 \end{center}
 \end{figure}

We have discussed at length the periodic quivers on $T^3$ associated with $2d$ $(0,2)$ toric theories. In analogy to the construction of brane tilings in $4d$, brane brick models can be obtained by dualizing the periodic quivers on $T^3$. This procedure is illustrated in \fref{fig:bcc2} for $\mathbb{C}^4$. The corresponding periodic quiver, which has been presented in section \sref{section_orbifold_exmaples0}, can be drawn in a manifestly symmetric way such that it takes the form of the \textit{body centered cubic (bcc) lattice}.\footnote{Notice that in order to make the bcc symmetry of the lattice on the covering space manifest, the region displayed in \fref{fig:bcc2} (b) has twice the volume of the unit cell in the periodic quiver.} The graph dual to the bcc lattice, which is also known as the Voronoi tessellation, gives the bitruncated cubic honeycomb. It is a space-filling tessellation of $T^3$ that is composed of \textit{truncated octahedra}, as shown in \fref{fbrick}. 

A truncated octahedron consists of 8 hexagonal and 6 square faces as illustrated in \fref{fbrick}. They map respectively to the chiral fields and Fermi fields of the $\mathbb{C}^4$ theory. The interior of the truncated octahedron corresponds to the single gauge group of the theory. Following our convention for $2d$ quiver diagrams, we have colored faces corresponding to chiral fields and Fermi fields respectively in black and red. As discussed in section \sref{section_orbifold_exmaples}, the periodic quivers for abelian orbifolds of $\mathbb{C}^4$ can be constructed by stacking together copies of the one for $\mathbb{C}^4$. As a result, the brane brick model for such an orbifold is the bitruncated cubic honeycomb with several truncated octahedra corresponding to nodes of the quiver diagram. The brane brick dictionary for abelian orbifolds of $\mathbb{C}^4$ is given in \tref{tbrick}.

\begin{table}[h]
\centering
\begin{tabular}{l|c|c}
\hline
\ \ \ \ {\bf Brane Brick} & {\bf Brane Brick for $\mathbb{C}^4/\Gamma$} & {\bf Gauge Theory}
\\
\hline\hline
Solid Brick & truncated octahedron & Gauge group \\
\hline
Brick Face (chiral) & hexagon & bifundamental or adjoint \\
& & chiral field \\
\hline
Brick Face (Fermi) & square & bifundamental or adjoint \\
& & Fermi field \\
\hline
\end{tabular}
\vspace{.5cm}\caption{
\textit{Dictionary for Brane Brick Models.}
The table gives the dictionary for a general brane brick model, the brane brick model for the $\mathbb{C}^4/\Gamma$ theory and the corresponding quiver gauge theory. 
\label{tbrick}
}
\end{table}
 
A brane brick model represents a Type IIA configuration consisting of an NS5-brane and D4-branes. The NS5-brane extends in the 01 directions and wraps a holomorphic surface (with four real dimensions) embedded in the 234567 directions (with 3, 5 and 7 compactified in a $T^3$). The D4-branes span 01357 and are suspended from the NS5-brane. 

Although in this section we have concentrated on orbifolds, the brane brick construction is fully generalizable to non-orbifold theories. This is the subject of a forthcoming paper \cite{topub1}.

\bigskip

\paragraph{Amoeba and Coamoeba.} 
A direct connection between toric geometry of the Calabi-Yau 4-fold, the brane brick model and the periodic quiver can be established in terms of the \textit{coamoeba}. It is useful to start by reviewing similar ideas that have been exploited in the context of brane tilings \cite{Feng:2005gw}. The toric diagram for a CY$_3$ cone is associated with a complex curve defined by the Newton polynomial as follows
\begin{align}
\sum_{(a,b) \in V} c_{(a,b)}\, x^a y^b  = 0~,~
\end{align}
where $V$ is the set of all vertices of the toric diagram, 
and $(x,y)$ take values in $(\mathbb{C}^*)^2$. This is the curve on which the NS5-brane of the brane tiling is wrapped. The projection of the curve onto the radial part $(\log|x|, \log|y|) \in \mathbb{R}^2$ defines the amoeba, which is a thickened 
version of the $(p,q)$-web dual to the toric diagram 
\cite{Aharony:1997ju,Aharony:1997bh,Leung:1997tw}.
The projection of the same curve onto the angular part $(\mathrm{arg}(x), \mathrm{arg}(y)) \in T^2$ defines the coamoeba. To get an idea of the structure of the coamoeba, 
it suffices to examine the asymptotics of the Newton polynomial. This is achieved by considering lines that are normal to the segments connecting external points in the toric diagram. 

\begin{figure}[h]
\begin{center}
\resizebox{0.6\hsize}{!}{
\includegraphics[trim=0cm 0cm 0cm 0cm,width=12 cm]{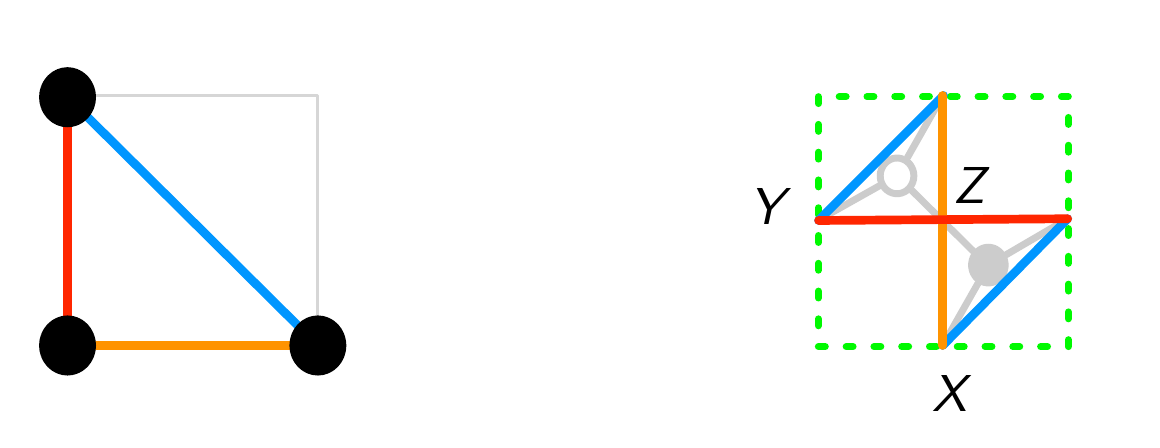}
}
\caption{
Left: toric diagram for $C^4$. Right: the lines on $T^2$ normal to the external edges of the toric diagram become the boundary of the coamoeba.
\label{c3coamoeba}}
 \end{center}
 \end{figure}

\begin{figure}[h]
\begin{center}
\resizebox{0.7\hsize}{!}{
\includegraphics[trim=0cm 0cm 0cm 0cm,totalheight=10 cm]{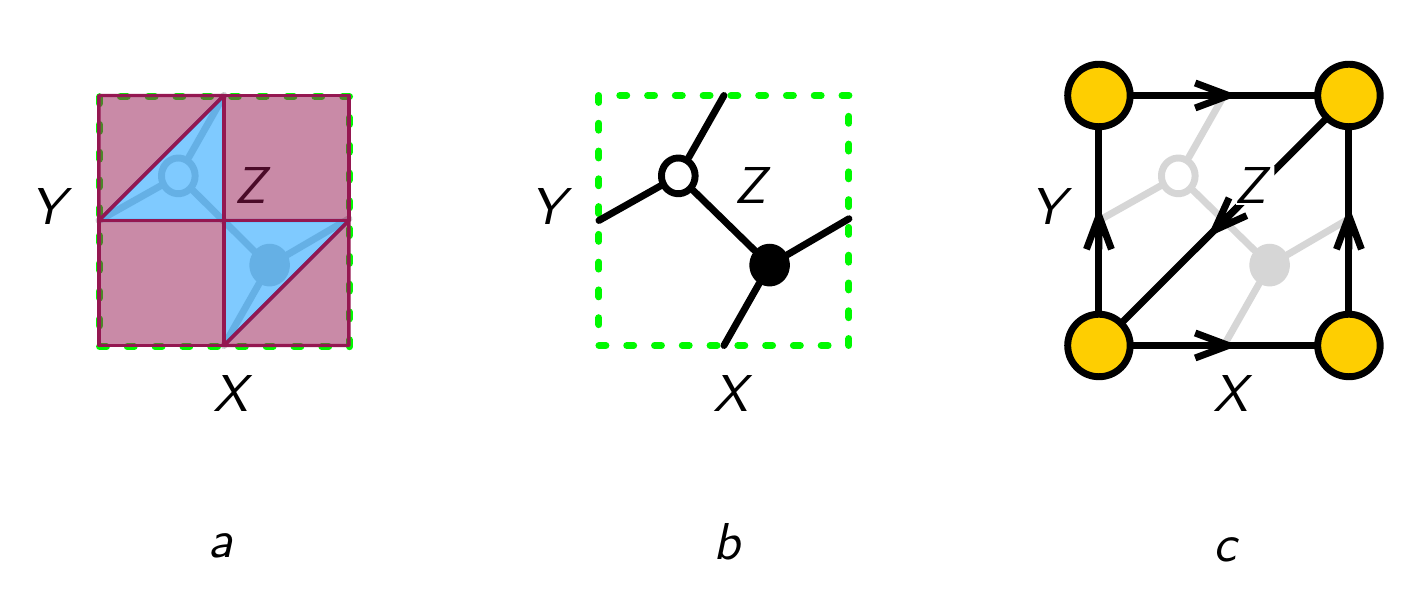}
}  
\caption{
\textit{Coamoeba and Brane Tiling for $\mathbb{C}^3$.} (a) The coamoeba and its complement are indicated in blue and red, respectively. (b) The brane tiling is the skeleton of the coamoeba. (c) The periodic quiver on $T^2$ is obtained by dualizing the brane tiling.
\label{fig:C3-coamoeba}}
 \end{center}
 \end{figure}

Let us illustrate this construction for $\mathbb{C}^3$. Its toric diagram and the corresponding normal lines on $T^2$ are shown in \fref{c3coamoeba}. The resulting coamoeba is shown in \fref{fig:C3-coamoeba}. The complement of the coamoeba is a disjoint union of domains in $T^2$. The brane tiling is the skeleton of the coamoeba.

It is straightforward to generalize the notions of amoeba and coamoeba 
to the Calabi-Yau 4-fold setup. We consider the complex surface defined by the Newton polynomial, 
\begin{align}
\sum_{(a,b,c) \in V} c_{(a,b,c)}\, x^a y^b z^c = 0~,~
\end{align}
where $V$ is the set of all vertices of the toric diagram
and $(x,y,z)$ take values in $(\mathbb{C}^*)^3$. This is the surface wrapped by the NS5-brane. For illustration, let us consider the example of $\mathbb{C}^4$, 
for which all coefficients can be removed by rescalings. Hence, we have 
\begin{align}
1 + x + y + z = 0 ~.~
\label{Newton_polynomial_C4}
\end{align}

\begin{figure}[h]
\begin{center}
\resizebox{0.23\hsize}{!}{
\includegraphics[trim=0cm 0cm 0cm 0cm,width=5cm]{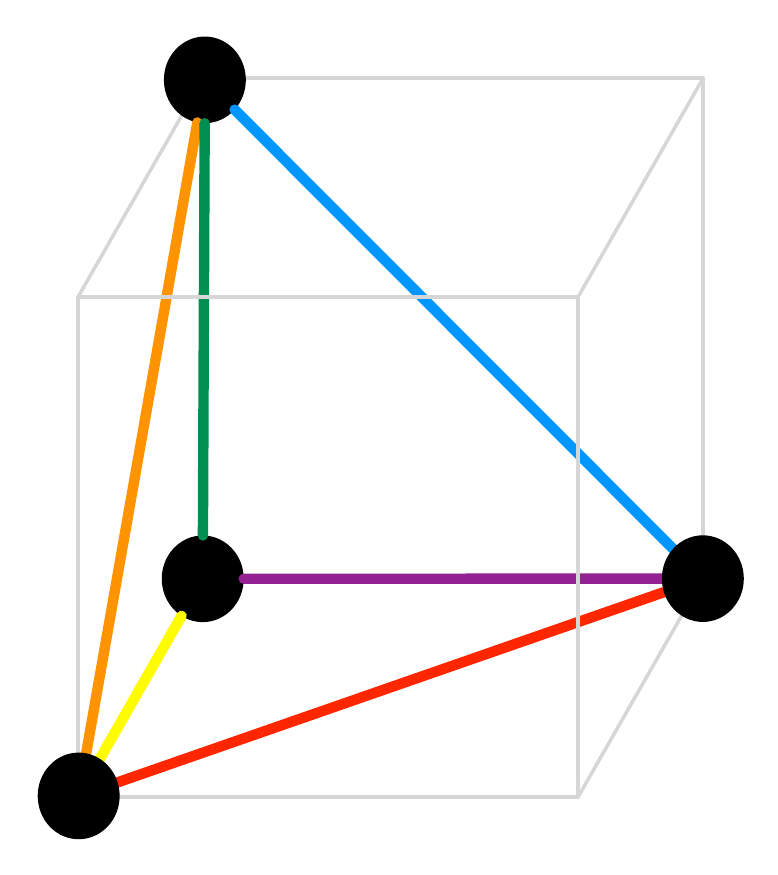}
}
\vspace{0.5cm}\caption{Toric diagram for $\mathbb{C}^4$. We have colored its external edges to identify the normal 2-planes in the coamoeba.
\label{toric_C4_color}}
 \end{center}
 \end{figure}

Once again, it is sufficient to study the asymptotic behavior of the Newton polynomial. In this case, it amounts to considering the 2-planes that are normal to the external edges of the toric diagram.\footnote{More generally, we should consider 2-cycles in $T^3$ whose homology is determined by the external edges of the toric diagram.} \fref{toric_C4_color} shows the toric diagram for $\mathbb{C}^4$.

\begin{figure}[h]
\begin{center}
\resizebox{0.7\hsize}{!}{
\includegraphics[trim=0cm 0cm 0cm 0cm,width=12 cm]{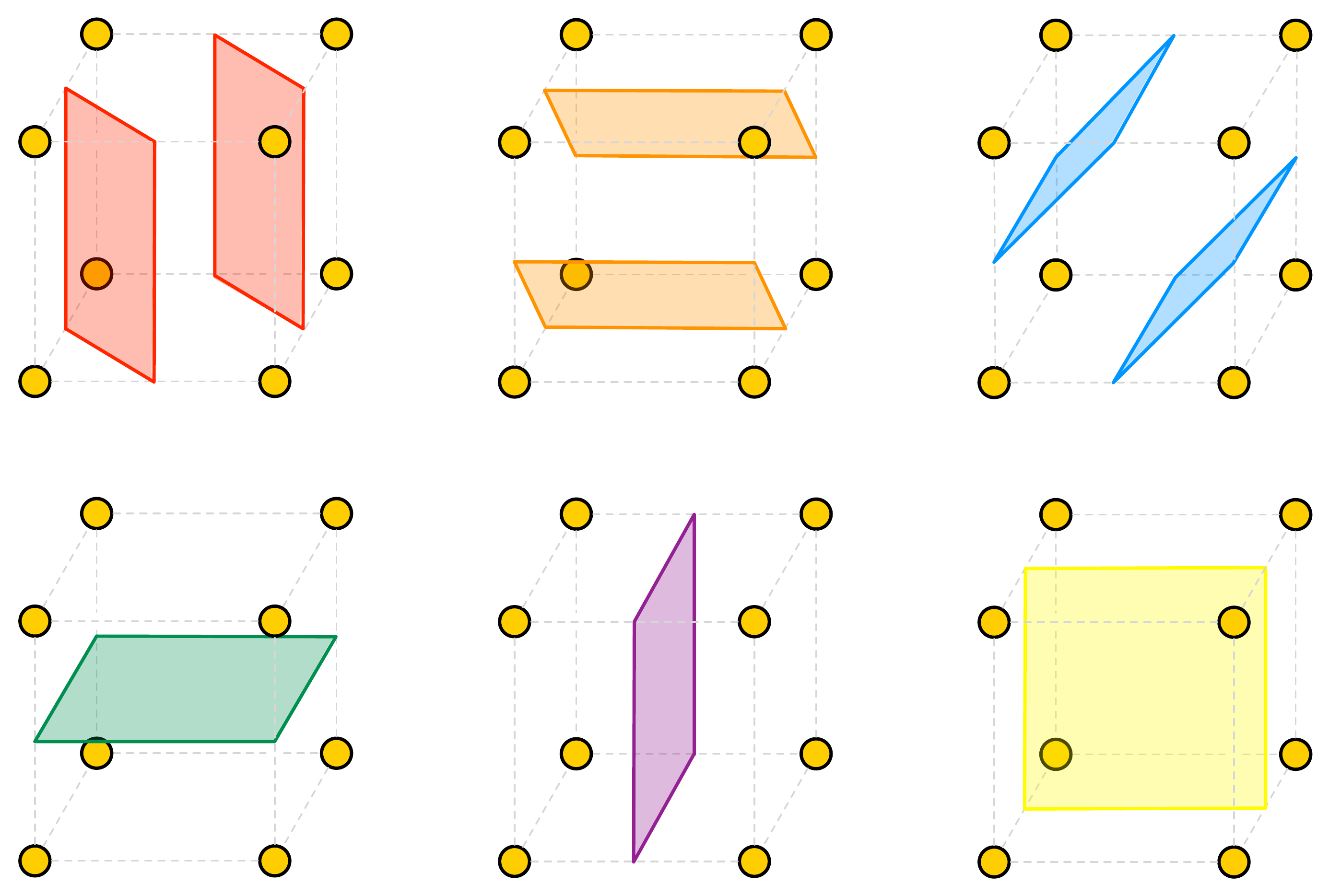}
}
\vspace{0.5cm}\caption{\textit{Coamoeba for Brane Brick Model for $\mathbb{C}^4$.}
The six 2-planes in $T^3$ corresponding to the six edges of the toric diagram of $\mathbb{C}^4$ and the asymptotic boundary of the coamoeba. We use the same colors for the planes and their normal vectors in \fref{toric_C4_color}.  The planes cut out a rhombic-dodecahedron in $T^3$ that is the complement of the coamoeba.
\label{coamoeba_planes_color}}
 \end{center}
 \end{figure}

\begin{figure}[h]
\begin{center}
\resizebox{1\hsize}{!}{
\includegraphics[trim=0cm 0cm 0cm 0cm,totalheight=10 cm]{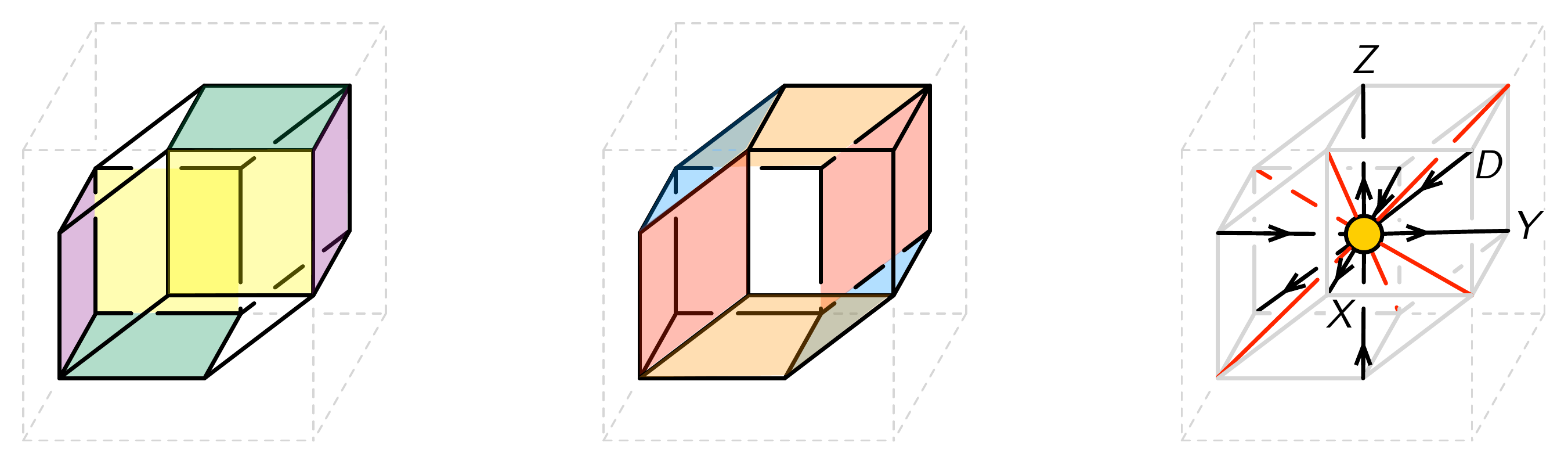}
}  
\caption{
\textit{Rhombic-dodecahedron.} The six 2-planes in \fref{coamoeba_planes_color} cut out a rhombic dodecahedron in $T^3$. It has eight 3-valent vertices and six 4-valent vertices, which correspond to the 4 chiral and 3 Fermi fields of the $\mathbb{C}^4$ theory, respectively. We precisely recover the periodic quiver in \fref{fig:bcc2}.
\label{fig:RD}}
 \end{center}
 \end{figure}

\fref{coamoeba_planes_color} shows the six 2-planes normal to the edges of the toric diagram. When combined, they carve out a single {\em rhombic-dodecahedron} (RD) in $T^3$ 
as the complement of the coamoeba. In analogy with brane tilings, we identify the bulk of the RD with the gauge group, and its vertices with the matter fields. 
The RD has eight 3-valent vertices and six 4-valent vertices, which nicely matches 
the fact that the $\mathbb{C}^4$ theory has four chiral fields and three Fermi fields as illustrated in \fref{fig:RD}.

The same ideas can be applied to produce the brane brick models corresponding to other orbifold and non-orbifold toric CY$_4$ geometries \cite{topub1}. 

\bigskip

\section{Conclusions \label{section_conclusions}}

We have initiated a comprehensive investigation of the $2d$ $(0,2)$ quiver gauge theories arising on D1-branes probing toric CY$_4$ cones, at the classical level. This setup can also give rise to theories with enhanced SUSY. 

The CY$_4$ transverse to the D1-branes arises as the mesonic moduli space of the worldvolume gauge theory. In order to efficiently calculate the mesonic moduli spaces of the class of gauge theories under consideration, we developed the forward algorithm. We applied our ideas to a variety of geometries, including abelian orbifolds of $\mathbb{C}^4$, CY$_3\times \mathbb{C}$ cones and generic toric singularities.

We also introduced a systematic procedure for constructing gauge theories associated with arbitrary toric singularities by means of partial resolution, which translates to higgsing in the gauge theory.  We showed how the gauge theories for several geometries are connected by RG flows triggered by VEVs for bifundamental scalars and presented two explicit examples of theories for singularities that are neither orbifolds nor of the form CY$_3\times \mathbb{C}$. We also explained how to use the $P$-matrix of the parent theory to identify the set of VEVs producing a desired partial resolution. At each stage, we used the forward algorithm to verify that the classical mesonic moduli space of the gauge theory agrees with the Calabi-Yau 4-fold under consideration.

We discussed how toric $2d$ gauge theories are fully captured by periodic quivers on $T^3$, which were originally introduced in \cite{GarciaCompean:1998kh} in the context of orbifolds. Periodic quivers not only encode the gauge symmetry and matter content of the theory, but also its $J$- and $E$-terms. In theories corresponding to toric geometries, these terms have a special structure involving contributions coming from pairs of plaquettes in the quiver. In the case of the $2d$ $(2,2)$ theories for toric CY$_3\times \mathbb{C}$ geometries, we introduced a lifting algorithm that produces the periodic quiver on $T^3$ from the periodic quiver on $T^2$ associated with the $4d$ $\mathcal{N}=1$ theory on D3-branes over the corresponding CY$_3$.

Partial resolution is an efficient method for obtaining gauge theories for arbitrary toric singularities, but it becomes considerably involved for complicated geometries. Similarly, determining the probed geometry as the mesonic moduli space of the corresponding gauge theory by means of the forward algorithm also turns computationally intensive as the complexity of the gauge theory is increased. It is thus desirable to establish a more direct connection between geometry and gauge theory. For this purpose, we introduced brane brick models, which are T-dual to the D1-CY$_4$ system. A brane brick model consists of stacks of D4-branes suspended from an NS5-brane wrapping a holomorphic surface, tessellating a 3-torus. Bricks correspond to gauge groups and their faces represent chiral or Fermi fields. Brane brick models can be obtained from the periodic quivers by graph dualization. In addition, we previewed an algorithm for constructing brane brick models directly from geometric data in terms of the coamoeba. A thorough study of brane brick models, including additional combinatorial tools for connecting geometry to gauge theory, will be presented in an upcoming work \cite{topub1}.

We conclude mentioning a few topics for future investigation. An obvious question is how the quantum behavior of the gauge theories is captured by branes. Another interesting direction is to establish how triality \cite{Gadde:2013lxa} is realized in terms of brane bricks. We will report on this issue in \cite{topub2}. More generally, it would also be interesting to establish to what extent different gauge theories associated with the same underlying CY$_4$ are related.

\bigskip

\section*{Acknowledgements}

We thank A. Amariti, A. Craw, T. Dimofte, A. Gadde, S. Kim, M. Yamazaki and especially A. Uranga for helpful discussions. This work was initiated in the ICMS workshop ``Gauge theories: quivers, tilings and Calabi-Yaus" in May 2014. The work of DG and SL is supported by Samsung Science and Technology Foundation under Project Number SSTF-BA1402-08. DY acknowledges support by the ERC Starting Grant N. 304806, ``The Gauge/Gravity Duality and Geometry in String Theory". DG is also supported by a POSCO TJ Park Science Fellowship. 


\newpage

\appendix

\section{Additional Examples \label{sappc4}} 

Here we collect detailed information on the gauge theories and the application of the forward algorithm for three additional orbifolds, two of which have been used as starting points for the partial resolutions leading to the theories presented in section \sref{section_beyond_orbifolds}. Early studies of some of these orbifolds were carried out in \cite{Ahn:1999iu, Sarkar:2000iz}.

\bigskip

\subsection{$\mathbb{C}^4/\mathbb{Z}_{2}\times\mathbb{Z}_{2}$ $(1,1,0,0)(1,0,1,0)$ \label{sc4z2z2a}}

\begin{figure}[H]
\begin{center}
\resizebox{0.4\hsize}{!}{
\includegraphics[trim=0cm 0cm 0cm 0cm,totalheight=10 cm]{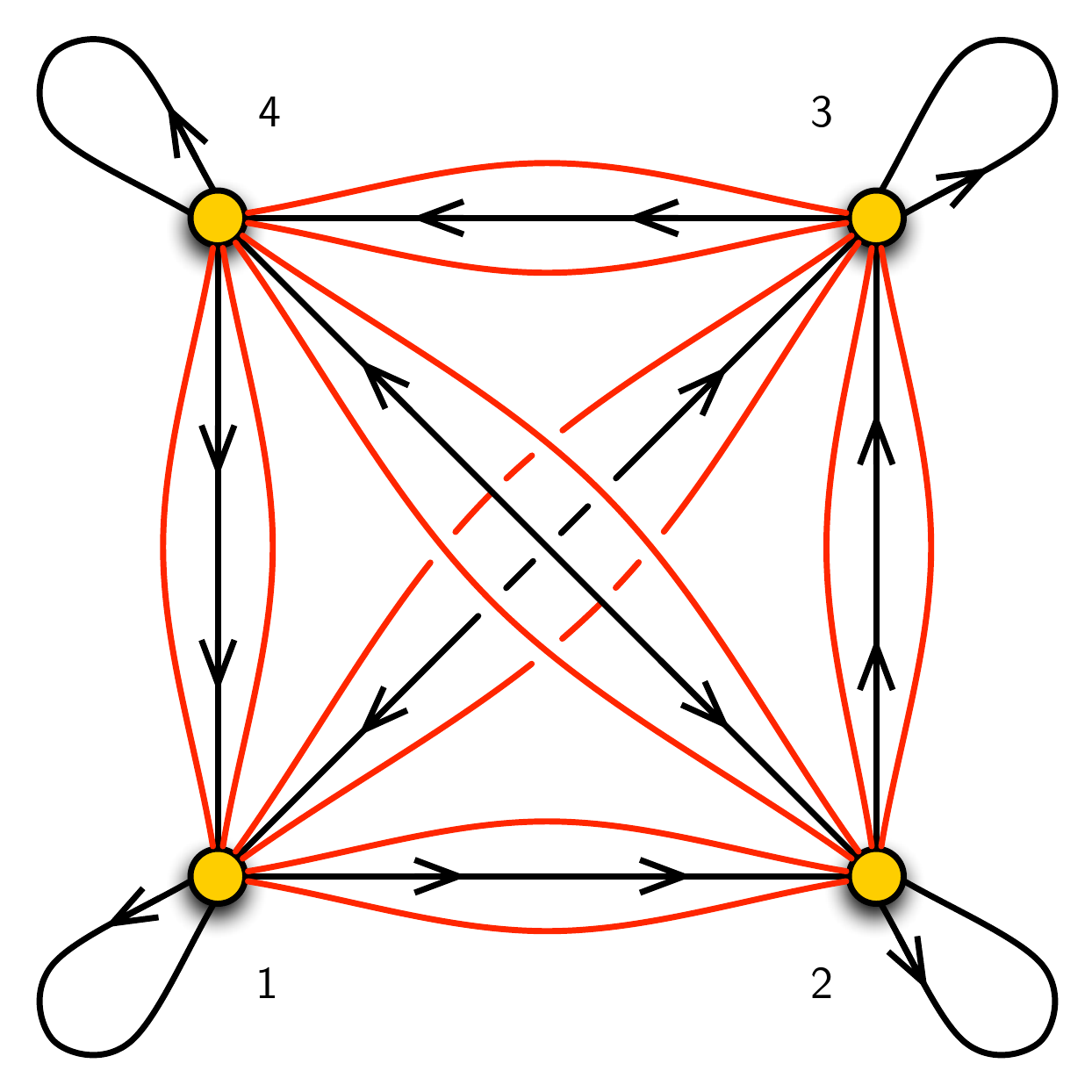}
}  
\caption{
Quiver diagram for $\mathbb{C}^4/\mathbb{Z}_2\times\mathbb{Z}_2$ $(1,1,0,0)(1,0,1,0)$.
\label{fz2z201010011}}
 \end{center}
 \end{figure}

\fref{fz2z201010011} shows the quiver diagram for $\mathbb{C}^4/\mathbb{Z}_{2}\times\mathbb{Z}_{2}$ $(1,1,0,0)(1,0,1,0)$. 

The $J$- and $E$-terms are 
\beq
\begin{array}{rclccclcc}
& & \ \ \ \ \ \ \ \ \ \ \ \ \ \ \ J  & & & & \ \ \ \ \ \ \ \ \ \ \ \ \ E & & \\
\Lambda_{12} : & \ \ \ & X_{23} \cdot Y_{31} - Y_{24} \cdot X_{41}& = & 0 & \ \ \ \ &D_{11} \cdot Z_{12} - Z_{12} \cdot D_{22}&= & 0 \\ 
\Lambda_{21} : & \ \ \ & X_{14} \cdot Y_{42} - Y_{13} \cdot X_{32}& = & 0 & \ \ \ \ & D_{22} \cdot Z_{21} - Z_{21} \cdot D_{11}&= & 0 \\ 
\Lambda_{13} : & \ \ \ & Z_{34} \cdot X_{41} - X_{32} \cdot Z_{21}& = & 0 & \ \ \ \ &D_{11} \cdot Y_{13} - Y_{13} \cdot D_{33}&= & 0 \\ 
\Lambda_{31} : & \ \ \ &  Z_{12} \cdot X_{23} -X_{14} \cdot Z_{43}& = & 0 & \ \ \ \ &D_{33} \cdot Y_{31} - Y_{31} \cdot D_{11}&= & 0 \\ 
\Lambda_{14} : & \ \ \ & Y_{42} \cdot Z_{21} - Z_{43} \cdot Y_{31}& = & 0 & \ \ \ \ &D_{11} \cdot X_{14} - X_{14} \cdot D_{44}&= & 0 \\ 
\Lambda_{41} : & \ \ \ & Y_{13} \cdot Z_{34} - Z_{12} \cdot Y_{24}& = & 0 & \ \ \ \ &D_{44} \cdot X_{41} - X_{41} \cdot D_{11}&= & 0 \\ 
\Lambda_{23} : & \ \ \ & Y_{31} \cdot Z_{12} - Z_{34} \cdot Y_{42}& = & 0 & \ \ \ \ &D_{22} \cdot X_{23} - X_{23} \cdot D_{33}&= & 0 \\ 
\Lambda_{32} : & \ \ \ & Y_{24} \cdot Z_{43} - Z_{21} \cdot Y_{13}& = & 0 & \ \ \ \ &D_{33} \cdot X_{32} - X_{32} \cdot D_{22}&= & 0 \\ 
\Lambda_{24} : & \ \ \ & Z_{43} \cdot X_{32} -X_{41} \cdot Z_{12}& = & 0 & \ \ \ \ &D_{22} \cdot Y_{24} - Y_{24} \cdot D_{44}&= & 0 \\ 
\Lambda_{42} : & \ \ \ & Z_{21} \cdot X_{14} -X_{23} \cdot Z_{34}& = & 0 & \ \ \ \ &D_{44} \cdot Y_{42} - Y_{42} \cdot D_{22}&= & 0 \\ 
\Lambda_{34} : & \ \ \ & X_{41} \cdot Y_{13} - Y_{42} \cdot X_{23}& = & 0 & \ \ \ \ &D_{33} \cdot Z_{34} - Z_{34} \cdot D_{44}&= & 0 \\
\Lambda_{43} : & \ \ \ & X_{32} \cdot Y_{24} - Y_{31} \cdot X_{14}& = & 0 & \ \ \ \ &D_{44} \cdot Z_{43} - Z_{43} \cdot D_{33}&= & 0
\end{array}
\label{es115e1}
\eeq

\begin{figure}[ht!!]
\begin{center}
\resizebox{0.4\hsize}{!}{
\includegraphics[trim=0cm 0cm 0cm 0cm,totalheight=10 cm]{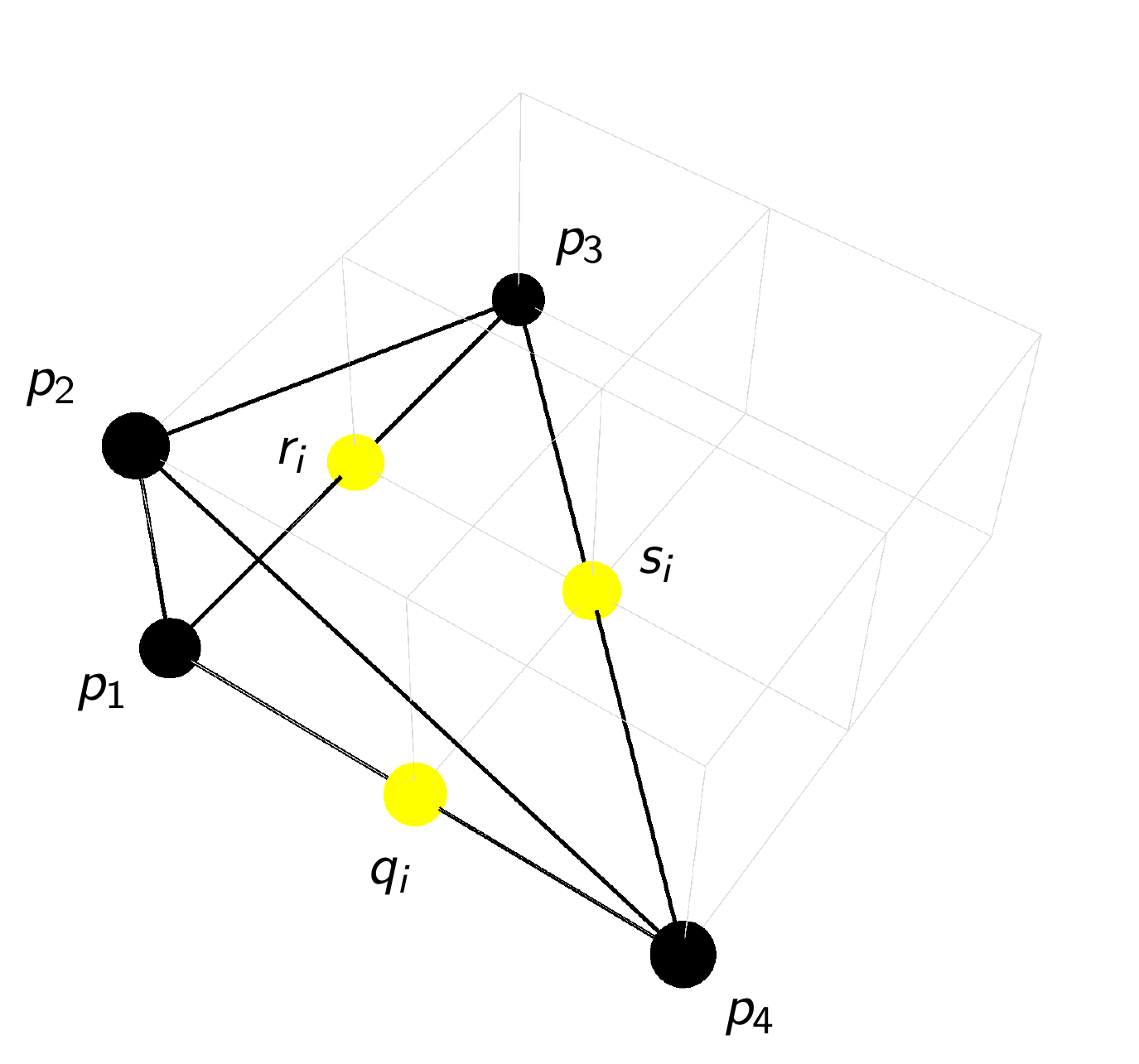}
}  
\caption{
Toric diagram for $\mathbb{C}^4/\mathbb{Z}_2\times\mathbb{Z}_2$ $(1,1,0,0)(1,0,1,0)$. This geometry has been obtained as the mesonic moduli space of the corresponding gauge theory.
\label{ftoricZ2Z201010011}}
 \end{center}
 \end{figure}

The quiver incidence matrix is
\beal{es115e2}
d=
\left(
\begin{array}{c|cccccccccccccccc}
\; & D_{11} & D_{22} & D_{33} & D_{44} & X_{14} & X_{23} & X_{32} & X_{41} & Y_{13} & Y_{24} & Y_{31} & Y_{42} & Z_{12} & Z_{21} & Z_{34} & Z_{43} \\
\hline
\encircle{1} &  0 & 0 & 0 & 0 & 1 & 0 & 0 & -1 & 1 & 0 & -1 & 0 & 1 & -1 & 0 & 0 \\
\encircle{2} &  0 & 0 & 0 & 0 & 0 & 1 & -1 & 0 & 0 & 1 & 0 & -1 & -1 & 1 & 0 & 0 \\
\encircle{3} &  0 & 0 & 0 & 0 & 0 & -1 & 1 & 0 & -1 & 0 & 1 & 0 & 0 & 0 & 1 & -1 \\
\encircle{4} &  0 & 0 & 0 & 0 & -1 & 0 & 0 & 1 & 0 & -1 & 0 & 1 & 0 & 0 & -1 & 1 \\
\end{array}
\right)
~.~
\nn\\
\eea
Following the forward algorithm, we obtain the $K$-matrix
\beal{es115e3}
K=
\left(
\begin{array}{c|cccccccccccccccc}
\; & D_{11} & D_{22} & D_{33} & D_{44} & X_{14} & X_{23} & X_{32} & X_{41} & Y_{13} & Y_{24} & Y_{31} & Y_{42} & Z_{12} & Z_{21} & Z_{34} & Z_{43} \\
\hline
D_{11} &  1 & 1 & 1 & 1 & 0 & 0 & 0 & 0 & 0 & 0 & 0 & 0 & 0 & 0 & 0 & 0 \\
X_{14} &  0 & 0 & 0 & 0 & 1 & 0 & 0 & -1 & 0 & 0 & -1 & -1 & 0 & -1 & 0 & -1 \\
X_{23} &  0 & 0 & 0 & 0 & 0 & 1 & 0 & 1 & 0 & 0 & 0 & 0 & 0 & 1 & 0 & 1 \\
X_{32} &  0 & 0 & 0 & 0 & 0 & 0 & 1 & 1 & 0 & 0 & 1 & 1 & 0 & 0 & 0 & 0 \\
Y_{13} &  0 & 0 & 0 & 0 & 0 & 0 & 0 & 0 & 1 & 0 & 0 & 1 & 0 & -1 & -1 & 0 \\
Y_{24} &  0 & 0 & 0 & 0 & 0 & 0 & 0 & 0 & 0 & 1 & 1 & 0 & 0 & 1 & 1 & 0 \\
Z_{12} &  0 & 0 & 0 & 0 & 0 & 0 & 0 & 0 & 0 & 0 & 0 & 0 & 1 & 1 & 1 & 1 \\
\end{array}
\right)
~,~
\nn\\
\eea
and the $P$-matrix
\beal{es115e4}
P=
\left(
\begin{array}{c|cccccccccc}
\; & p_1& p_2& p_3& p_4& q_1& q_2& r_1 & r_2 &s_1 & s_2\\
\hline
D_{11} &  0 & 1 & 0 & 0 & 0 & 0 & 0 & 0 & 0 & 0 \\
D_{22} &  0 & 1 & 0 & 0 & 0 & 0 & 0 & 0 & 0 & 0 \\
D_{33} &  0 & 1 & 0 & 0 & 0 & 0 & 0 & 0 & 0 & 0 \\
D_{44} &  0 & 1 & 0 & 0 & 0 & 0 & 0 & 0 & 0 & 0 \\
X_{14} &  1 & 0 & 0 & 0 & 0 & 1 & 1 & 0 & 0 & 0 \\
X_{23} &  1 & 0 & 0 & 0 & 0 & 1 & 0 & 1 & 0 & 0 \\
X_{32} &  1 & 0 & 0 & 0 & 1 & 0 & 1 & 0 & 0 & 0 \\
X_{41} &  1 & 0 & 0 & 0 & 1 & 0 & 0 & 1 & 0 & 0 \\
Y_{13} &  0 & 0 & 0 & 1 & 0 & 1 & 0 & 0 & 0 & 1 \\
Y_{24} &  0 & 0 & 0 & 1 & 0 & 1 & 0 & 0 & 1 & 0 \\
Y_{31} &  0 & 0 & 0 & 1 & 1 & 0 & 0 & 0 & 1 & 0 \\
Y_{42} &  0 & 0 & 0 & 1 & 1 & 0 & 0 & 0 & 0 & 1 \\
Z_{12} &  0 & 0 & 1 & 0 & 0 & 0 & 1 & 0 & 0 & 1 \\
Z_{21} &  0 & 0 & 1 & 0 & 0 & 0 & 0 & 1 & 1 & 0 \\
Z_{34} &  0 & 0 & 1 & 0 & 0 & 0 & 1 & 0 & 1 & 0 \\
Z_{43} &  0 & 0 & 1 & 0 & 0 & 0 & 0 & 1 & 0 & 1 \\
\end{array}
\right)~.~
\eea
The GLSM charge matrices are
\beal{es115e5}
Q_{JE} &=&
\left(
\begin{array}{cccccccccc}
p_1& p_2& p_3& p_4& q_1& q_2& r_1 & r_2 &s_1 & s_2\\
\hline
 1 & 0 & 1 & 0 & 0 & 0 & -1 & -1 & 0 & 0 \\
 0 & 0 & 2 & 0 & 1 & 1 & -1 & -1 & -1 & -1 \\
 0 & 0 & 1 & 1 & 0 & 0 & 0 & 0 & -1 & -1 \\
\end{array}
\right)~,~
\nn\\
Q_D&=&
\left(
\begin{array}{cccccccccc}
p_1& p_2& p_3& p_4& q_1& q_2& r_1 & r_2 &s_1 & s_2\\
\hline
 0 & 0 & 0 & 0 & -1 & 1 & -1 & 1 & 1 & -1 \\
 0 & 0 & 0 & 0 & 1 & -1 & 1 & -1 & 1 & -1 \\
 0 & 0 & 0 & 0 & 1 & -1 & -1 & 1 & -1 & 1 \\
\end{array}
\right)
~.~
\eea
Using the above charges, the toric diagram is captured by the matrix
\beal{es115e6}
G=
\left(
\begin{array}{cccccccccc}
p_1& p_2& p_3& p_4& q_1& q_2& r_1 & r_2 &s_1 & s_2\\
\hline
 1 & 1 & 1 & 1 & 1 & 1 & 1 & 1 & 1 & 1 \\
 0 & 0 & 0 & 2 & 1 & 1 & 0 & 0 & 1 & 1 \\
 0 & 0 & 2 & 0 & 0 & 0 & 1 & 1 & 1 & 1 \\
 0 & 1 & 0 & 0 & 0 & 0 & 0 & 0 & 0 & 0 \\
\end{array}
\right)~.~
\eea
as shown in \fref{ftoricZ2Z201010011}.

\bigskip

\subsection{$\mathbb{C}^4/\mathbb{Z}_{2}\times\mathbb{Z}_{2}$ $(0,0,1,1)(1,1,1,1)$}

\begin{figure}[H]
\begin{center}
\resizebox{0.4\hsize}{!}{
\includegraphics[trim=0cm 0cm 0cm 0cm,totalheight=10 cm]{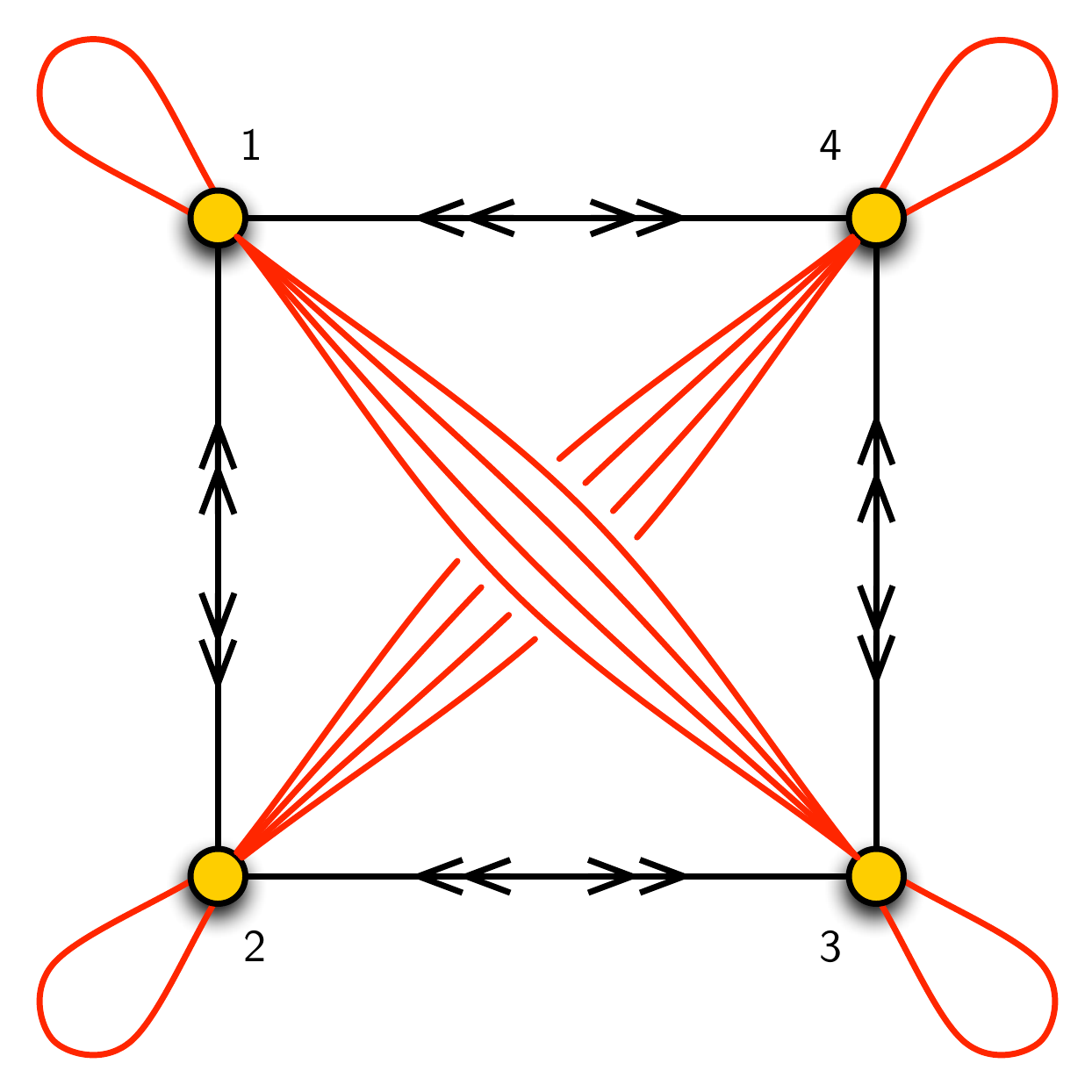}
}  
\caption{
Quiver diagram for $\mathbb{C}^4/\mathbb{Z}_2\times\mathbb{Z}_2$ $(0,0,1,1)(1,1,1,1)$.
\label{fz2z200111111}}
 \end{center}
 \end{figure}

\fref{fz2z200111111} shows the quiver diagram for $\mathbb{C}^4/\mathbb{Z}_{2}\times\mathbb{Z}_{2}$ $(0,0,1,1)(1,1,1,1)$. 

The $J$- and $E$-terms are 
\beq
\begin{array}{rclccclcc}
& & \ \ \ \ \ \ \ \ \ \ \ \ \ \ \ J  & & & & \ \ \ \ \ \ \ \ \ \ \ \ \ E & & \\
 \Lambda_{13}^{1} : & \ \ \ &  Y_{34} \cdot Z_{41} - Z_{32} \cdot Y_{21} & = & 0 & \ \ \ \ &D_{14} \cdot X_{43} - X_{12} \cdot D_{23}&= & 0 \\
 \Lambda_{31}^{1} : & \ \ \ &  Y_{12} \cdot Z_{23} - Z_{14} \cdot Y_{43} & = & 0 & \ \ \ \ &D_{32} \cdot X_{21} - X_{34} \cdot D_{41}&= & 0 \\
 \Lambda_{13}^{2} : & \ \ \ &  Z_{32} \cdot X_{21} - X_{34} \cdot Z_{41} & = & 0 & \ \ \ \ & D_{14} \cdot Y_{43} - Y_{12} \cdot D_{23}&= & 0 \\
 \Lambda_{31}^{2} : & \ \ \ &  Z_{14} \cdot X_{43} -X_{12} \cdot Z_{23} & = & 0 & \ \ \ \ & D_{32} \cdot Y_{21} - Y_{34} \cdot D_{41}&= & 0 \\
 \Lambda_{24}^{1} : & \ \ \ &  Y_{43} \cdot Z_{32} - Z_{41} \cdot Y_{12} & = & 0 & \ \ \ \ &D_{23} \cdot X_{34} - X_{21} \cdot D_{14}&= & 0 \\
 \Lambda_{42}^{1} : & \ \ \ &  Y_{21} \cdot Z_{14} - Z_{23} \cdot Y_{34} & = & 0 & \ \ \ \ &D_{41} \cdot X_{12} - X_{43} \cdot D_{32}&= & 0 \\
 \Lambda_{24}^{2} : & \ \ \ &  Z_{41} \cdot X_{12} -X_{43} \cdot Z_{32} & = & 0 & \ \ \ \ & D_{23} \cdot Y_{34} - Y_{21} \cdot D_{14}&= & 0 \\
 \Lambda_{42}^{2} : & \ \ \ &  Z_{23} \cdot X_{34} -X_{21} \cdot Z_{14} & = & 0 & \ \ \ \ & D_{41} \cdot Y_{12} - Y_{43} \cdot D_{32}&= & 0 \\
 \Lambda_{11} : & \ \ \ &  X_{12} \cdot Y_{21} - Y_{12} \cdot X_{21}& = & 0 & \ \ \ \ & D_{14} \cdot Z_{41} - Z_{14} \cdot D_{41}&= & 0 \\
 \Lambda_{22} : & \ \ \ &  X_{21} \cdot Y_{12} - Y_{21} \cdot X_{12}& = & 0 & \ \ \ \ & D_{23} \cdot Z_{32} - Z_{23} \cdot D_{32}&= & 0 \\
 \Lambda_{33} : & \ \ \ &  X_{34} \cdot Y_{43} - Y_{34} \cdot X_{43}& = & 0 & \ \ \ \ & D_{32} \cdot Z_{23} - Z_{32} \cdot D_{23}&= & 0 \\
 \Lambda_{44} : & \ \ \ &  X_{43} \cdot Y_{34} - Y_{43} \cdot X_{34}& = & 0 & \ \ \ \ &D_{41} \cdot Z_{14} - Z_{41} \cdot D_{14}&= & 0 
\end{array}
\label{es116e1}
\eeq

\begin{figure}[ht!!]
\begin{center}
\resizebox{0.4\hsize}{!}{
\includegraphics[trim=0cm 0cm 0cm 0cm,totalheight=10 cm]{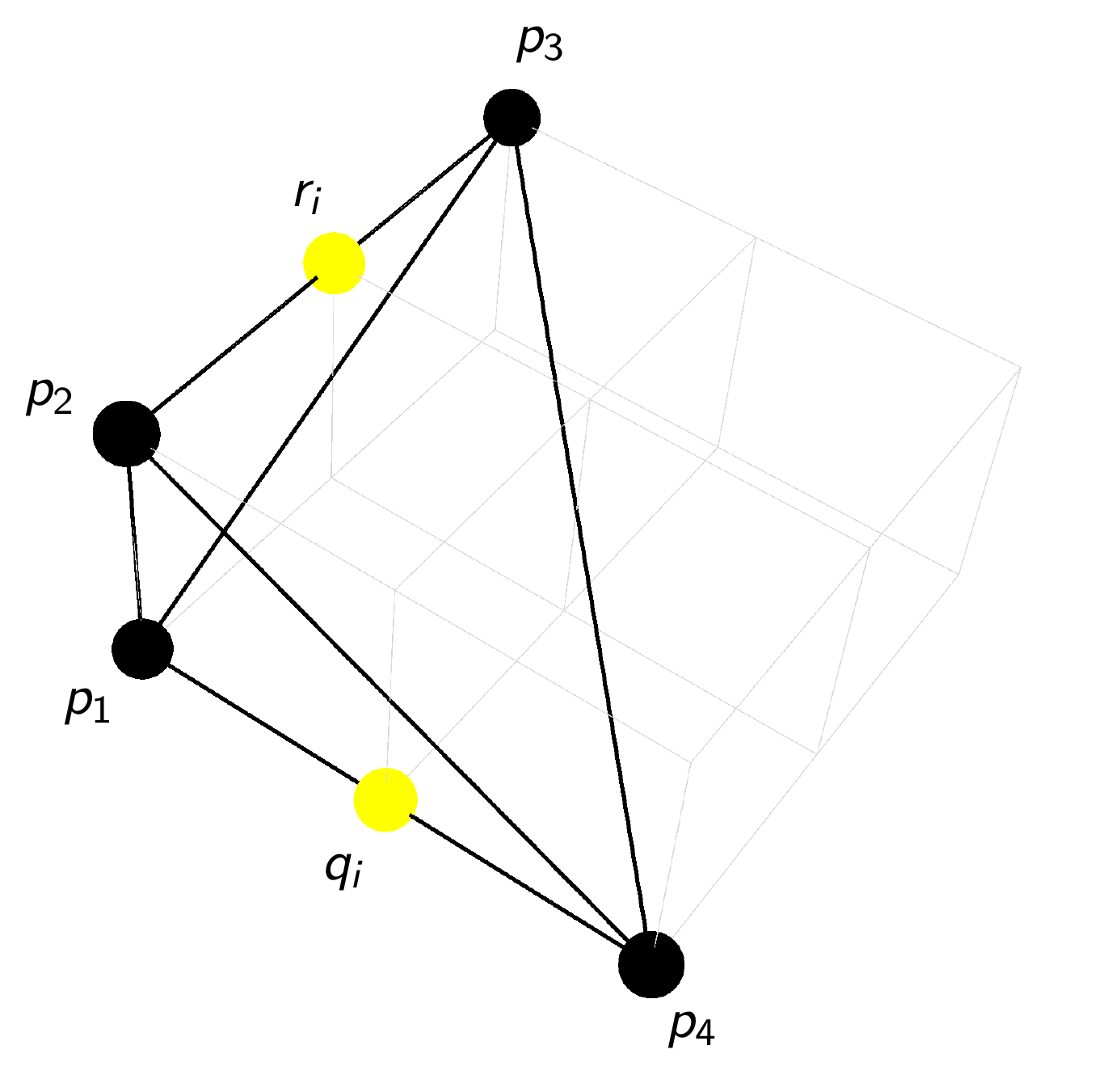}
}  
\caption{
Toric diagram for $\mathbb{C}^4/\mathbb{Z}_2\times\mathbb{Z}_2$ $(1,0,0,1)(1,1,1,1)$, obtained after removing the extra GLSM fields $s_1$ and $s_2$. 
\label{ftoricZ2Z210011111}}
 \end{center}
 \end{figure}

The corresponding incidence matrix is
\beal{es116e2}
d=
\left(
\begin{array}{c|cccccccccccccccc}
\; &D_{14} & D_{23} & D_{32} & D_{41} & X_{12} & X_{21} & X_{34} & X_{43} & Y_{12} & Y_{21} & Y_{34} & Y_{43} & Z_{14} & Z_{23} & Z_{32} & Z_{41}
\\
\hline
\encircle{1} & 1 & 0 & 0 & -1 & 1 & -1 & 0 & 0 & 1 & -1 & 0 & 0 & 1 & 0 & 0 & -1 \\
\encircle{2} & 0 & 1 & -1 & 0 & -1 & 1 & 0 & 0 & -1 & 1 & 0 & 0 & 0 & 1 & -1 & 0 \\
\encircle{3} & 0 & -1 & 1 & 0 & 0 & 0 & 1 & -1 & 0 & 0 & 1 & -1 & 0 & -1 & 1 & 0 \\
\encircle{4} & -1 & 0 & 0 & 1 & 0 & 0 & -1 & 1 & 0 & 0 & -1 & 1 & -1 & 0 & 0 & 1 \\
\end{array}
\right)
~.~
\nn\\
\eea
As part of the forward algorithm, we find the $K$-matrix
\beal{es116e3}
K=
\left(
\begin{array}{c|cccccccccccccccc}
\; &D_{14} & D_{23} & D_{32} & D_{41} & X_{12} & X_{21} & X_{34} & X_{43} & Y_{12} & Y_{21} & Y_{34} & Y_{43} & Z_{14} & Z_{23} & Z_{32} & Z_{41}
\\
\hline
D_{14} & 1 & 1 & 0 & 0 & 0 & 0 & 0 & 0 & 0 & 0 & 0 & 0 & 0 & 0 & -1 & -1 \\
D_{32} & 0 & 0 & 1 & 1 & 0 & 0 & 0 & 0 & 0 & 0 & 0 & 0 & 0 & 0 & 1 & 1 \\
X_{12} & 0 & 0 & 0 & 0 & 1 & 0 & 0 & 1 & 1 & 0 & 0 & 1 & 0 & 0 & 0 & 0 \\
X_{21} & 0 & 1 & 0 & 1 & 0 & 1 & 0 & 1 & 0 & 1 & 0 & 1 & 0 & 1 & 0 & 1 \\
X_{34} & 0 & -1 & 0 & -1 & 0 & 0 & 1 & -1 & -1 & -1 & 0 & -2 & 0 & -1 & 0 & -1 \\
Y_{34} & 0 & 0 & 0 & 0 & 0 & 0 & 0 & 0 & 1 & 1 & 1 & 1 & 0 & 0 & 0 & 0 \\
Z_{14} & 0 & 0 & 0 & 0 & 0 & 0 & 0 & 0 & 0 & 0 & 0 & 0 & 1 & 1 & 1 & 1 \\
\end{array}
\right)
~,~
\nn\\
\eea
and the $P$-matrix
\beal{es116e4}
P=
\left(
\begin{array}{c|cccccccccc}
\; & p_1 & p_2 & p_3 & p_4 & q_1 & q_2 & r_1 & r_2 & s_1 & s_2 \\
\hline
D_{14} & 1 & 0 & 0 & 0 & 1 & 0 & 0 & 0 & 1 & 0 \\
D_{23} & 1 & 0 & 0 & 0 & 1 & 0 & 0 & 0 & 0 & 1 \\
D_{32} & 1 & 0 & 0 & 0 & 0 & 1 & 0 & 0 & 1 & 0 \\
D_{41} & 1 & 0 & 0 & 0 & 0 & 1 & 0 & 0 & 0 & 1 \\
X_{12} & 0 & 1 & 0 & 0 & 0 & 0 & 1 & 0 & 1 & 0 \\
X_{21} & 0 & 1 & 0 & 0 & 0 & 0 & 0 & 1 & 0 & 1 \\
X_{34} & 0 & 1 & 0 & 0 & 0 & 0 & 0 & 1 & 1 & 0 \\
X_{43} & 0 & 1 & 0 & 0 & 0 & 0 & 1 & 0 & 0 & 1 \\
Y_{12} & 0 & 0 & 1 & 0 & 0 & 0 & 1 & 0 & 1 & 0 \\
Y_{21} & 0 & 0 & 1 & 0 & 0 & 0 & 0 & 1 & 0 & 1 \\
Y_{34} & 0 & 0 & 1 & 0 & 0 & 0 & 0 & 1 & 1 & 0 \\
Y_{43} & 0 & 0 & 1 & 0 & 0 & 0 & 1 & 0 & 0 & 1 \\
Z_{14} & 0 & 0 & 0 & 1 & 1 & 0 & 0 & 0 & 1 & 0 \\
Z_{23} & 0 & 0 & 0 & 1 & 1 & 0 & 0 & 0 & 0 & 1 \\
Z_{32} &  0 & 0 & 0 & 1 & 0 & 1 & 0 & 0 & 1 & 0 \\
Z_{41} & 0 & 0 & 0 & 1 & 0 & 1 & 0 & 0 & 0 & 1 \\
\end{array}
\right)
~.~
\eea
The GLSM charge matrices become
\beal{es116e5}
Q_{JE}
&=&
\left(
\begin{array}{cccccccccc}
p_1 & p_2 & p_3 & p_4 & q_1 & q_2 & r_1 & r_2 & s_1 & s_2 \\
\hline
 1 & 0 & 0 & 1 & 0 & 0 & 1 & 1 & -1 & -1 \\
 0 & 0 & 0 & 0 & 1 & 1 & 1 & 1 & -1 & -1 \\
 0 & 1 & 1 & 0 & 0 & 0 & -1 & -1 & 0 & 0 \\
\end{array}
\right)
~,~
\nn\\
Q_D
&=&
\left(
\begin{array}{cccccccccc}
p_1 & p_2 & p_3 & p_4 & q_1 & q_2 & r_1 & r_2 & s_1 & s_2 \\
\hline
 0 & 0 & 0 & 0 & 0 & -1 & -1 & 0 & 0 & 1 \\
 0 & 0 & 0 & 0 & 0 & 1 & 0 & 1 & 0 & -1 \\
 0 & 0 & 0 & 0 & 0 & 1 & 1 & 0 & -1 & 0 \\
\end{array}
\right)
~.~
\eea
From them we obtain
\beal{es116e6}
G=
\left(
\begin{array}{cccccccccc}
p_1 & p_2 & p_3 & p_4 & q_1 & q_2 & r_1 & r_2 & s_1 & s_2 \\
\hline
 1 & 1 & 1 & 1 & 1 & 1 & 1 & 1 & 2 & 2 \\
 0 & 0 & 0 & 2 & 1 & 1 & 0 & 0 & 1 & 1 \\
 0 & 0 & 2 & 0 & 0 & 0 & 1 & 1 & 1 & 1 \\
 0 & 1 & 1 & 0 & 0 & 0 & 1 & 1 & 1 & 1 \\
\end{array}
\right)
~.~
\eea
It can be verified that $s_1$ and $s_2$ are extra GLSM fields. The remaining GLSM fields correspond to points on a 3-dimensional hyperplane and give rise to the expected toric diagram as illustrated in \fref{ftoricZ2Z210011111}.

\bigskip

\subsection{$\mathbb{C}^4/\mathbb{Z}_{2}\times\mathbb{Z}_{2}\times\mathbb{Z}_{2}$ $(1,0,0,1)(0,1,0,1)(0,0,1,1)$ \label{sc4z2z2z2}}

\begin{figure}[ht!]
\begin{center}
\resizebox{0.5\hsize}{!}{
\includegraphics[trim=0cm 0cm 0cm 0cm,totalheight=10 cm]{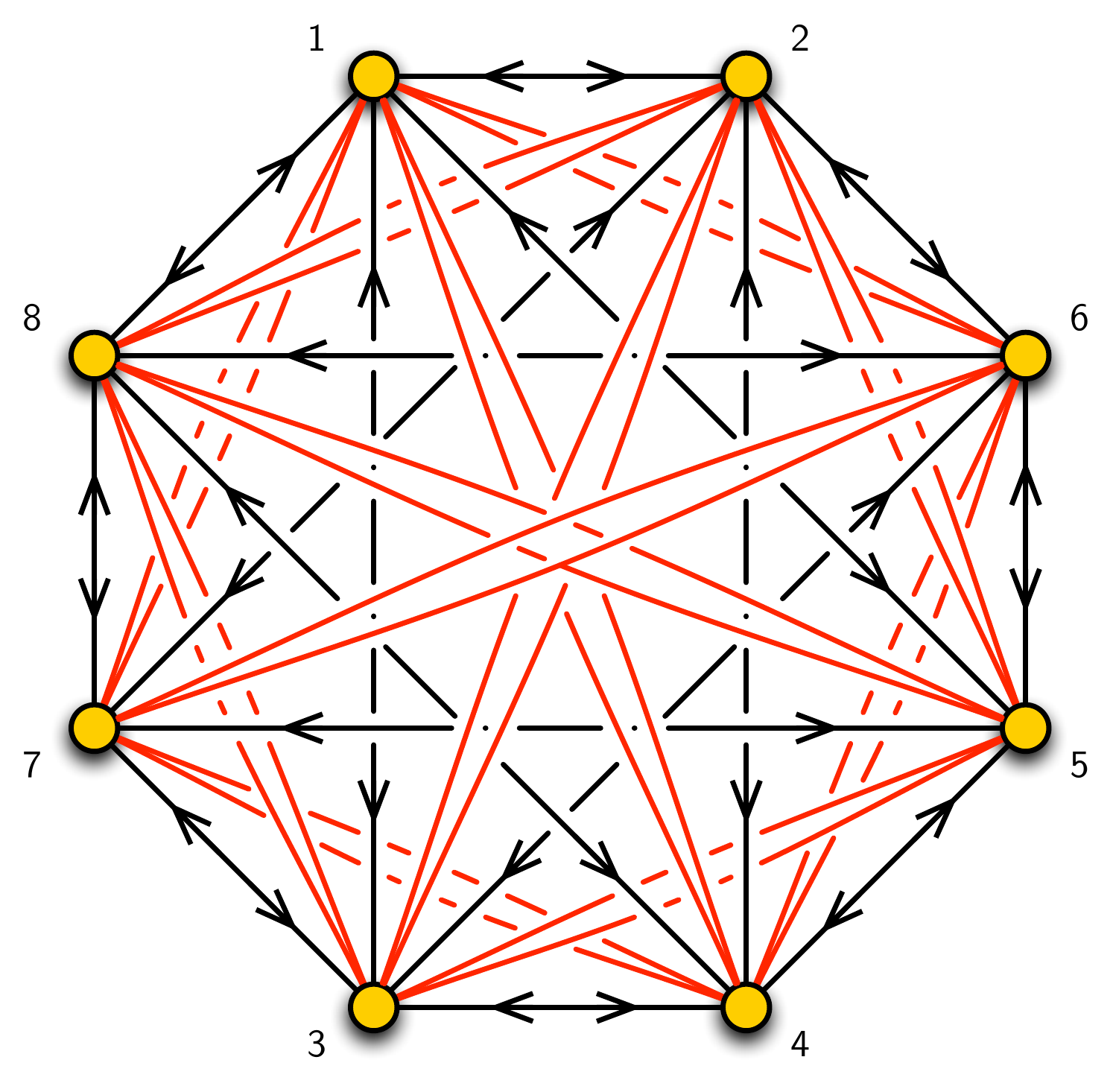}
}  
\caption{
Quiver diagram for $\mathbb{C}^4/\mathbb{Z}_2\times\mathbb{Z}_2\times \mathbb{Z}_2$ $(1,0,0,1)(0,1,0,1)(0,0,1,1)$.
\label{fc4z2z2z2quiver}}
 \end{center}
 \end{figure}

The quiver diagram for $\mathbb{C}^4/\mathbb{Z}_{2}\times\mathbb{Z}_{2}\times\mathbb{Z}_{2}$ $(1,0,0,1)(0,1,0,1)(0,0,1,1)$ is shown in \fref{fc4z2z2z2quiver}.
The $J$- and $E$-terms are
\beq
\begin{array}{rclccclcc}
& & \ \ \ \ \ \ \ \ \ \ \ \ \ \ \ J  & & & & \ \ \ \ \ \ \ \ \ \ \ \ \ E & & \\
 \Lambda_{14} : & \ \ \ & Y_{43}\cdot  Z_{31} - Z_{42}\cdot  Y_{21} & = & 0 & \ \ \ \ & D_{18}\cdot  X_{84} - X_{15}\cdot  D_{54} & = & 0 \\
 \Lambda_{41} : & \ \ \ & Y_{12}\cdot  Z_{24} - Z_{13}\cdot  Y_{34} & = & 0 & \ \ \ \ & D_{45}\cdot  X_{51} - X_{48}\cdot  D_{81} & = & 0 \\
 \Lambda_{16} : & \ \ \ & X_{62}\cdot  Y_{21} - Y_{65}\cdot  X_{51} & = & 0 & \ \ \ \ & D_{18}\cdot  Z_{86} - Z_{13}\cdot  D_{36} & = & 0 \\
 \Lambda_{61} : & \ \ \ & X_{15}\cdot  Y_{56} - Y_{12}\cdot  X_{26} & = & 0 & \ \ \ \ & D_{63}\cdot  Z_{31} - Z_{68}\cdot  D_{81} & = & 0 \\
 \Lambda_{17} : & \ \ \ &  X_{73}\cdot  Z_{31} - Z_{75}\cdot  X_{51} & = & 0 & \ \ \ \ & D_{18}\cdot  Y_{87} - Y_{12}\cdot  D_{27} & = & 0 \\
 \Lambda_{71} : & \ \ \ &  X_{15}\cdot  Z_{57} - Z_{13}\cdot  X_{37} & = & 0 & \ \ \ \ & D_{72}\cdot  Y_{21} - Y_{78}\cdot  D_{81} & = & 0 \\
 \Lambda_{23} : & \ \ \ & Y_{34}\cdot  Z_{42} - Z_{31}\cdot  Y_{12} & = & 0 & \ \ \ \ & D_{27}\cdot  X_{73} - X_{26}\cdot  D_{63} & = & 0 \\
 \Lambda_{32} : & \ \ \ & Y_{21}\cdot  Z_{13} - Z_{24}\cdot  Y_{43} & = & 0 & \ \ \ \ & D_{36}\cdot  X_{62} - X_{37}\cdot  D_{72} & = & 0 \\
 \Lambda_{25} : & \ \ \ & X_{51}\cdot  Y_{12} - Y_{56}\cdot  X_{62} & = & 0 & \ \ \ \ & D_{27}\cdot  Z_{75} - Z_{24}\cdot  D_{45} & = & 0 \\
 \Lambda_{52} : & \ \ \ & X_{26}\cdot  Y_{65} - Y_{21}\cdot  X_{15} & = & 0 & \ \ \ \ & D_{54}\cdot  Z_{42} - Z_{57}\cdot  D_{72} & = & 0 \\
 \Lambda_{28} : & \ \ \ &  X_{84}\cdot  Z_{42} - Z_{86}\cdot  X_{62} & = & 0 & \ \ \ \ & D_{27}\cdot  Y_{78} - Y_{21}\cdot  D_{18} & = & 0 \\
 \Lambda_{82} : & \ \ \ &  X_{26}\cdot  Z_{68} - Z_{24}\cdot  X_{48} & = & 0 & \ \ \ \ & D_{81}\cdot  Y_{12} - Y_{87}\cdot  D_{72} & = & 0 \\
 \Lambda_{35} : & \ \ \ &  X_{51}\cdot  Z_{13} - Z_{57}\cdot  X_{73} & = & 0 & \ \ \ \ & D_{36}\cdot  Y_{65} - Y_{34}\cdot  D_{45} & = & 0 \\
 \Lambda_{53} : & \ \ \ &  X_{37}\cdot  Z_{75} - Z_{31}\cdot  X_{15} & = & 0 & \ \ \ \ & D_{54}\cdot  Y_{43} - Y_{56}\cdot  D_{63} & = & 0 \\
 \Lambda_{38} : & \ \ \ & X_{84}\cdot  Y_{43} - Y_{87}\cdot  X_{73} & = & 0 & \ \ \ \ & D_{36}\cdot  Z_{68} - Z_{31}\cdot  D_{18} & = & 0 \\
 \Lambda_{83} : & \ \ \ & X_{37}\cdot  Y_{78} - Y_{34}\cdot  X_{48} & = & 0 & \ \ \ \ & D_{81}\cdot  Z_{13} - Z_{86}\cdot  D_{63} & = & 0 \\
 \Lambda_{46} : & \ \ \ &  X_{62}\cdot  Z_{24} - Z_{68}\cdot  X_{84} & = & 0 & \ \ \ \ & D_{45}\cdot  Y_{56} - Y_{43}\cdot  D_{36} & = & 0 \\
 \Lambda_{64} : & \ \ \ &  X_{48}\cdot  Z_{86} - Z_{42}\cdot  X_{26} & = & 0 & \ \ \ \ & D_{63}\cdot  Y_{34} - Y_{65}\cdot  D_{54} & = & 0 \\
 \Lambda_{47} : & \ \ \ & X_{73}\cdot  Y_{34} - Y_{78}\cdot  X_{84} & = & 0 & \ \ \ \ & D_{45}\cdot  Z_{57} - Z_{42}\cdot  D_{27} & = & 0 \\
 \Lambda_{74} : & \ \ \ & X_{48}\cdot  Y_{87} - Y_{43}\cdot  X_{37} & = & 0 & \ \ \ \ & D_{72}\cdot  Z_{24} - Z_{75}\cdot  D_{54} & = & 0 \\
 \Lambda_{58} : & \ \ \ & Y_{87}\cdot  Z_{75} - Z_{86}\cdot  Y_{65} & = & 0 & \ \ \ \ & D_{54}\cdot  X_{48} - X_{51}\cdot  D_{18} & = & 0 \\
 \Lambda_{85} : & \ \ \ & Y_{56}\cdot  Z_{68} - Z_{57}\cdot  Y_{78} & = & 0 & \ \ \ \ & D_{81}\cdot  X_{15} - X_{84}\cdot  D_{45} & = & 0 \\
 \Lambda_{67} : & \ \ \ & Y_{78}\cdot  Z_{86} - Z_{75}\cdot  Y_{56} & = & 0 & \ \ \ \ & D_{63}\cdot  X_{37} - X_{62}\cdot  D_{27} & = & 0 \\
 \Lambda_{76} : & \ \ \ & Y_{65}\cdot  Z_{57} - Z_{68}\cdot  Y_{87} & = & 0 & \ \ \ \ & D_{72}\cdot  X_{26} - X_{73}\cdot  D_{36} & = & 0 
 \end{array}
 \label{es117a1}
\eeq

\begin{figure}[ht!!]
\begin{center}
\resizebox{0.5\hsize}{!}{
\includegraphics[trim=0cm 0cm 0cm 0cm,totalheight=10 cm]{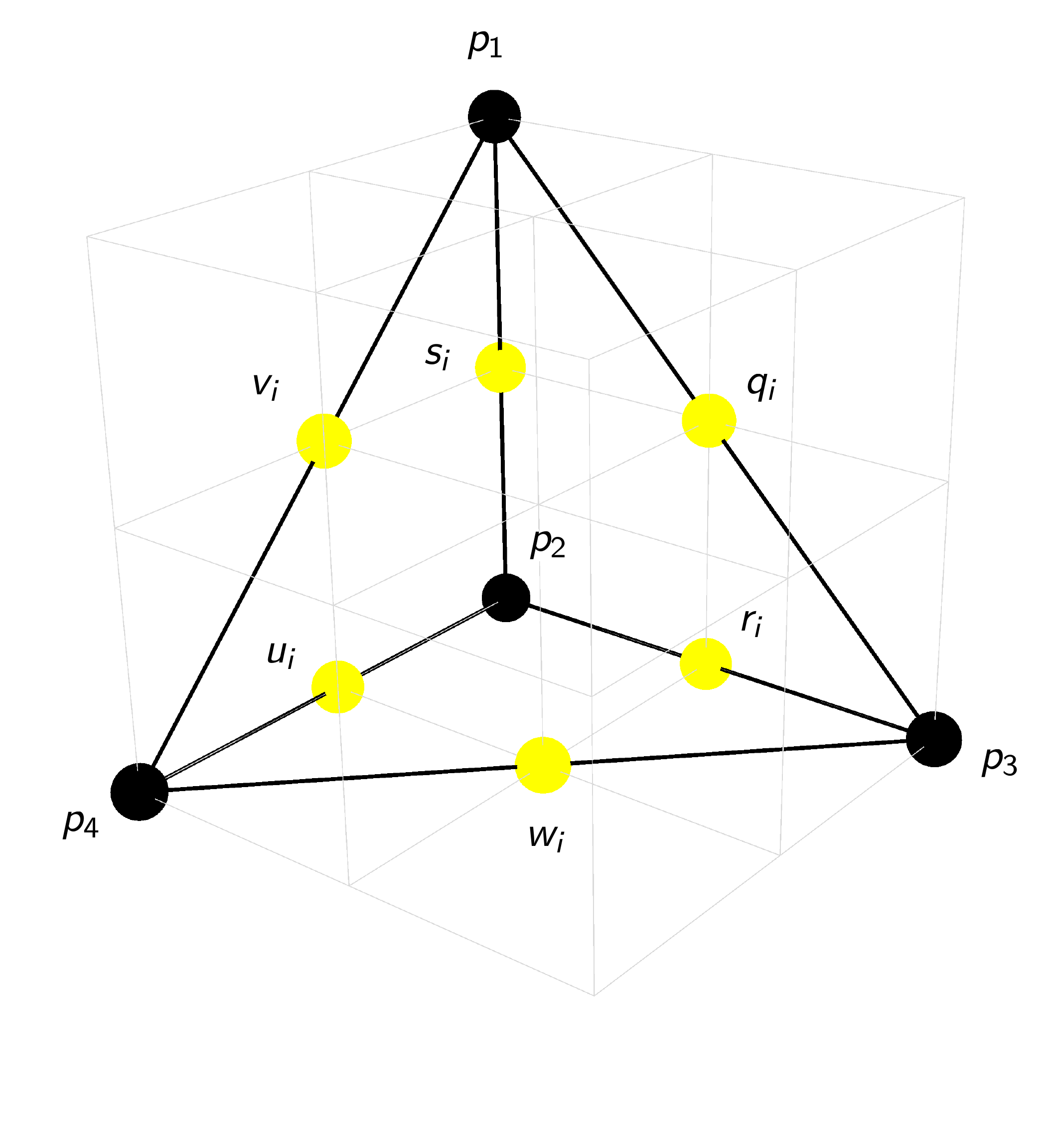}
}  
\caption{
Toric diagram for $\mathbb{C}^4/\mathbb{Z}_2\times\mathbb{Z}_2\times \mathbb{Z}_2$ $(1,0,0,1)(0,1,0,1)(0,0,1,1)$, obtained after removing the extra GLSM fields $e_i$.
\label{fc4z2z2z2toric}}
 \end{center}
 \end{figure}

Using the forward algorithm, the $P$-matrix can be found to be
{\tiny
\beal{es117aaa2}
&&
P=
\nn\\
&&
\left(
\begin{array}{c|cccccccccccccccccccccccccccccccccc}
\; & p_1 & p_2 & p_3 & p_4 & q_1 & q_2 & r_1 & r_2 & s_1 & s_2 & u_1 & u_2 & v_1 & v_2 & w_1 & w_2 & e_1 & e_2 & e_3 & e_4 & e_5 & e_6 & e_7 & e_8 & e_9 & e_{10} & e_{11} & e_{12} & e_{13} & e_{14} & e_{15} & e_{16} & e_{17} & e_{18}
\\
\hline
X_{15} & 1 & 0 & 0 & 0 & 1 & 0 & 0 & 0 & 0 & 1 & 0 & 0 & 0 & 1 & 0 & 0 & 0 & 0 & 0 & 0 & 0 & 0 & 0 & 1 & 1 & 0 & 0 & 1 & 1 & 1 & 1 & 1 & 1 & 1 \\
X_{51} & 1 & 0 & 0 & 0 & 0 & 1 & 0 & 0 & 1 & 0 & 0 & 0 & 1 & 0 & 0 & 0 & 1 & 1 & 1 & 1 & 1 & 1 & 1 & 0 & 0 & 1 & 1 & 0 & 0 & 0 & 0 & 0 & 0 & 0 \\
X_{26} & 1 & 0 & 0 & 0 & 1 & 0 & 0 & 0 & 1 & 0 & 0 & 0 & 0 & 1 & 0 & 0 & 1 & 1 & 1 & 1 & 0 & 0 & 0 & 1 & 0 & 1 & 0 & 1 & 1 & 1 & 0 & 0 & 0 & 0 \\
X_{62} & 1 & 0 & 0 & 0 & 0 & 1 & 0 & 0 & 0 & 1 & 0 & 0 & 1 & 0 & 0 & 0 & 0 & 0 & 0 & 0 & 1 & 1 & 1 & 0 & 1 & 0 & 1 & 0 & 0 & 0 & 1 & 1 & 1 & 1 \\
X_{37} & 1 & 0 & 0 & 0 & 0 & 1 & 0 & 0 & 0 & 1 & 0 & 0 & 0 & 1 & 0 & 0 & 1 & 1 & 0 & 0 & 1 & 1 & 0 & 1 & 0 & 1 & 0 & 1 & 0 & 0 & 1 & 1 & 0 & 0 \\
X_{73} & 1 & 0 & 0 & 0 & 1 & 0 & 0 & 0 & 1 & 0 & 0 & 0 & 1 & 0 & 0 & 0 & 0 & 0 & 1 & 1 & 0 & 0 & 1 & 0 & 1 & 0 & 1 & 0 & 1 & 1 & 0 & 0 & 1 & 1 \\
X_{48} & 1 & 0 & 0 & 0 & 0 & 1 & 0 & 0 & 1 & 0 & 0 & 0 & 0 & 1 & 0 & 0 & 1 & 0 & 0 & 0 & 0 & 0 & 1 & 0 & 0 & 1 & 1 & 0 & 1 & 1 & 1 & 1 & 1 & 0 \\
X_{84} & 1 & 0 & 0 & 0 & 1 & 0 & 0 & 0 & 0 & 1 & 0 & 0 & 1 & 0 & 0 & 0 & 0 & 1 & 1 & 1 & 1 & 1 & 0 & 1 & 1 & 0 & 0 & 1 & 0 & 0 & 0 & 0 & 0 & 1 \\
Y_{12} & 0 & 1 & 0 & 0 & 0 & 0 & 0 & 1 & 0 & 1 & 0 & 1 & 0 & 0 & 0 & 0 & 0 & 0 & 0 & 0 & 0 & 1 & 0 & 0 & 1 & 0 & 1 & 1 & 0 & 1 & 1 & 1 & 1 & 1 \\
Y_{21} & 0 & 1 & 0 & 0 & 0 & 0 & 1 & 0 & 1 & 0 & 1 & 0 & 0 & 0 & 0 & 0 & 1 & 1 & 1 & 1 & 1 & 0 & 1 & 1 & 0 & 1 & 0 & 0 & 1 & 0 & 0 & 0 & 0 & 0 \\
Y_{34} & 0 & 1 & 0 & 0 & 0 & 0 & 1 & 0 & 0 & 1 & 0 & 1 & 0 & 0 & 0 & 0 & 0 & 1 & 0 & 1 & 1 & 1 & 0 & 1 & 0 & 1 & 0 & 1 & 0 & 0 & 0 & 1 & 0 & 1 \\
Y_{43} & 0 & 1 & 0 & 0 & 0 & 0 & 0 & 1 & 1 & 0 & 1 & 0 & 0 & 0 & 0 & 0 & 1 & 0 & 1 & 0 & 0 & 0 & 1 & 0 & 1 & 0 & 1 & 0 & 1 & 1 & 1 & 0 & 1 & 0 \\
Y_{56} & 0 & 1 & 0 & 0 & 0 & 0 & 0 & 1 & 1 & 0 & 0 & 1 & 0 & 0 & 0 & 0 & 1 & 1 & 1 & 1 & 0 & 1 & 0 & 0 & 0 & 1 & 1 & 1 & 0 & 1 & 0 & 0 & 0 & 0 \\
Y_{65} & 0 & 1 & 0 & 0 & 0 & 0 & 1 & 0 & 0 & 1 & 1 & 0 & 0 & 0 & 0 & 0 & 0 & 0 & 0 & 0 & 1 & 0 & 1 & 1 & 1 & 0 & 0 & 0 & 1 & 0 & 1 & 1 & 1 & 1 \\
Y_{78} & 0 & 1 & 0 & 0 & 0 & 0 & 1 & 0 & 1 & 0 & 0 & 1 & 0 & 0 & 0 & 0 & 0 & 0 & 0 & 1 & 0 & 0 & 1 & 0 & 0 & 1 & 1 & 0 & 1 & 1 & 0 & 1 & 1 & 1 \\
Y_{87} & 0 & 1 & 0 & 0 & 0 & 0 & 0 & 1 & 0 & 1 & 1 & 0 & 0 & 0 & 0 & 0 & 1 & 1 & 1 & 0 & 1 & 1 & 0 & 1 & 1 & 0 & 0 & 1 & 0 & 0 & 1 & 0 & 0 & 0 \\
Z_{13} & 0 & 0 & 1 & 0 & 1 & 0 & 0 & 1 & 0 & 0 & 0 & 0 & 0 & 0 & 0 & 1 & 0 & 0 & 1 & 0 & 0 & 0 & 0 & 0 & 1 & 0 & 1 & 1 & 1 & 1 & 1 & 0 & 1 & 1 \\
Z_{31} & 0 & 0 & 1 & 0 & 0 & 1 & 1 & 0 & 0 & 0 & 0 & 0 & 0 & 0 & 1 & 0 & 1 & 1 & 0 & 1 & 1 & 1 & 1 & 1 & 0 & 1 & 0 & 0 & 0 & 0 & 0 & 1 & 0 & 0 \\
Z_{24} & 0 & 0 & 1 & 0 & 1 & 0 & 1 & 0 & 0 & 0 & 0 & 0 & 0 & 0 & 0 & 1 & 0 & 1 & 1 & 1 & 1 & 0 & 0 & 1 & 0 & 1 & 0 & 1 & 1 & 0 & 0 & 0 & 0 & 1 \\
Z_{42} & 0 & 0 & 1 & 0 & 0 & 1 & 0 & 1 & 0 & 0 & 0 & 0 & 0 & 0 & 1 & 0 & 1 & 0 & 0 & 0 & 0 & 1 & 1 & 0 & 1 & 0 & 1 & 0 & 0 & 1 & 1 & 1 & 1 & 0 \\
Z_{57} & 0 & 0 & 1 & 0 & 0 & 1 & 0 & 1 & 0 & 0 & 0 & 0 & 0 & 0 & 0 & 1 & 1 & 1 & 1 & 0 & 1 & 1 & 0 & 0 & 0 & 1 & 1 & 1 & 0 & 0 & 1 & 0 & 0 & 0 \\
Z_{75} & 0 & 0 & 1 & 0 & 1 & 0 & 1 & 0 & 0 & 0 & 0 & 0 & 0 & 0 & 1 & 0 & 0 & 0 & 0 & 1 & 0 & 0 & 1 & 1 & 1 & 0 & 0 & 0 & 1 & 1 & 0 & 1 & 1 & 1 \\
Z_{68} & 0 & 0 & 1 & 0 & 0 & 1 & 1 & 0 & 0 & 0 & 0 & 0 & 0 & 0 & 0 & 1 & 0 & 0 & 0 & 0 & 1 & 0 & 1 & 0 & 0 & 1 & 1 & 0 & 1 & 0 & 1 & 1 & 1 & 1 \\
Z_{86} & 0 & 0 & 1 & 0 & 1 & 0 & 0 & 1 & 0 & 0 & 0 & 0 & 0 & 0 & 1 & 0 & 1 & 1 & 1 & 1 & 0 & 1 & 0 & 1 & 1 & 0 & 0 & 1 & 0 & 1 & 0 & 0 & 0 & 0 \\
D_{18} & 0 & 0 & 0 & 1 & 0 & 0 & 0 & 0 & 0 & 0 & 0 & 1 & 0 & 1 & 0 & 1 & 0 & 0 & 0 & 0 & 0 & 0 & 0 & 0 & 0 & 1 & 1 & 1 & 1 & 1 & 1 & 1 & 1 & 1 \\
D_{18} & 0 & 0 & 0 & 1 & 0 & 0 & 0 & 0 & 0 & 0 & 1 & 0 & 1 & 0 & 1 & 0 & 1 & 1 & 1 & 1 & 1 & 1 & 1 & 1 & 1 & 0 & 0 & 0 & 0 & 0 & 0 & 0 & 0 & 0 \\
D_{27} & 0 & 0 & 0 & 1 & 0 & 0 & 0 & 0 & 0 & 0 & 1 & 0 & 0 & 1 & 0 & 1 & 1 & 1 & 1 & 0 & 1 & 0 & 0 & 1 & 0 & 1 & 0 & 1 & 1 & 0 & 1 & 0 & 0 & 0 \\
D_{72} & 0 & 0 & 0 & 1 & 0 & 0 & 0 & 0 & 0 & 0 & 0 & 1 & 1 & 0 & 1 & 0 & 0 & 0 & 0 & 1 & 0 & 1 & 1 & 0 & 1 & 0 & 1 & 0 & 0 & 1 & 0 & 1 & 1 & 1 \\
D_{36} & 0 & 0 & 0 & 1 & 0 & 0 & 0 & 0 & 0 & 0 & 0 & 1 & 0 & 1 & 1 & 0 & 1 & 1 & 0 & 1 & 0 & 1 & 0 & 1 & 0 & 1 & 0 & 1 & 0 & 1 & 0 & 1 & 0 & 0 \\
D_{63} & 0 & 0 & 0 & 1 & 0 & 0 & 0 & 0 & 0 & 0 & 1 & 0 & 1 & 0 & 0 & 1 & 0 & 0 & 1 & 0 & 1 & 0 & 1 & 0 & 1 & 0 & 1 & 0 & 1 & 0 & 1 & 0 & 1 & 1 \\
D_{45} & 0 & 0 & 0 & 1 & 0 & 0 & 0 & 0 & 0 & 0 & 1 & 0 & 0 & 1 & 1 & 0 & 1 & 0 & 0 & 0 & 0 & 0 & 1 & 1 & 1 & 0 & 0 & 0 & 1 & 1 & 1 & 1 & 1 & 0 \\
D_{54} & 0 & 0 & 0 & 1 & 0 & 0 & 0 & 0 & 0 & 0 & 0 & 1 & 1 & 0 & 0 & 1 & 0 & 1 & 1 & 1 & 1 & 1 & 0 & 0 & 0 & 1 & 1 & 1 & 0 & 0 & 0 & 0 & 0 & 1 \\
\end{array}
\right)~.~
\nn\\
\eea}
Following further the forward algorithm, we obtain
{\tiny
\beal{es117aaa3}
G=
\left(
\begin{array}{cccccccccccccccccccccccccccccccccc}
p_1 & p_2 & p_3 & p_4 & q_1 & q_2 & r_1 & r_2 & s_1 & s_2 & u_1 & u_2 & v_1 & v_2 & w_1 & w_2 & e_1 & e_2 & e_3 & e_4 & e_5 & e_6 & e_7 & e_8 & e_9 & e_{10} & e_{11} & e_{12} & e_{13} & e_{14} & e_{15} & e_{16} & e_{17} & e_{18}
\\
\hline
1 & 1 & 1 & 1 & 1 & 1 & 1 & 1 & 1 & 1 & 1 & 1 & 1 & 1 & 1 & 1 & 2 & 2 & 2 & 2 & 2 & 2 & 2 & 2 &
  2 & 2 & 2 & 2 & 2 & 2 & 2 & 2 & 2 & 2 \\
0 & 0 & 0 & 2 & 0 & 0 & 0 & 0 & 0 & 0 & 1 & 1 & 1 & 1 & 1 & 1 & 1 & 1 & 1 & 1 & 1 & 1 & 1 & 1 &
  1 & 1 & 1 & 1 & 1 & 1 & 1 & 1 & 1 & 1 \\
1 & 1 & -1 & 1 & 0 & 0 & 0 & 0 & 1 & 1 & 1 & 1 & 1 & 1 & 0 & 0 & 1 & 1 & 1 & 1 & 1 & 1 & 1 & 1 &
  1 & 1 & 1 & 1 & 1 & 1 & 1 & 1 & 1 & 1 \\
-1 & 1 & 1 & 1 & 0 & 0 & 1 & 1 & 0 & 0 & 1 & 1 & 0 & 0 & 1 & 1 & 1 & 1 & 1 & 1 & 1 & 1 & 1 & 1 &
  1 & 1 & 1 & 1 & 1 & 1 & 1 & 1 & 1 & 1 \\
\end{array}
\right)~,~
\nn\\
\eea}
The GLSM fields $e_i$ are extra and do not play a defining role for the geometry. The remaining GLSM fields give rise to the expected toric diagram, as shown in \fref{fc4z2z2z2toric}.

\newpage

\bibliographystyle{JHEP}
\bibliography{mybib}


\end{document}